\documentclass[journal,table]{IEEEtran}

\usepackage{amsmath, amsthm, amssymb, bm}
\usepackage{epsfig}
\usepackage{amsfonts}
\usepackage{enumerate}

\usepackage{amsmath}
\usepackage{color}

\usepackage{cite}
\usepackage{epstopdf}

\newtheorem{theorem}{Theorem}

\newtheorem{proposition}{Proposition}

\newtheorem{lemma}{Lemma}

\newtheorem{assumption}{Assumption}

\usepackage{subfigure}
\usepackage[autostyle]{csquotes}

\usepackage{hyperref}

\makeatletter
\newcommand*\rel@kern[1]{\kern#1\dimexpr\macc@kerna}
\newcommand*\widebar[1]{%
  \begingroup
  \def\mathaccent##1##2{%
    \rel@kern{0.8}%
    \overline{\rel@kern{-0.8}\macc@nucleus\rel@kern{0.2}}%
    \rel@kern{-0.2}%
  }%
  \macc@depth\@ne
  \let\math@bgroup\@empty \let\math@egroup\macc@set@skewchar
  \mathsurround\z@ \frozen@everymath{\mathgroup\macc@group\relax}%
  \macc@set@skewchar\relax
  \let\mathaccentV\macc@nested@a
  \macc@nested@a\relax111{#1}%
  \endgroup
}
\makeatother

\newlength\figureheight
\newlength\figurewidth

\long\def\comment#1{}

\newfont{\bbb}{msbm10 scaled 700}

\newfont{\bb}{msbm10 scaled 1100}

\newcommand{\av}{{\bf a}}

\newcommand{\cv}{{\bf c}}
\newcommand{\dv}{{\bf d}}

\newcommand{\fv}{{\bf f}}
\newcommand{\gv}{{\bf g}}
\newcommand{\hv}{{\bf h}}

\newcommand{\jv}{{\bf j}}
\newcommand{\kv}{{\bf k}}

\newcommand{\nv}{{\bf n}}

\newcommand{\qv}{{\bf q}}
\newcommand{\rv}{{\bf r}}
\newcommand{\sv}{{\bf s}}
\newcommand{\tv}{{\bf t}}
\newcommand{\uv}{{\bf u}}
\newcommand{\wv}{{\bf w}}
\newcommand{\vv}{{\bf v}}
\newcommand{\xv}{{\bf x}}
\newcommand{\yv}{{\bf y}}
\newcommand{\zv}{{\bf z}}

\newcommand{\Am}{{\bf A}}
\newcommand{\Bm}{{\bf B}}
\newcommand{\Cm}{{\bf C}}
\newcommand{\Dm}{{\bf D}}
\newcommand{\Em}{{\bf E}}
\newcommand{\Fm}{{\bf F}}
\newcommand{\Gm}{{\bf G}}
\newcommand{\Hm}{{\bf H}}
\newcommand{\Id}{{\bf I}}
\newcommand{\Jm}{{\bf J}}

\newcommand{\Lm}{{\bf L}}
\newcommand{\Mm}{{\bf M}}

\newcommand{\Om}{{\bf O}}

\newcommand{\Qm}{{\bf Q}}
\newcommand{\Rm}{{\bf R}}
\newcommand{\Sm}{{\bf S}}
\newcommand{\Tm}{{\bf T}}
\newcommand{\Um}{{\bf U}}
\newcommand{\Wm}{{\bf W}}

\newcommand{\Zm}{{\bf Z}}

\newcommand{\alphav}{\hbox{\boldmath$\alpha$}}
\newcommand{\betav}{\hbox{\boldmath$\beta$}}
\newcommand{\gammav}{\hbox{\boldmath$\gamma$}}

\newcommand{\lambdav}{\hbox{\boldmath$\lambda$}}

\newcommand{\Thetam}{\hbox{\boldmath$\Theta$}}

\begin{document}

\title{\vspace{-25pt} Blind Two-Dimensional Super-Resolution and Its Performance Guarantee (Extended Version)}
%
% Blind two dimensional super-resolution and its performance guarantee,

\author{ Mohamed~A.~Suliman and Wei Dai% <-this % stops a space

\thanks{This work is supported by ONRG (Award N62909-16-1-2051).

M. A. Suliman and W. Dai are with the Department of Electrical and Electronic Engineering, Imperial College London, London, SW7 2AZ, United Kingdom. Emails:$\{$m.suliman17, wei,dai1$\}$@imperial.ac.uk.  Part of this work is presented at the IEEE International Conference on Acoustics, Speech, and Signal Processing (ICASSP), Brighton, UK, May 2019 \cite{suliman2019blind}.}}

\maketitle

\begin{abstract}
We study the problem of identifying the parameters of a linear system from its response to multiple \emph{unknown} waveforms. We assume that the system response is a scaled superposition of time-delayed and frequency-shifted versions of the unknown waveforms. Such kind of problem is severely ill-posed and does not yield a unique solution without introducing further constraints. To fully characterize the system, we assume that the unknown waveforms lie in a common known low-dimensional subspace that satisfies certain properties. Then, we develop a blind two-dimensional (2D) super-resolution framework that applies to a large number of applications. In this framework, we show that under a minimum separation between the time-frequency shifts, all the unknowns that characterize the system can be recovered precisely and with high probability provided that a lower bound on the number of the observed samples is satisfied. The proposed framework is based on a 2D atomic norm minimization problem, which is shown to be reformulated and solved via semidefinite programming. Simulation results that confirm the theoretical findings of the paper are provided.
\end{abstract}

\section{Introduction}
\label{sec:intro}
\subsection{Background}
Throughout the years, researchers have paid close attention to acquire various ways of breaking the physical limits in sensing systems with the aim of enhancing their resolution. Generally speaking, super-resolution techniques are those mechanisms that address the problem of recovering high-resolution information from coarse-scale data. Interests in such field come from the fact that those techniques afford colossal performance improvement in many applications such as radar imaging \cite{mi2008radar}, medical imaging \cite{kennedy2007improved}, microscopy \cite{mccutchen1967superresolution}, astronomy \cite{puschmann2005super}, and communication systems, to mention a few.

In this paper, we study the problem of identifying the parameters of a linear system from its response to multiple \emph{unknown signals}. More precisely, we consider a continuous-time linear system in which the observed signal $y\left(t\right)$ is a weighted sum of $R$ different versions of time-delayed and frequency-shifted unknown signals $s_{j}\left(t\right), j=1,\dots, R$. i.e., 
\begin{equation}
\label{eq: model}
y \left(t\right)= \sum_{j=1}^{R} c_{j} s_{j}\left(t-\tilde{\tau}_{j}\right) e^{i2\pi \tilde{f}_{j}t}.
\end{equation}
Here, the unknown scaling factor $c_{j} \in \mathbb{C}$ has an amplitude $|c_{j}|>0$ and phase $[0,2\pi)$ while the pair $\left(\tilde{\tau}_{j}, \tilde{f}_{j}\right)$ represents the unknown \emph{continuous} time-frequency shift. Finally, we assume that both $R$ and $\{s_{j}\left(t\right)\}_{j=1}^{R}$ are unknown. Thus, our question is given the received signal $y\left(t\right)$ can we retrieve precisely the unknown quintuple $\left(R, c_{j}, \tilde{\tau}_{j}, \tilde{f}_{j}, s_{j}\left(t\right)\right)$.

{{The formulation in (\ref{eq: model}) arises in various signal and image processing and communication applications. In passive indoor source localization, as \cite{amar2004direct} shows, the transmitted waveforms $s_{j}\left(t\right)$ by the $R$ unknown moving objects are unknown. The locations of such objects are obtained by estimating $\tilde{\tau}_{j}$ and $\tilde{f_{j}})$, which refer to the time delay and Doppler shift, respectively. Following that, we integrate this information with the base station location to obtain the locations of the objects in a certain space \cite{amar2004direct}. Note that $\tilde{\tau}_{j}$ and $\tilde{f_{j}}$ can lie anywhere in a \emph{certain continuous domain} and are not constrained to be on a grid. The same also applies to the problem of parameter estimation in passive radar imaging, where the goal is to estimate distances (i.e., $\tilde{\tau}_{j}$) and velocities (i.e., $\tilde{f_{j}}$) of multiple targets relative to the radar. Furthermore, the exact formulation also appears in the problem of underwater acoustics localization, as shown in \cite{xerri2002passive}. Moreover, the work in \cite{rust2006sub} applies the formulation in (\ref{eq: model}) to the problem of blind super-resolution of a 2D point source in microscopy, whereas in \cite{chavanne2004target}, (\ref{eq: model}) is applied to formulate the problem of target detection using blind channel equalization.

On the other hand, the problem formulation in (\ref{eq: model}) is widely applied in different imaging-related problems such as medical imaging and astronomy, as shown in \cite{campisi2017blind, starck2002deconvolution}. In such applications, the goal is to estimate the unknown shifts $\left(\tilde{\tau}_{j}, \tilde{f}_{j}\right)$ as well as the coefficients $c_{j}$ from the observed signal $y\left(t\right)$, with $s_{j}\left(t\right)$ being the unknown point-spread functions (PSFs) of the system. Such functions are unknown for multiple reasons, such as the inaccuracy in the lens focus, due to the camera movement, or because they are space/time-varying functions. As another example, the model in (\ref{eq: model})  appears in the problem of blind calibration of uniform linear arrays where the goal is to calibrate the antennas' gains blindly, as shown in \cite{weiss1990eigenstructure}.}}

\subsection{Related Work}
The recent approach for super-resolution is based on the atomic norm minimization \cite{chandrasekaran2012convex} which provides a general convex framework to recover a set of data. The work shows that we can retrieve a set of frequencies, in the noiseless scenario, provided that a certain separation exits between them.

Cand\`es and Fernandez-Granda in \cite{candes2014towards} apply the framework in \cite{chandrasekaran2012convex} to super-resolve a number of locations in the continuous domain $[0,1]$ from equally spaced consecutive samples. Their work is groundbreaking, and abundant literature soon followed after for various settings. The result in \cite{candes2014towards} shows that we can recover, with infinite precision, the exact locations of multiple points by solving a convex total-variation (TV) norm minimization problem, which can be reformulated and solved via semidefinite programming (SDP). The work guarantees this exact recovery provided that a minimum separation between the points is satisfied. Related convex framework has been developed to recover unknown frequencies from noisy models in \cite{candes2013super}. On the other hand, the work in \cite{tang2013compressed} studies super-resolving a set of frequencies in $[0,1]$ from a randomly selected set of samples using the atomic norm. The work concludes that by randomly selecting a subset of the observed samples, that exact recovery of the frequencies is assured with high probability provided that they are well-separated. This work is extended in \cite{li2016off} for off-grid line spectrum denoising and estimation from multiple spectacularly-sparse signals. 

With all previous work being based on 1D super-resolution, some other work is also performed on multidimensional (MD) super-resolution. For instance, \cite{heckel2016super} studies super-resolving time-frequency shifts simultaneously in radar applications where the recovery problem is formulated as a 2D line spectrum estimation problem using the atomic norm. The received signal is modeled as a superposition of delayed and frequency-shifted versions of a single transmitted signal. Thus, it has the same formulation as in (\ref{eq: model}); however, as we will discuss later, the \emph{single} transmitted waveform in \cite{heckel2016super} is known with its samples being Gaussian. An SDP relaxation for the dual is then obtained using the results in \cite{xu2014precise, dumitrescu2017positive}. The exact recovery of the shifts is shown to exist, given that their number is linear with a log-factor in the observed samples' number.

Furthermore, the work in \cite{heckel2016supermimo} extends \cite{heckel2016super} to MIMO radars upon applying the same settings in \cite{heckel2016super}. The authors in \cite{bendory2015super} study super-resolving ensemble of Diracs on a sphere from their low-resolution measurements. The problem is formed as a 2D atomic norm minimization and then solved using the results in \cite{xu2014precise, dumitrescu2017positive}. {{On the other hand, the work in \cite{chi2014compressive} provides a 2D atomic norm super-resolution framework to estimate a set of 2D frequencies from a random subset of samples gathered from a mixture of 2D sinusoids. The work shows that all the unknown frequencies can be recovered under a minimum separation condition upon solving an atomic norm minimization problem. In contrast to our general framework, the recovery problem in \cite{chi2014compressive} is simplified by addressing the case where the observed data is assumed to be represented using two 2D square matrices.}} Finally,\cite{yang2016vandermonde} addresses the MD super-resolution with compressive measurements where an exact reformulation for the atomic norm recovery problem is obtained and solved using a proposed Vandermonde decomposition. We point out that all the above-mentioned theories address the non-blind case where the waveforms/PSFs are known.  
 
From another point of view, much work tackled the problem of blind deconvolution \cite{friedlander1988eigenstructure}, where an unknown sparse signal is assumed to be convolved with an unknown PSF. In general, blind deconvolution is an ill-posed problem that does not yield a unique solution without imposing further constraints \cite{li2017identifiability}. A survey on multichannel blind deconvolution methods in communications is provided in \cite{tong1998multichannel}, while a review about classical blind deconvolution methods is given in \cite{campisi2017blind}.

The authors in \cite{ahmed2014blind} develop an algorithm to blindly deconvolve two signals lying in known low-dimensional subspaces. The deconvolution problem is transformed into a low-rank matrix recovery problem by using the so-called lifting trick and then solved. This result is extended in \cite{ling2015self} by allowing one of the two signals to be sparse in a known dictionary and in \cite{lee2015stability} by assuming that both signals are sparse in a known dictionary. It should be noted that these works apply the $\ell_{1}$ norm minimization, which is different from ours, as we will discuss later. Finally, a convex optimization framework for estimating a single PSF and a spike signal is introduced in \cite{chi2016guaranteed}, where the PSF is assumed to lie in a known low-dimensional subspace. The work shows that recovering the spike signal is assured under mild randomness assumptions on the subspace and a separation condition on the spike signal. 

The authors in \cite{yang2016super} study the problem of estimating the parameters of complex exponentials from their modulations with unknown waveforms. To convert the ill-posed recovery problem into a well-posed one, the waveforms are assumed to lie in a known low-dimensional subspace. Then, an atomic norm convex framework is formulated to super-resolve the points and to recover the waveforms. The atomic norm minimization problem is reformulated and then solved efficiently via SDP. The work shows that when the number of the measurements is proportional to the degrees of freedom in the problem, the 1D blind super-resolution recovery problem provides exact recovery for the unknowns with high probability given that a minimum separation between the points exist.

\subsection{Contributions with Connections to Prior Art}
The contributions of this paper are as follows. First, we propose a general mathematical framework for blind 2D super-resolution that applies to a large number of applications. The blindness of this framework comes from the fact that the waveforms $\{s_{j}\left(t\right)\}_{j=1}^{R}$ are unknown while the \textquote{2D super-resolution} term is because we are super-resolving two continuous unknowns ($\tilde{\tau}_{j}$ and $\tilde{f_{j}}$) simultaneously. The superiority of this framework, as we will discuss later, is that most of the recent approaches in super-resolution can be shown as a special case of it. Since the recovery problem is severely ill-posed, and inspired by \cite{ahmed2014blind, chi2016guaranteed, yang2016super}, we assume that the unknown waveforms live in a common known low-dimensional subspace that satisfies certain conditions. Second, we show that with high probability, the unknown quintuple $\left(R, c_{j}, \tilde{\tau}_{j}, \tilde{f}_{j}, s_{j}\left(t\right)\right)$ in (\ref{eq: model}) can be recovered precisely from the samples of $y\left(t\right)$ upon using the atomic norm. The recovery problem is formulated as an atomic norm minimization problem and then reformulated and solved via SDP. The exact recovery of all the unknowns is guaranteed provided that the number of the observed samples satisfies certain lower bound, which is found to be of the same order as the number of unknowns. This bound is derived using random kernels under a minimum separation condition between the time-frequency shifts.

The work in this paper is inspired by the recent work in \cite{chi2016guaranteed, yang2016super, heckel2016super}. The model in \cite{heckel2016super} has the same formulation in (\ref{eq: model}); however, as opposed to what we have, \cite{heckel2016super} assumes a single known transmitted waveform. Furthermore, the samples of this waveform are assumed to be independent and identically distributed (i.i.d.) from a Gaussian distribution of zero-mean and a known variance. Moreover, the pioneering work in \cite{yang2016super} can be viewed as a special case of our framework based on (\ref{eq: model}). That can be upon assuming that either $s_{j}\left(t\right)$ or $\tilde{\tau}_{j}$ is known. Considering $s_{j}\left(t-\tilde{\tau}_{j}\right)$ as a single unknown makes the approach in \cite{yang2016super} fail to resolve the ambiguity between $s_{j}\left(t\right)$ and $\tilde{\tau}_{j}$ in its final solution. The fact that $s_{j}\left(t-\tilde{\tau}_{j}\right)$ has to be considered as two unknowns converts the super-resolution problem in \cite{yang2016super} from being a 1D estimation problem to a 2D one and makes most of the proof techniques and the performance guarantee conditions in \cite{yang2016super} invalid. Finally, \cite{chi2016guaranteed} is a special case of \cite{yang2016super} by assuming identical waveforms.

{{From a technical perspective, our proposed framework comes with significant mathematical differences and additional contributions to existing work in the literature. For example, to prove the existence of the solution of the 1D super-resolution problem in \cite{yang2016super}, a 1D polynomial is formulated using shifted versions of a \emph{single} kernel. Such formulation fails to be generalized beyond the 1D case. That comes from the fact that, for example, in our case, our 2D trigonometric vector polynomial has to satisfy multiple constraints that a single kernel cannot cover to represent them. As a result, we introduce using multiple kernel matrices. The formulation of those kernels is based on using probabilistic approaches and some matrix theories and is not merely based on solving a weighted least energy minimization problem as in \cite{yang2016super}. Such formulation is obtained in a way that should guarantee the existence of the solution to our super-resolution recovery problem, as we will show in Section~\ref{sec: polynomial}. In addition to that, our newly developed proof techniques also allow us to impose less restricted assumptions on the low-dimensional subspace those in \cite{chi2016guaranteed, yang2016super}, as we will discuss in Section~\ref{sec: recovery cind}. On the other hand, the non-blindness and the Gaussianity assumption imposed on the transmitted signal in \cite{heckel2016super} simplify the scalar polynomial formulation used to guarantee the existence of the super-resolution recovery problem solution and make most of their proof methodologies inapplicable for our case. Compared to that, our new proof methodology does not impose any assumption on the signal distribution.}}

\subsection{Paper Organization}
In Section~\ref{sec: model}, we discuss the system model, and we formulate our blind 2D super-resolution recovery problem. In Section~\ref{sec: recovery cind}, we present the main theorem of the paper, which provides sufficient conditions for the exact recovery of the unknowns, and we discuss its associated assumptions. In Section~\ref{sec: solution sec}, we study the dual formulation of our recovery problem, and we propose an SDP relaxation for it. Moreover, we show how the unknowns can be retrieved from the solution of the dual. Section~\ref{sec: results} is dedicated to validate the performance of our framework using extensive simulations. In Section~\ref{sec: polynomial}, we provide the proof of the theorem in Section~\ref{sec: recovery cind}. Concluding remarks and future work directions are given in Section~\ref{sec: conc}.
  
\subsection{Notations}
Boldface lower-case symbols are used for column vectors (i.e., $\sv$) and upper-case for matrices (i.e., $\Sm$). $[\sv]_{i}$ denotes the $i$-th element of $\sv$ while $[\Sm]_{(i,j)}$ indicates the element in the $(i,j)$ entry. $\left(\cdot\right)^{T}$, $\left(\cdot\right)^{H}$, $\text{Tr}\left(\cdot\right)$, and $\text{det}\left(\cdot\right)$ denote the transpose, the Hermitian, the trace, and the determinant, respectively. The notation $\Id_{\text{M}}$ denotes the $M \times M$ identity matrix while $\bm{0}$ refers to the zero matrix/vector. $\Sm \succeq \bm{0}$ signifies that $\Sm$ is a positive semidefinite (PSD) matrix. When we use a two-dimensional index for vectors or matrices such as $[\sv]_{((k,l),1)},\  k,l=-N,\dots, N$, we mean that $\sv = [s_{(-N,-N)}, s_{(-N,-N+1)}, \dots , s_{(-N,N)}, \dots\dots, s_{(N,N)}]^{T}$. Moreover, we refer to the Kronecker product by $\otimes$. The notation $||\cdot||_{2}$ designates the spectral norm for matrices and the Euclidean norm for vectors while $||\cdot||_{F}$ is for Frobenius norm. The infinity norm is denoted by $||\cdot||_{\infty}$. diag $\left(\sv\right)$ represents a diagonal matrix whose diagonal entries are the elements of $\sv$. $\left\langle\cdot,\cdot\right\rangle$ stands for the inner product whilst $\langle\cdot, \cdot\rangle_{\mathbb{R}}$ denotes the real inner product. The notation $\text{Re}\left[\cdot\right]$ stands for the real part of a scalar or a vector with the real parts of the entries of a vector. $\mathbb{E} [\cdot]$ denotes the expectation operator while $\text{Pr}[\cdot]$ indicates the probability of an event. The set of real numbers is denoted by $\mathbb{R}$ while that of the complex numbers is denoted by $\mathbb{C}$. For a given set $\mathcal{S}$, $|\mathcal{S}|$ is the cardinality of the set, i.e., the number of the elements. Finally, $C$, $C_{1}$, $C^{*}$, $C^{*}_{1}$, $\hat{C}$, $\bar{C}, \dots$ are used to denote {{fixed universal constants.}}

\section{System Model and Recovery Problem Formulation}
\label{sec: model}

In this section, we discuss our system model and its associated underlying assumptions. Then, we formulate our super-resolution problem using the atomic norm framework. 

%Consider a continuous-time linear system where the received signal is a weighted sum of $R$ different time-delayed and frequency-shifted versions of unknown signals $s_{j}\left(t\right)$, i.e.,
%\begin{equation}
%\label{eq: model}
%y \left(t\right)= \sum_{j=1}^{R} c_{j} s_{j}\left(t-\tilde{\tau}_{j}\right) e^{i2\pi \tilde{f}_{j}t}.
%\end{equation}
Our ambition in this paper is to fully characterize (\ref{eq: model}) by retrieving  $\left(R, c_{j}, \tilde{\tau}_{j}, \tilde{f}_{j}, s_{j}\left(t\right)\right)$ using the observed signal $y\left(t\right)$ over a certain period of time. For that, it is important to first address the principal assumptions on $s_{j}\left(t\right)$ and $y\left(t\right)$.

To start with, we assume that $\{s_{j}\left(t\right)\}_{j=1}^{R}$ are band-limited periodic signals with a bandwidth of $W$ and a period of $T$ and that $y\left(t\right)$ is observed over an interval of length $T$. Such assumptions are quite common in many applications such as wireless communication, array processing, and radar imaging. Based on that, we can assume that the time-frequency shifts $\left(\tilde{\tau}_{j},\tilde{f}_{j} \right)$ lie in the domain $\left(\left[-T/2, T/2\right], \left[-W/2,W/2\right]\right)$.  

Now, based on the $2WT$-Theorem, we can fully characterize $y\left(t\right)$ by sampling it at a rate of $1/W$ samples-per-second to gather a total of $L:=WT$ samples. For simplicity, we assume that $L$ is an odd number. By sampling (\ref{eq: model}) at a rate of $1/W$, then applying the discrete Fourier transform (DFT) and the inverse DFT (IDFT) to (\ref{eq: model}), we can easily show that the sampled version of $y\left(t\right)$, i.e., $y\left(p/W\right)$ can be written as
\begin{align}
\label{eq: model sampled}
&y\left(p\right):= y\left(p/W\right)=  \nonumber\\
&\frac{1}{L}\sum_{j=1}^{R} c_{j} \left( \sum_{k=-N}^{N} \hspace{-1pt}\left[\hspace{-1pt}\left(\sum_{l=-N}^{N} s_{j}\left(l\right) e^{\frac{-i2\pi k l}{L}}\right)\hspace{-1pt} e^{-i2\pi \tau_{j} k}\right] e^{\frac{i2\pi k p}{L}} \right)\times\nonumber\\
& e^{i2\pi f_{j} p},   \ \hspace{30pt} p=-N,\dots, N,  \    \hspace{5pt} N := \frac{L-1}{2},
\end{align}
where we set ${\tau_{j}} := \frac{\tilde{\tau}_{j}}{T}$ and ${f_{j}} := \frac{\tilde{f}_{j}}{W}$. Note that the samples $s_{j}\left(l\right)$ are now $L$ periodic and that based on $\tau_{j}$ and $f_{j}$ we have $\left(\tau_{j}, f_{j}\right) \in [-1/2,1/2]^{2}$. Due to the periodicity property, we can assume without loss of generality that $\left(\tau_{j}, f_{j}\right) \in [0,1]^{2}$. In this paper, we refer to $\left(\tau_{j}, f_{j}\right)$ by delay-Doppler shift pair.

Before proceeding, we provide a delightful connection between  (\ref{eq: model sampled}) and compressed sensing theory. As opposed to what in (\ref{eq: model sampled}), assume that $s_{j}\left(l\right)$ are known. Then, if $(\tilde{\tau}_{j}, \tilde{f}_{j})$ are lying on a set of grid points defined by $(\frac{1}{W},\frac{1}{T})$, recovering $(\tau_{j}, f_{j})$ becomes sparse recovery problem which can be solved using compressed sensing algorithms. Nonetheless, in our problem, and even if $s_{j}\left(l\right)$ are known, there will be gridding error as $(\tau_{j}, f_{j})$ could lie anywhere in $[0,1]^{2}$. Moreover, fine discretization leads to dictionaries with highly correlated columns which collides with compressed sensing theories. Now, given that $s_{j}\left(l\right)$ are unknown and that $(\tau_{j}, f_{j})$ could lie anywhere, compressed sensing algorithms cannot be applied.
 
Going back to (\ref{eq: model sampled}), we see that the number of the unknowns is $RL+3R+1$, i.e., $\mathcal{O} \left(RL\right)$ which is much greater than the number of samples $L$. Hence, our recovery problem is severely ill-posed and cannot be solved without structural assumption on $\sv_{j} :=[s_{j}\left(-N\right), \dots, s_{j}\left(N\right)]^{T}$. Inspired by \cite{ahmed2014blind, chi2016guaranteed, yang2016super}, we assume that $\left\lbrace\sv_{j}\right\rbrace_{j=1}^{R}$ belong to a common known low-dimensional subspace spanned by the columns of a known matrix $\Dm\in \mathbb{C}^{L\times K}$ such that $\sv_{j}= \Dm \hv_{j}$ where $K \leq L$ and
\begin{equation}
\label{eq:matrix formulation 2}
\Dm=  \begin{bmatrix} \dv_{-N},  \hdots, \dv_{N} \end{bmatrix}^{H} \in \mathbb{C}^{L \times K}, \ \dv_{l} \in \mathbb{C}^{K\times 1}.
\end{equation}
Here, the unknown orientation vectors $\hv_{j}, j=1,\dots, R$ are to be estimated and are assumed, without loss of generality, to satisfy $||\hv_{j}||_{2}=1$. Thus, recovering $\hv_{j}$ is equivalent to estimating $\sv_{j}$. Based on the above discussion, the number of degrees of freedom in the problem reduces to $\mathcal{O} \left(RK\right)$ which can be less than $L$ when $R, K \ll L$. 
 
{{As mentioned above, the random subspace assumption is a core condition in many blind super-resolution-related frameworks, e.g., \cite{yang2016super, chi2016guaranteed, ahmed2014blind}. The work in \cite{ahmed2014blind} shows that this assumption exists in applications such as image deblurring and in the framework of channel coding for transmitting a message over an unknown multipath channel. Moreover, \cite{yang2016super} illustrates that the random subspace assumption appears in super-resolution imaging. Finally, in multi-user communication systems \cite{luo2006low}, transmitters may send out a random signal for security and privacy reasons. In such a case, the transmitted signal can be represented in a known low-dimensional random subspace.}}

Substituting $s_{j}\left(l\right) = \dv_{l}^{H}\hv_{j}$ in (\ref{eq: model sampled}) and manipulating we get
\begin{equation}
\label{eq: model 2}
y\left(p\right) = \sum_{j=1}^{R} c_{j} \frac{1}{L} \sum_{k,l=-N}^{N} \dv_{l}^{H}\hv_{j} e^{\frac{i2\pi k \left(p-l\right)}{L}} e^{i2 \pi \left(pf_{j}-k\tau_{j}\right)}.
\end{equation}
Now, we consider writing (\ref{eq: model 2}) in matrix-vector form. Starting from the definition of the Dirichlet kernel
\begin{equation}
\label{eq: diric kernel}
D_{N}\left(t\right) := \frac{1}{L} \sum_{r=-N}^{N} e^{i2\pi t r}
\end{equation}
we can rewrite (\ref{eq: model 2}) as
\begin{align}
\label{eq: model 3}
y\left(p\right)= \sum_{j=1}^{R} c_{j} \sum_{k,l=-N}^{N} &D_{N}\left(\frac{k}{L}-f_{j}\right) D_{N}\left(\frac{l}{L}-\tau_{j}\right) \times \nonumber\\
&\dv^{H}_{\left(p-l\right)} \hv_{j} e^{\frac{i 2 \pi p k}{L}}.
\end{align}
The proof of the equivalence between (\ref{eq: model 2}) and (\ref{eq: model 3}) is given in Appendix \ref{sec:proof A}. Now, let us define for convenience the vector $\rv:= [\tau, f]^{T}$ and the atoms $\av \left(\rv_{j}\right) \in \mathbb{C}^{L^{2}\times 1}$ such that 
\begin{eqnarray}
\label{eq: vec 1}
[\av\left(\rv_{j}\right)]_{\left((k,l),1\right)} = D_{N}\left(\frac{l}{L}-\tau_{j}\right) D_{N}\left(\frac{k}{L}-f_{j}\right),
\end{eqnarray}
where $ k,l = -N, \dots, N.$ Moreover, consider the matrices $\widetilde{\Dm}_{p} \in \mathbb{C}^{L^{2} \times K}, \ p=-N,\dots, N$ such that
\begin{equation}
\label{eq: matrix D}
[\widetilde{\Dm}_{p}]_{\left((k,l),1 \to K\right)} = e^{\frac{i 2 \pi p k}{L}} \dv_{\left(p-l\right)}^{H}, \ \ \ k,l = -N, \dots, N. 
\end{equation}
Based on (\ref{eq: vec 1}) and (\ref{eq: matrix D}) we can rewrite (\ref{eq: model 3}) as
\begin{align}
\label{eq: final model}
&y\left(p\right) = \sum_{j=1}^{R} c_{j} \av\left(\rv_{j}\right)^{H} \widetilde{\Dm}_{p} \hv_{j}  = \text{Tr}\left(\widetilde{\Dm}_{p} \sum_{j=1}^{R} c_{j} \hv_{j} \av\left(\rv_{j}\right)^{H}\right) \nonumber\\
& = \left\langle \sum_{j=1}^{R} c_{j} \hv_{j} \av\left(\rv_{j}\right)^{H}, \widetilde{\Dm}_{p}^{H} \right\rangle = \left\langle \Um, \widetilde{\Dm}_{p}^{H} \right\rangle=:[\mathcal{X}\left(\Um\right)]_{p}, 
\end{align}
where $p=-N, \dots, N$, $\Um:=\sum_{j=1}^{R} c_{j} \hv_{j} \av\left(\rv_{j}\right)^{H}$, whereas the linear operator $\mathcal{X}: \mathbb{C}^{K \times L^{2}} \to \mathbb{C}^{L}$ is defined as
\begin{equation}
\label{eq: operator}
[\mathcal{X}\left(\Um\right)]_{p}= \text{Tr}\left(\widetilde{\Dm}_{p}\Um\right), \ \ \ p=-N, \dots, N.
\end{equation}
Using (\ref{eq: operator}) we can relate $\Um$ to $\yv:= [y\left(-N\right),\dots,y\left(N\right)]^{T}$ by
\begin{equation}
\label{eq: sensing to vector}
\yv= \mathcal{X}\left(\Um\right).
\end{equation}
%In practical scenarios, the number of the shifts $R$ is very small compared to $L$. Therefore, $\Um$ is a sparse linear combination of different versions of the atoms $\av\left(\rv_{j}\right)$ in the set
In practice, $R$ is very small compared to $L$, thus, $\Um$ is a sparse linear combination of different versions of $\av\left(\rv_{j}\right)$ in the set
$
\mathcal{A} = \left\lbrace \hv \av\left(\rv\right)^{H}: \ \rv \in [0,1]^{2}, ||\hv||_{2}=1, \hv \in \mathbb{C}^{K\times 1} \right\rbrace.
$
By estimating $\Um$, we can recover all the unknowns. To promote this sparsity when we estimate $\Um$, we apply the atomic norm given in \cite{chandrasekaran2012convex}. The atomic norm is defined by
% Starting from (\ref{eq: atoms}), we can define the problem of obtaining the smallest number of atoms that formulate the decomposition of $\Um$ as 
%\begin{align}
%%\label{eq: atomic norm original}
%&||\Um||_{\mathcal{A},0}  = \nonumber\\
%& \inf_{R}  \left\lbrace \Um=\sum_{j=1}^{R} c_{j} \hv_{j} \av\left(\rv_{j}\right)^{H}: c_{j} \in \mathbb{C}, \rv_{j}\in [0,1]^{2}, ||\hv_{j}||_{2}=1 \right\rbrace \nonumber
%\end{align}
%Since solving the previous problem is difficult, its convex relaxation \cite{chandrasekaran2012convex, li2016off}, which refers to as the atomic norm of $\Um$, is frequently used instead. The atomic norm of $\Um$ is defined by
\begin{align}
%\label{eq: atomic norm}
&||\Um||_{\mathcal{A}} = \inf \left\lbrace t>0: \Um \in t \ \text{conv}\left(\mathcal{A}\right) \right\rbrace \nonumber\\
&= \inf_{c_{j} \in \mathbb{C}, \rv_{j}\in [0,1]^{2}, ||\hv_{j}||_{2}=1} \left\lbrace \sum_{j} |c_{j}| : \Um = \sum_{j} c_{j} \hv_{j} \av\left(\rv_{j}\right)^{H} \right\rbrace, \nonumber
\end{align}
where $\text{conv}\left(\mathcal{A}\right)$ denotes the convex hull of $\mathcal{A}$. Now, we can formulate our blind 2D super-resolution recovery problem as
\begin{align}
\label{eq: optimization atomic} 
\mathcal{P}_{1}: \ \ &\underset{\widetilde{\Um}}{\text{minimize}} \ ||\widetilde{\Um}||_{\mathcal{A}} \nonumber\\
 &\text{subject to} : y\left(p\right) =   \big \langle \widetilde{\Um}, \widetilde{\Dm}_{p}^{H} \big\rangle,  \  \ p = -N, \dots, N.
\end{align}
The problem in (\ref{eq: optimization atomic}) can be used to \emph{recover precisely} $R$ as well as $(\tau_{j}, f_{j})$ $\forall j$. This process will be followed by recovering the unknown waveforms and $\{c_{j}\}_{j=1}^{R}$. Looking at (\ref{eq: optimization atomic}) we can see that recovering the unknowns is achieved by seeking $\widetilde{\Um}$ with a minimal atomic norm that satisfies the observation constraints. %The exact recovery of all the unknowns is granted with very high probability provided that a lower bound on the total number of the observed samples $L$ in conjunction with a minimum separation condition between the delay-Doppler shifts are satisfied. 

We remark that finding a solution for (\ref{eq: optimization atomic}) is intimidating as it includes taking the infimum over infinitely many variables. In Section~\ref{sec: solution sec}, we discuss how to solve (\ref{eq: optimization atomic}) using its dual. Before that, we provide in Section~\ref{sec: recovery cind} the sufficient conditions under which $\Um$ is guaranteed to be the unique solution to (\ref{eq: optimization atomic}).

%We remark that finding a solution for (\ref{eq: optimization atomic}) is intimidating as it includes taking the infimum over infinitely many variables. In Section~\ref{sec: solution sec}, we discuss how to solve (\ref{eq: optimization atomic}) by using its dual. Before that, we provide in the next section the sufficient conditions under which $\Um$ is guaranteed to be the unique solution to (\ref{eq: optimization atomic}), and we summarize the central assumptions of this paper.
%%%%%%%%%%%%%%%%%%%%%%%%%%%%%%%%%%%%%%%%%%%%%%%%%%%%%%%%%%%%%%%%%%%%%%%%%%%%%%%%%
\section{Recovery Conditions and Main Result}
\label{sec: recovery cind}
In this section, we provide our main theorem, which provides sufficient conditions under which (\ref{eq: optimization atomic}) recovers $\Um$. We start first by giving the main assumptions of this theorem. 
\subsection{Main Assumptions}
\begin{assumption}\normalfont
\label{as 1}
The columns of $\Dm^{H}$, i.e., $\dv_{l}$, are independent and drawn from any distribution. Moreover, the entries of $\dv_{l}$ are independent with independent real and imaginary parts and
\begin{align}
\label{eq: G assumption 1}
&\mathbb{E}[\dv_{l}]= \bm{0},\ \ \ \ \ \ \ l=-N, \dots, N\\
\label{eq: G assumption 2}
&\mathbb{E}[\dv_{l} \dv_{l}^{H}]= \Id_{\text{K}}, \ \ l=-N, \dots, N.
\end{align}
\end{assumption}
\begin{assumption}\normalfont
\label{as 2}
(Concentration property) We assume that the rows of $\Dm^{H}$, refer to its column form by $\hat{\dv}_{i} \in \mathbb{C}^{L \times 1}$ where $ i=1,\dots, K$, are $\widetilde{K}$-concentrated with $\widetilde{K} \geq 1$. That is, there exist two constants $\widetilde{C}^{*}_{1}$ and $\widetilde{C}^{*}_{2}$ such that for any 1-Lipschitz function $\varphi : \mathbb{C}^{K} \to \mathbb{R}$ and any $t_{\widetilde{K}} > 0$, it holds
\begin{eqnarray}
\label{eq: concetration x}
\text{Pr}\left[ \big| \varphi(\hat{\dv}_{i})  - \mathbb{E}[ \varphi(\hat{\dv}_{i})] \big| \geq t_{\widetilde{K}} \right] \leq \widetilde{C}^{*}_{1} \exp\left( -\widetilde{C}^{*}_{2}t^{2}_{\widetilde{K}}/\widetilde{K}^{2}\right).
\end{eqnarray}
\end{assumption}

\begin{assumption}\normalfont
\label{as 3}
The entries of $\hv_{j}$ are i.i.d. that are drawn from a uniform distribution with $||\hv_{j}||_{2}=1$.
\end{assumption}

\begin{assumption}\normalfont (Minimum separation)
\label{as 4}
We assume that $\left(\tau_{j}, f_{j}\right) \in [0,1]^{2}, j=1, \dots, R$ satisfy the following separation
%\begin{align}
%\label{eq: seperation condition}
%\Delta_{\text{min}}& = \min_{ \forall j,j': j \neq j'} ||\rv_{j} -\rv_{j'}||_{\infty} \nonumber \\
%&=\min_{ \forall j,j': j \neq j'} \max \left(|\tau_{j}-\tau_{j'}|,|f_{j}-f_{j'}|\right) \geq \frac{2.38}{N}, \nonumber\\
%& \hspace{20pt} \forall [\tau_{j}, f_{j}]^{T}, [\tau_{j'}, f_{j'}]^{T} \in \{\rv_{1}, \dots, \rv_{R}\},
%\end{align}
\begin{eqnarray}
\label{eq: seperation condition}
&\min_{ \forall j,j': j \neq j'} \max \left(|\tau_{j}-\tau_{j'}|,|f_{j}-f_{j'}|\right) \geq \frac{2.38}{N}, \nonumber\\
& \hspace{20pt} \forall [\tau_{j}, f_{j}]^{T}, [\tau_{j'}, f_{j'}]^{T} \in \{\rv_{1}, \dots, \rv_{R}\}, 
\end{eqnarray}
where $|a-b|$ is the wrap-around distance on the unit circle.
\end{assumption}
\subsection{Remarks on the Assumptions}
First, we point out that many random vectors in practice satisfy the concentration property. For example, if the entries of $\hat{\dv}_{i}$ are i.i.d. from standard complex Gaussian distribution, then $\hat{\dv}_{i}$ is a 1-concentrated vector, whereas if each element in $\hat{\dv}_{i}$ is upper bounded by a constant $C$, then $\hat{\dv}_{i}$ is a $C$-concentrated vector \cite[Theorem F.5]{tao2010random}. Thus, the concentration assumption is a more general assumption than the incoherence assumption imposed on the elements of the low-dimensional subspace matrix in \cite{chi2016guaranteed, yang2016super}. For more details about the concentration property, the interested reader is referred to \cite{ledoux2005concentration}.

{{On the other hand, and as discussed in \cite{yang2016super, chi2016guaranteed}, the randomness assumptions on $\dv_{l}$ and $\hv_{j}$, as given by Assumptions~\ref{as 1} and \ref{as 3}, do not appear to be crucial in practice and are doubtless to be artifacts for our proofs. Looking from a different perspective, and based on (\ref{eq: operator}), the random subspace assumption on $\Dm$ can be viewed as a way to obtain random measurement results from $\Um$. Generally speaking, random measurements are crucial in the derivation of theoretical and empirical results \cite{candes2011probabilistic}. The elimination of such a condition is left for the future extension of this work. Ideas on that can be based on modifying our proposed proof methodology or introducing a different technique based on the dual analysis of the atomic norm. }}

Finally, we point out that the separation between the shifts is essential for a precise and stable recovery. The existence of a certain separation between the shifts has appeared in all existing super-resolution theories, e.g., \cite{candes2014towards, yang2016super, tang2013compressed, heckel2016super}. This follows from the fact that the recovery problem becomes very ill-conditioned when the shifts are close to each other \cite{candes2014towards}. Nevertheless, we stress that (\ref{eq: seperation condition}) is not a necessary condition, and a smaller separation (with a constant less than 2.38) is expected to be enough (see \cite[Section~1.3]{candes2014towards} ). We leave tackling this issue for the future work.

\subsection{Performance Guarantee Theorem}
We are now ready to provide our main theorem as follows:
\begin{theorem}
\label{th: main the}
Let $ y \left(p\right) \in \mathbb{C}$ be as in (\ref{eq: model 2}) with $p = -N, \dots, N$ and $N \geq 512$. Additionally, assume that $\left\lbrace\sv_{j}\right\rbrace_{j=1}^{R}$ can be written as $\sv_{j}= \Dm \hv_{j}$, $\Dm \in \mathbb{C}^{L \times K}$ where $\Dm$ satisfies Assumptions~\ref{as 1} and \ref{as 2} while $\hv_{j}$ are satisfying Assumption~\ref{as 3}. Moreover, let $\rv_{j} = [\tau_{j},f_{j}]^{T}$ and define the set $\mathcal{R} := \{\rv_{1}, \dots, \rv_{R}\}$, where the elements of $\mathcal{R}$ are assumed to satisfy Assumption~\ref{as 4}. Then, there exist two numerical constants $C_{1}^{*}$ and $C_{2}^{*}$ such that when
\begin{eqnarray}
\label{eq: L fianl}
L \geq  \hspace{-2pt} C_{1}^{*} R K \widetilde{K}^{4} \log^{2}\left(\hspace{-2pt}\frac{  C_{2}^{*} R^{2} K^{2} L^{3} }{\delta}\right)\hspace{-2pt}\log^{2}\left(\hspace{-2pt}\frac{ C_{2}^{*} (K+1)  L^{3} }{\delta}\hspace{-1pt}\right)
\end{eqnarray}
is satisfied with $\delta >0$, $\Um=\sum_{j=1}^{R} c_{j} \hv_{j} \av\left(\rv_{j}\right)^{H}$ is the optimal minimizer of $\mathcal{P}_{1}$ in (\ref{eq: optimization atomic}) with probability at least $1- \delta$.
\end{theorem}
\subsection{Remarks on Theorem~\ref{th: main the}}
{{Theorem~\ref{th: main the} states the minimum number of samples $L$ that guarantees the exact recovery of $\Um$ upon solving the super-resolution recovery problem in (\ref{eq: optimization atomic}). The bound on $L$ suggests that the more concentrated are the rows of $\Dm^{H}$, the fewer number of samples required for the exact recovery. Moreover, for a given $\widetilde{K}$, (\ref{eq: L fianl}) states that having $L = \mathcal{O} \left(RK\right)$ provides a sufficient condition for recovering the unknowns. This fact coincides with the number of degrees of freedom in the problem and follows the sufficient condition for stable recovery in both 1D and 2D non-blindness super-resolution ($L = \mathcal{O} \left(R\right)$) as \cite{candes2014towards} and \cite{heckel2016super} show, respectively. The work on 1D blind super-resolution in \cite{li2019atomic} has a bound of $\mathcal{O}\left(R^{2}K\right)$ on the sample complexity, but without the random assumption on $\hv$. It will be interesting to see how dropping the random assumption on $\hv$, or some of our other assumptions, will affect our sample complexity bound in the future extension of this work.

On the other hand, $N\geq 512$ is a requirement that is made to facilitate some of our proofs upon following the steps in \cite{candes2014towards}. However, as \cite{candes2014towards} shows, this assumption can be discarded at the cost of having a more significant separation. Our simulations show that the exact recovery exists even when this condition is not met. Finally, in contrast to the non-blind 2D super-resolution work in \cite{heckel2016super}, where $\text{sign}\left(c_{j}\right):= \frac{c_{j}}{|c_{j}|}$ are assumed to be i.i.d. for all $j$, we do not impose any assumptions on $c_{j}$. 

The proof of Theorem~\ref{th: main the} is based on the dual of (\ref{eq: optimization atomic}). We show in Section~\ref{sec: solution sec} that this proof boils down to be a problem of formulating a 2D trigonometric random vector polynomial that satisfies certain interpolation conditions. The formulation of this polynomial requires using some random kernels in company with matrix theory and probability measures.}}

{{

Before concluding this part, we point out in practice, the samples $y\left(p\right)$ can be contaminated by noise. If we assume that the Euclidean norm on the noise vector is upper bounded by $\zeta >0$, then, the super-resolution recovery problem can be shown to take the form
\begin{align}
\label{eq: model noise }
&\hspace{-5pt}\mathcal{P}_{2}: \underset{\widetilde{\Um}}{\text{minimize}} \ \|\widetilde{\Um}\|_{\mathcal{A}} \nonumber\\
 &\text{subject to} : \|y\left(p\right) -  \big \langle \widetilde{\Um}, \widetilde{\Dm}_{p}^{H} \big\rangle\|_{2} \leq \zeta, p = -N, \dots, N
\end{align}
Addressing such a problem is beyond the scope of this paper and is the topic of our work in \cite{suliman2021mathematical}. However, it is worth mentioning that while exact recovery for the unknowns is guaranteed in this paper, such exactness does not exist in the presence of noise. Thus, an entirely different goal based on assessing the framework's robustness to noise and the existence of a good estimate for the unknowns (in terms of the mean-squared error) is addressed in \cite{suliman2021mathematical}. For such a goal, and as opposed to our recovery problem formulation in (\ref{eq: optimization atomic}), the work in \cite{suliman2021mathematical} considers reformulating the above recovery problem as a regularized least-square atomic norm minimization problem that controls the noise and enforces sparsity on the obtained solution simultaneously. In this paper, we provide a single simulation experiment that shows that our proposed framework is stable in the existence of noise and refers the reader to \cite{suliman2021mathematical} for detailed theoretical analysis and extensive simulation experiments. 
}} 
%%%%%%%%%%%%%%%%%%%%%%%%%%%%%%%%%%%%%%%%%%%%%%%%%%%%%%%%%%%%%%%%%%%%%%%%%%%%%%%%%%%%%%%%%%%%%
\section{Identifying the Unknowns: Problem Solution}
\label{sec: solution sec}
In this section, we discuss the solution of (\ref{eq: optimization atomic}) and the recovering of the unknowns. Moreover, we give some remarks about the optimality and the uniqueness of the solution and the complexity of the problem.
\subsection{Background}
Let us assume that $\hv_{j}$ are known, then, the building blocks of $\mathcal{A}$ become vectors. Now, if we further assume that we only have delay- or only Doppler shifts (i.e., 1D problem), the obtained atomic norm recovery problem and its dual can be solved efficiently via SDP upon characterizing them in terms of linear matrix inequalities. This characterization is based on the classical Vandermonde decomposition for PSD Toeplitz matrix by Carath\'eodory lemma \cite[Proposition 2.1]{tang2013compressed}. {{Now when both delay and Doppler shifts are unknown (i.e., 2D), the generalization of this lemma comes with a rank constraint on the Toeplitz matrix. Hence, it prohibits a characterization of the atomic norm similar to that in \cite[Proposition 2.1]{tang2013compressed}.}} Now, consider that $\hv_{j}$ are unknown and that we have a 1D problem. Here, the atomic set of the recovery problem is formulated by matrices. Such a scenario is tackled in \cite{li2016off, yang2014exact} where its shown that SDP can characterize the atomic norm problem.
  
In this paper, and to address our case where both $\hv_{j}$ and $\left(\tau_{j},f_{j}\right)$ are unknown, we follow the path of obtaining an SDP relaxation to the dual problem. Our formulation is inspired by that in \cite[Section~4]{candes2014towards}, \cite[Section~2.2]{tang2013compressed}, and \cite[Section~6.1]{heckel2016super}, and is built on top of the results in \cite[Equation 3.3]{xu2014precise} and \cite[Corollary 4.25]{dumitrescu2017positive}. The main idea is to express the constraint of the dual (\ref{eq: optimization atomic}) using linear matrix inequalities. The relaxation comes from the fact that the matrices used to express the dual constraint are of unspecified dimensions, and an approximation for their dimensions is required. We show later that this SDP relaxation leads to the optimal solution in practice. 

\subsection{Dual Problem Formulation}
%Starting from (\ref{eq: optimization atomic}), we can show that the dual certificate of this optimization problem can be written as \cite[Section~5.1.6]{boyd2004convex}
The dual of (\ref{eq: optimization atomic}) can be written as \cite[Section~5.1.6]{boyd2004convex}
\begin{eqnarray}
\label{eq: optimzation dual}
\mathcal{P}_{3}:  \ \underset{\qv}{\text{maximize}} \ \langle\qv, \yv\rangle_{\mathbb{R}}, \ \text{subject to} : ||\mathcal{X}^{*}\left(\qv\right) ||_{\mathcal{A}}^{*} \leq 1,
\end{eqnarray}
where $\mathcal{X}^{*}: \mathbb{C}^{L} \to \mathbb{C}^{K \times L^{2}} $ is the adjoint of $\mathcal{X}$, i.e., $\mathcal{X}^{*}\left(\qv \right)= \sum_{p=-N}^{N} [\qv]_{p} \widetilde{\Dm}_{p}^{H}$ while $||\cdot||_{\mathcal{A}}^{*}$ is the dual atomic norm, i.e.,
\begin{eqnarray}
\label{eq: dual atomic for}
||\Cm||_{\mathcal{A}}^{*} = \hspace{-2pt}\sup_{||\Um ||_{\mathcal{A}}\leq 1} \big\langle\Cm,\Um \big\rangle_{\mathbb{R}} \hspace{-2pt} = \hspace{-4pt}\sup_{\rv \in [0,1]^{2}, ||\hv||_{2}=1} \hspace{-4pt}\big\langle\Cm,\hv \av\left(\rv \right)^{H}\hspace{-2pt}\big\rangle_{\mathbb{R}}
\end{eqnarray}
Since (\ref{eq: optimization atomic}) has only equality constraints, Slater's condition is satisfied \cite[Chapter 5]{boyd2004convex}, and strong duality holds between (\ref{eq: optimization atomic}) and (\ref{eq: optimzation dual}). Consequently, the optimal of (\ref{eq: optimization atomic}) is equal to that of (\ref{eq: optimzation dual}). If we refer to the solution of (\ref{eq: optimization atomic}) by $\widehat{\Um}$ and that of (\ref{eq: optimzation dual}) by $\qv$, then, this equality only holds if and only if $\widehat{\Um}$ is the primal optimal and $\qv$ is the dual optimal. In Proposition~\ref{pro: main pro} below, we use this strong duality to discuss when $\widehat{\Um} = \Um$. Before that, we can write the constraint of (\ref{eq: optimzation dual}) using (\ref{eq: dual atomic for}) as
\begin{eqnarray}
%\label{eq: dual cons equivalent}
||\mathcal{X}^{*}\left(\qv\right)\hspace{-2pt} ||_{\mathcal{A}}^{*}\hspace{-3pt}= \hspace{-5pt}\sup_{\substack{\rv \in [0,1]^{2}\\ ||\hv||_{2}=1}} \hspace{-2pt}\big| \big\langle \hv, \mathcal{X}^{*}\hspace{-2pt}\left(\qv\right)  \av\left(\rv\right) \rangle\big|\hspace{-2pt} =\hspace{-2pt} \hspace{-3pt}\sup_{\rv \in [0,1]^{2}} \hspace{-2pt}|| \mathcal{X}^{*}\hspace{-2pt}\left(\qv\right)  \av\left(\rv\right)\hspace{-2pt}||_{2} \nonumber
\end{eqnarray}
Now, define a vector polynomial function $\fv\left(\rv\right) \in \mathbb{C}^{K \times 1}$ as 
\begin{eqnarray}
\label{eq: dual polynomail equa}
\fv\left(\rv\right) \triangleq \mathcal{X}^{*}\left(\qv\right)  \av\left(\rv\right) = \sum_{p=-N}^{N} [\qv]_{p}  \widetilde{\Dm}_{p}^{H} \av\left(\rv\right).
\end{eqnarray}
Looking at (\ref{eq: dual polynomail equa}), we can see that the constraint in (\ref{eq: optimzation dual}) is equivalent in demand that the norm of a 2D vector polynomial $\fv\left(\rv\right)$ is upper bounded by one. The existence of such polynomial combined with some other conditions and that strong duality holds between (\ref{eq: optimization atomic}) and (\ref{eq: optimzation dual}) all serve as sufficient conditions to recover $\Um$ from (\ref{eq: optimization atomic}). In the following proposition, we state the sufficient conditions under which (\ref{eq: optimization atomic}) is assured to obtain its \emph{unique} optimal solution based on the dual problem constraint.
\begin{proposition}
\label{pro: main pro}
\normalfont
Let $\mathcal{R}= \left\lbrace \rv_{j}\right\rbrace_{j=1}^{R}, \rv_{j} = \left[\tau_{j}, f_{j}\right]^{T}$ and refer to the solution of (\ref{eq: optimization atomic}) by $\widehat{\Um}$. Then, $\widehat{\Um}=\Um$ is the \emph{unique} optimal solution of (\ref{eq: optimization atomic}) if the following two conditions are satisfied:
\begin{enumerate}
\item There exists a 2D vector polynomial in $\tau$ and $f$ as in (\ref{eq: dual polynomail equa})
%
%
%, i.e., 
%\begin{equation}
%\label{eq: dual poly eq}
%\fv\left(\rv\right) = \mathcal{X}^{*}\left(\qv\right)  \av\left(\rv\right) = \sum_{p=-N}^{N} [\qv]_{p} \widetilde{\Dm}_{p}^{H} \av\left(\rv\right) \in \mathbb{C}^{K \times 1} \nonumber
%\end{equation} 
with $\qv = \begin{bmatrix} q\left(-N\right),  \hdots ,q\left(N\right) \end{bmatrix}^{T} \in \mathbb{C}^{L \times 1}$ such that: 
\begin{equation}
\label{eq: hold assump 1}
\fv\left(\rv_{j}\right)= \text{sign}\left(c_{j}\right) \hv_{j}, \ \ \forall \rv_{j} \in \mathcal{R}
\end{equation}
\begin{equation}
\label{eq: hold assump 2}
||\fv\left(\rv\right)||_{2} < 1 , \ \ \forall \rv \in [0,1]^{2}\setminus \mathcal{R},
\end{equation}
\item $\left\lbrace\begin{bmatrix}\av\left(\rv_{j}\right)^{H} \widetilde{\Dm}_{-N} \\  \vdots \\ \av\left(\rv_{j}\right)^{H} \widetilde{\Dm}_{N} \end{bmatrix}\right\rbrace_{j=1}^{R}$ is a linearly independent set.
\end{enumerate}
\end{proposition}
The proof of Proposition~\ref{pro: main pro}, which is in Appendix \ref{A2}, follows\cite[Proposition~1]{yang2016super}, and is based on strong duality. 
\subsection{SDP Relaxation of the Dual Problem} 
\label{sub: dual relax}
In this section, we obtain an equivalent SDP for (\ref{eq: optimzation dual}) based on Proposition~\ref{pro: tri go} below. Before we proceed, note that (\ref{eq: dual polynomail equa}) can be written as (see Appendix \ref{A3})
%We start by showing in Appendix~\ref{A3} that $\fv\left(\rv\right)$ in (\ref{eq: dual polynomail equa}) can be expressed equivalently by
\begin{eqnarray}
\label{eq: simple polynomail}
\fv\left(\rv\right)= \sum_{p,k=-N}^{N} \left(\frac{1}{L}[\qv]_{p} \sum_{l=-N}^{N} \dv_{l}e^{\frac{i2\pi k \left(p-l\right)}{L}}\right) e^{-i2\pi \left(k\tau+pf\right)}. 
\end{eqnarray}
\begin{proposition}\normalfont \cite{xu2014precise} (special case \cite[Chapter 3]{dumitrescu2017positive})
\label{pro: tri go}
Let $K\left(\lambdav\right)$ be a $d$-variate trigonometric polynomial with variables $\lambdav= \left[\lambda_{1}, \dots, \lambda_{d}\right]$, i.e., 
%\begin{align}
%\label{eq: d variate poly}
$K\left(\lambdav\right) = \sum_{\jv} k_{\jv} e^{-i2 \pi \lambdav^{T} \jv}, \nonumber$
%\end{align} 
where $\jv = \left\lbrace j_{1}, \cdots, j_{d}\right\rbrace, 0 \leq j_{p} \leq l_{p}-1, 1 \leq  p \leq d  $. Then, if 
%\begin{align}
%\label{eq: d variate poly 2}
$\sup_{\lambdav \in [0,1]^{d}} \big| K\left(\lambdav\right)\big| \leq 1,$
%\end{align}
there exists a PSD matrix $\Qm$ with
\begin{align}
\label{eq: d variate poly 3}
\begin{bmatrix} \Qm  & \kv  \\ \kv^{H} &  1  \end{bmatrix} \succeq \bm{0}, \hspace{20pt} \text{Tr}\left(	\Thetam_{\nv}\Qm\right) =  \delta_{\nv},
\end{align}
where $\kv$ is a column vector that contains the elements of $k_{\jv}$ and is then padded with zeros to match the dimension of $\Qm$. Moreover, $\Thetam_{\nv}= \Thetam_{n_{d}} \otimes \cdots \otimes \Thetam_{n_{1}} $ with $\nv = [n_{1}, \cdots, n_{d}]^{T}$, where $-m_{p} \leq n_{p} \leq m_{p}$ for every $1 \leq p \leq d$ whereas $\Thetam_{n_{p}}$ is $\left(m_{p}+1\right) \times  \left(m_{p}+1\right)$ Toeplitz matrix with ones on its $n_{p}$ diagonal and zeros elsewhere. Finally, $\delta_{\nv}$ is the Dirac delta function, i.e., $\delta_{\bm{0}}=1$ and $\delta_{\nv}=0$ for $\nv \neq \bm{0}$.

Note that $\Qm$ is an $\prod_{p=1}^{d}(m_{p}+1) \times  \prod_{p=1}^{d}(m_{p}+1)$ matrix and that the exact value of $m_{p}$ is not recognized but satisfy $m_{p} \geq l_{p}$. Thus, $m_{p}=l_{p}$ provides a relaxation to the problem but is observed to yield the optimal solution in practice.\footnote{The relaxation is based on the so-called sum of square relaxation for a non-negative multivariate trigonometric polynomial. The discussion about this point is beyond the scope of the paper. The interested reader may consult \cite[Chapter 3]{dumitrescu2017positive} and \cite{xu2014precise} for $d\geq 2$.} Finally, the other way around is also true, i.e., having a PSD matrix $\Qm$ that satisfies (\ref{eq: d variate poly 3}) means that $\sup_{\lambdav \in [0,1]^{d}} \big| K\left(\lambdav\right)\big| \leq 1$ holds.
\end{proposition}

To formulate the SDP relaxation of (\ref{eq: optimzation dual}) using Proposition~\ref{pro: tri go} we first define a matrix $\widehat{\Qm} \in \mathbb{C}^{K \times L^2}$ based on (\ref{eq: simple polynomail}) such that 
%\begin{eqnarray}
%\label{eq: matrix temp }
%\left[\widehat{\Qm}\right]_{(i,(p,k))}  &:=  \left[\frac{1}{L}\qv\left(p\right) \sum_{l=-N}^{N} \dv_{l}e^{\frac{i2\pi k \left(p-l\right)}{L}}\right]_{i},\nonumber\\
%& \ \ \ i=1,\dots, K,\ \ \ p,k=-N,\dots, N.
%\end{eqnarray}
\begin{eqnarray}
\label{eq: matrix temp }
\left[\widehat{\Qm}\right]_{(i,(p,k))} \hspace{-2pt}:=  \left[\frac{1}{L}\qv\left(p\right) \hspace{-2pt}\sum_{l=-N}^{N} \hspace{-2pt}\dv_{l}e^{\frac{i2\pi k \left(p-l\right)}{L}}\right]_{i}\hspace{-2pt}, i=1,\dots, K,
\end{eqnarray}
where $p,k=-N,\dots, N$. By setting $d=2$ in Proposition~\ref{pro: tri go} and using (\ref{eq: matrix temp }), we formulate the SDP relaxation of (\ref{eq: optimzation dual}) as
\begin{align}
\label{eq: dual of the dual}
&\mathcal{P}_{4}:\  \ \ \underset{\qv, \Qm}{\text{maximize}} \ \langle\qv, \yv\rangle_{\mathbb{R}} \nonumber\\
&\text{subject to} : \Qm \succeq \bm{0}, \hspace{2pt}\begin{bmatrix} \Qm  & \widehat{\Qm}^{H}  \\ \widehat{\Qm} &  \Id_{\text{K}}  \end{bmatrix} \succeq \bm{0}, \hspace{2pt} \text{Tr}\left(	\Thetam_{\nv}\Qm\right) =  \delta_{\nv},
\end{align}
where $\Thetam_{\nv} =\Thetam_{\tilde{k}} \otimes \Thetam_{\tilde{l}}$ with $ -(L-1)\leq \tilde{k}, \tilde{l} \leq (L-1)$. Note that we take the main diagonal of $\Thetam_{i}$ as the 0-th diagonal. 
\subsection{Dual Problem Solution}
\label{sec: d p s}
The problem in (\ref{eq: dual of the dual}) can be solved using any SDP solver such as CVX. As (\ref{eq: dual of the dual}) shows, we set $m_{p}=L$. Using a larger value than $L$ will lead to better approximation to (\ref{eq: optimzation dual}). Our simulations show that $m_{p}=L$ yields the optimal solution in all scenarios. This is observed in other related work such as \cite{dumitrescu2017positive, heckel2016super}. Once we solve the problem, we proceed as follow:

\begin{itemize}
\item  We obtain $\fv\left(\rv\right)$ as a function of $\rv$ using $\qv$.

\item  {{Then, to acquire an estimate for $\rv_{j}$, i.e., $\hat{\rv}_{j}$, we can compute the roots of the polynomial $1-\|\fv\left(\rv\right)\|_{2}^{2}$ on the unit circle as in \cite[Section~4]{candes2014towards}  using standard 2D line spectral estimation approaches such as Prony’s method}} or we can discretize $[0,1]^{2}$ on a fine grid and then recover $\hat{\rv}_{j}$ at which $\|\fv\left(\hat{\rv}_{j}\right)\|_{2} = 1$ (based on (\ref{eq: hold assump 1}) and the fact that $\|\hv_{j}\|_{2}=1$). In this paper, we use the second approach to estimate the shifts.

\item Finally, we recover the atoms $\av \left(\hat{\rv}_{j}\right)$ and then formulate the following overdetermined linear system
\begingroup
\fontsize{9.pt}{9.5pt}
\begin{eqnarray}
\label{eq: ls for h}
\hspace{-6pt}\begin{bmatrix} \av\left(\hat{\rv}_{1}\right)^{H} \widetilde{\Dm}_{-N} &  \hdots & \av\left(\hat{\rv}_{R}\right)^{H} \widetilde{\Dm}_{-N} \\ \vdots &  \ddots & \vdots \\ \av\left(\hat{\rv}_{1}\right)^{H} \widetilde{\Dm}_{N} & \hdots & \av\left(\hat{\rv}_{R}\right)^{H} \widetilde{\Dm}_{N}  \end{bmatrix} \hspace{-4pt} \begin{bmatrix} c_{1}\hv_{1} \\ \vdots  \\ c_{R}\hv_{R} \end{bmatrix} \hspace{-4pt} = \hspace{-3pt}  \begin{bmatrix} y\left(-N\right) \\ \vdots \\  y\left(N\right) \end{bmatrix} \hspace{-4pt}
\end{eqnarray}
\endgroup
based on (\ref{eq: final model}). The above system can be solve using the LS algorithm to obtain the estimates $\hat{c}_{j}\hat{\hv}_{j}, j=1, \dots, R$. 
\end{itemize}
{{Note that the solution of the dual is not unique in general. However, the estimated set $\hat{\mathcal{R}}$ will always contain the $R$ required shifts. This is discussed in Appendix \ref{A2}. Therefore, generally speaking, $\hat{\mathcal{R}}$ might include false shifts; however, in most of the cases, and with very high probability, the SDP solvers will provide a solution such that $\hat{\mathcal{R}} = \mathcal{R}$. This fact is discussed in \cite[Section 4]{candes2014towards} and in further detail in \cite[Proposition 2.5 and Corollary 2.6]{tang2013compressed}. Hence, we will only highlight the main idea behind it here. First, recall (\ref{eq: dual polynomail equa}) and let $\left\lbrace\left(\qv, \Qm\right)\right\rbrace$ be the set of the optimal solutions to the dual SDP in (\ref{eq: dual of the dual}) that includes a solution such that $\fv\left(\rv\right)$ satisfies (\ref{eq: hold assump 1}) and (\ref{eq: hold assump 2}). Then, we can easily show that all $\left(\qv, \Qm\right)$ in the relative interior of $\left\lbrace\left(\qv, \Qm\right)\right\rbrace$ satisfy (\ref{eq: hold assump 1}) and (\ref{eq: hold assump 2}). This fact suggests that we can use any optimal solution in the relative interior of the set $\left\lbrace\left(\qv, \Qm\right)\right\rbrace$ to obtain the unknown shifts (see \cite[Appendix B]{tang2013compressed} for more details). On the other hand, we can also show that the dual central path converges to a point in the relative interior of $\left\lbrace\left(\qv, \Qm\right)\right\rbrace$, indicating that any primal-dual path obtained using an SDP solver will provide an interior point in the dual optimal set. Based on that,  we can conclude that the relative interior of $\left\lbrace\left(\qv, \Qm\right)\right\rbrace$ excludes all dual optimal solutions that contain false shifts. Moreover, the uniqueness of $\hat{c}_{j}\hat{\hv}_{j}$ is from the fact that the columns of the matrix in (\ref{eq: ls for h}) are linearly independent based on Proposition~\ref{pro: main pro}. Finally, it is impossible to resolve between $\hat{c}_{j}$ and $\hat{\hv}_{j}$ in the final solution. }}

\subsection{Computational Complexity}
{{As (\ref{eq: dual of the dual}) shows, estimating the unknowns involves solving a convex problem with an optimization variable of dimensions $L^{2} \times L^{2}$. Therefore, any algorithm that solves this problem will have a computational cost of order $\mathcal{O}\left(L\right)$ \cite{van1983matrix}.

For large values of $L$, addressing such high complexity becomes impossible, making this framework infeasible in real applications with large $L$. This fact prohibits us from evaluating our framework performance for large values of samples as well as exploring some of the framework characteristics, such as the trade-offs between sample complexity bound in (\ref{eq: L fianl}) and problem dimensions. A future extension to this work should look at reducing such high complexity. This could include, for example,  investigating alternative optimization techniques as proposed in \cite{rao2015forward}, solving the problem from a subset of the observed samples as in \cite{tang2013compressed},  or applying the alternating direction method of multipliers (ADMM) as proposed in \cite{ran2021fast}.}}
%%%%%%%%%%%%%%%%%%%%%%%%%%%%%%%%%%%%%%%%%%%%%%%%%%%%%%%%%%%%%%%%%%%%%%%%%%%%%%%%%%%%%%%%%%%%%
\section{Simulations Experiment}
\label{sec: results}
In this section, we validate the performance of the proposed framework using extensive simulations. In all the experiments, we use the CVX solver, which calls SDPT3, to solve (\ref{eq: dual of the dual}).

In the first experiment, we set $L=19, K=2, R=2$ and we let the entries of $\Dm$ to be i.i.d. from a complex Gaussian distribution of zero mean and unit variance, i.e., $\mathcal{CN}\left(0,1\right)$. Moreover, the elements of $\hv_{j}$ are generated as $\stackrel{ i.i.d.}{\sim} \mathcal{CN}\left(0,1\right)$ and then normalized to have $||\hv_{j}||_{2}=1$. The locations of $\left(\tau_{j}, f_{j}\right)$ are generated randomly from a uniform distribution in $[0,1]^{2}$ in accordance to (\ref{eq: seperation condition}) and found to be $\left(0.28,0.53\right)$ and $\left(0.94,0.42\right)$. Finally, the real and the imaginary parts of $c_{j}$ are generated from $\mathcal{N}\left(0,1\right)$ and normalized to have $|c_{j}|=1$.

\begin{figure}[h!]
\centering
\subfigure[]{\label{fig:out1}\includegraphics[width=1.72in]{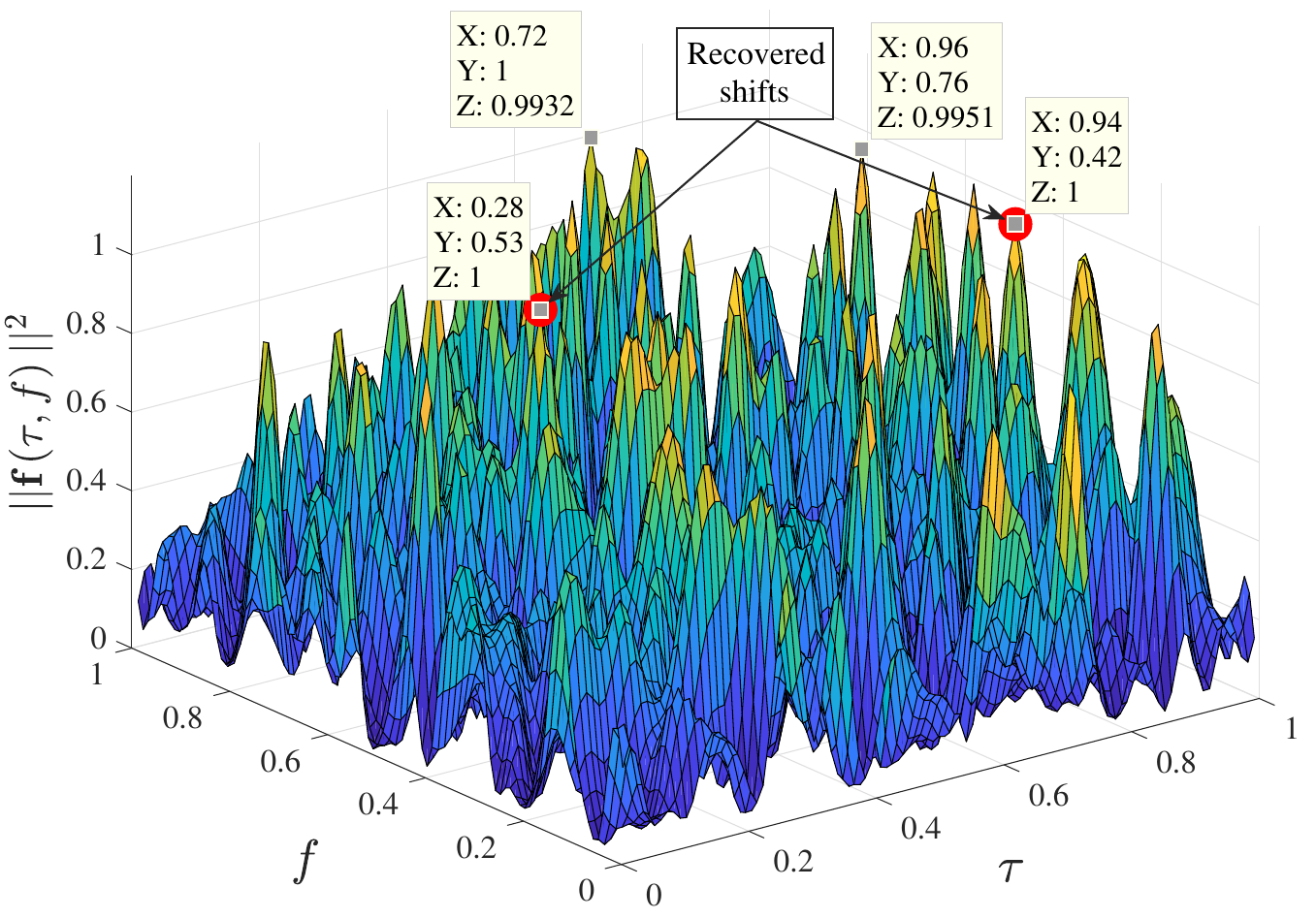}}
\subfigure[]{\label{fig: per 1_2}\includegraphics[width=1.72in]{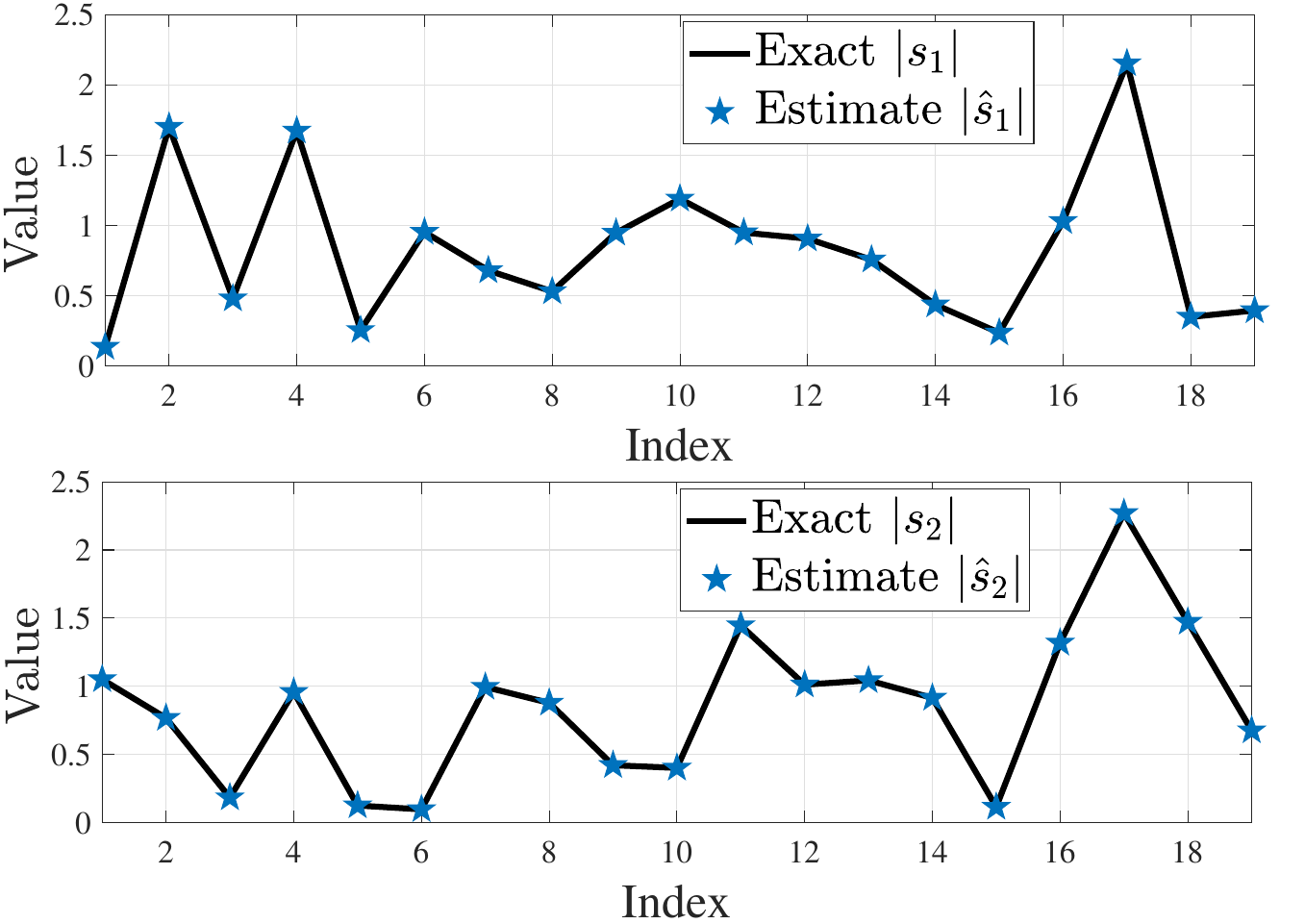}}
\caption{(a) Plot of $||\fv\left(\rv\right)||_{2}^{2}$ and the locations of the estimated shifts (denoted by red circles). (b) Comparing the estimated samples of the signals with the true ones.}
\label{fig:out1 all}
\end{figure}

In Fig~\ref{fig:out1}, we plot $||\fv\left(\rv\right)||_{2}^{2}$ for $\rv \in [0,1]^{2}$. To estimate the shifts, we first discretize the 2D grid with a step size of $10^{-3}$. Then, we locate the points at which $||\fv\left({\rv}\right)||_{2}^{2}=1$ as discussed in Section~\ref{sec: d p s}. From Fig~\ref{fig:out1}, we can observe that the two shifts are recovered perfectly, i.e., $\hat{\mathcal{R}}= \mathcal{R}$ as indicated by circle points, and that $||\fv\left(\rv\right)||_{2} < 1, \forall \rv \in [0,1]^{2}\setminus \mathcal{R}$. 

Once we estimate the shifts, we generate $\av\left(\hat{\rv}_{j}\right)$ using (\ref{eq: vec 1}) and then formulate (\ref{eq: ls for h}). To obtain $\hat{c}_{j} \hat{\hv}_{j}$, we solve (\ref{eq: ls for h}) using the LS algorithm. Given that we cannot retrieve the phases of $\hat{c}_{j}$, we plot in Fig~\ref{fig: per 1_2} the magnitudes of the estimated samples $\hat{s}_{j}\left(l\right)$ and we compare them with the true ones. Fig~\ref{fig: per 1_2} shows that we are able to retrieve the signals samples exactly. Finally, we compute $|\hv_{j}^{H}\hat{\hv}_{j}|$ to find that $|\hv_{1}^{H}\hat{\hv}_{1}|=1-10^{-8}$ and $|\hv_{2}^{H}\hat{\hv}_{2}|=1.0$ which confirms the superiority of the approach. 

In the second experiment, we generate the columns of $\Dm^{H}$ as \cite{chi2016guaranteed}
\begin{equation}
%\label{eq: new column b}
\dv_{l} = \begin{bmatrix}1, & e^{i2 \pi \sigma_{l}}, & \hdots, & e^{i2 \pi \left(K-1\right)\sigma_{l}}  \end{bmatrix}^{T}, \ \ \ l=-N, \dots, N, \nonumber
\end{equation}
where $\sigma_{l}$ is uniformly distributed in $[0,1]$. We let $L=21$, $K=3$, $R=1$, and we randomly generate the shift pair in $[0,1]^{2}$ which is found to be $\left(0.13,0.67\right)$. Finally, we use the same configurations for $\hv_{j}$ and $c_{j}$ as in the previous scenario.

\begin{figure}[h!]
\vspace{-5pt}
\centering
\subfigure[]{\label{fig: per 2_1}\includegraphics[width=1.72in]{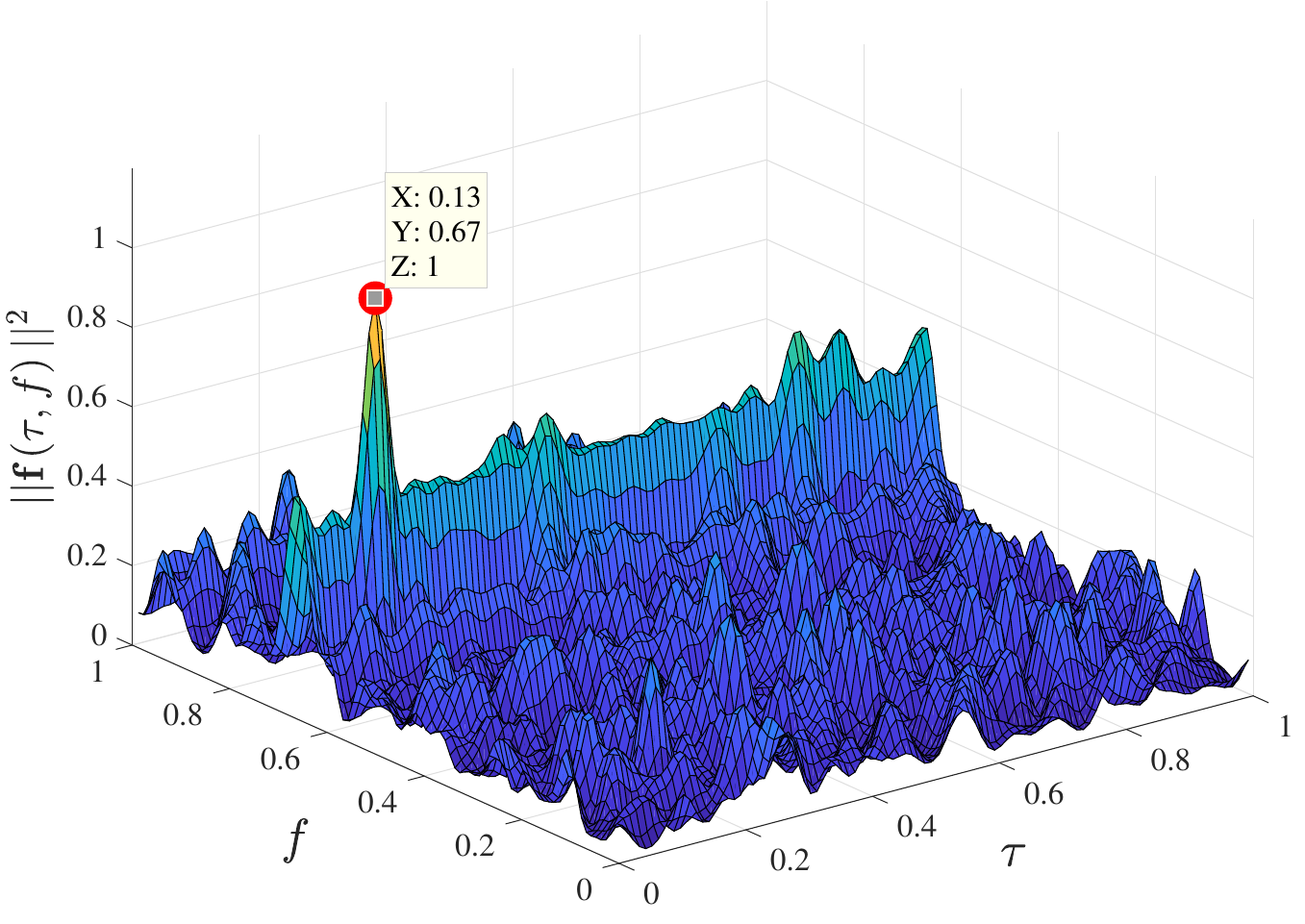}}
\subfigure[]{\label{fig: per 2_2}\includegraphics[width=1.72in]{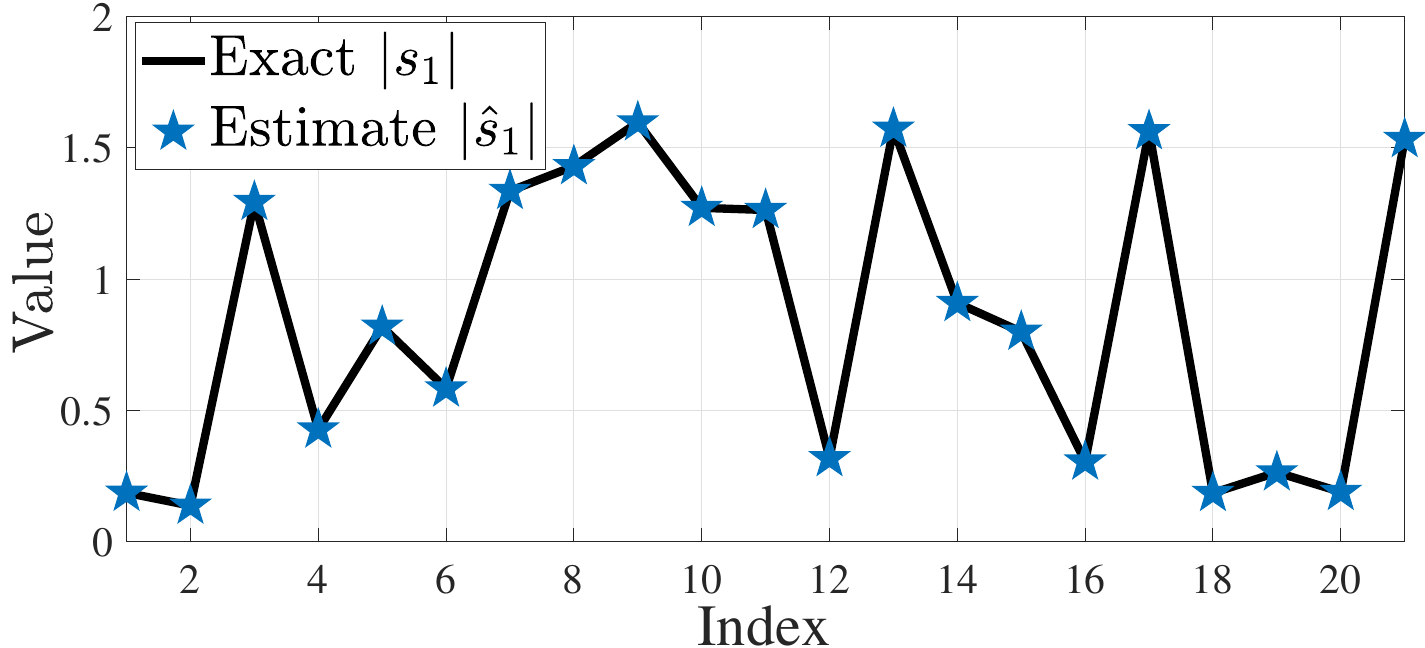}}
\caption{(a) Plot of $||\fv\left(\rv\right)||_{2}^{2}$ and the location of the estimated shift. (b) Comparing the estimated samples of the signal with the true ones.}
\label{fig: per 2_1 all}
\end{figure}

In Fig~\ref{fig: per 2_1}, we plot $||\fv\left(\rv\right)||_{2}^{2}$ in $[0,1]^{2}$. From Fig~\ref{fig: per 2_1}, we can observe that $||\fv\left(\rv\right)||_{2}^{2}=1$ at the true shift. Moreover, we plot in Fig~\ref{fig: per 2_2} the magnitudes of $\hat{s}_{1}\left(l\right)$ and we compare them with the actual ones. Fig~\ref{fig: per 2_2} shows that the estimated samples coincide with the true ones over all the index range. Finally, we find that $|\hv_{1}^{H}\hat{\hv}_{1}|=1.0$.

\begin{figure}[h!]
\vspace{-5pt}
\centering
\subfigure[]{\label{fig: per 3_1}\includegraphics[width=1.72in]{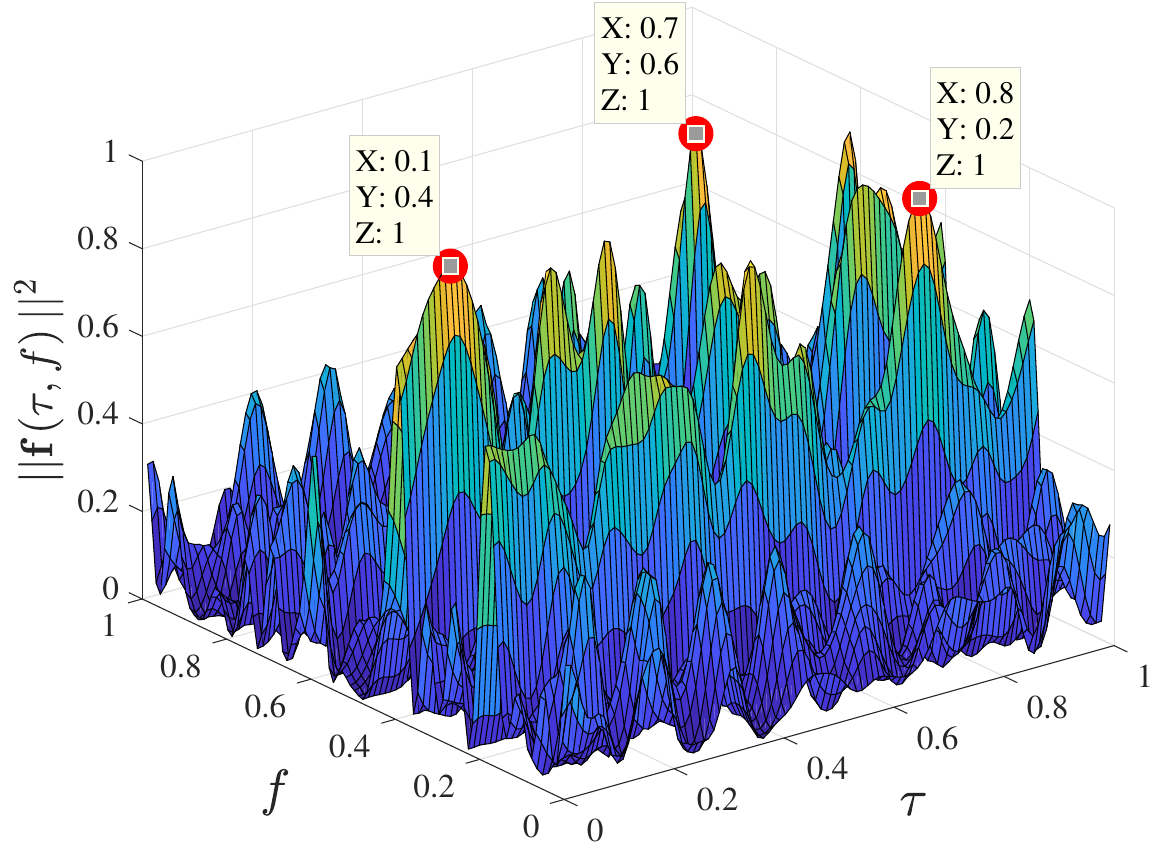}}
\subfigure[]{\label{fig: per 3_2}\includegraphics[width=1.72in]{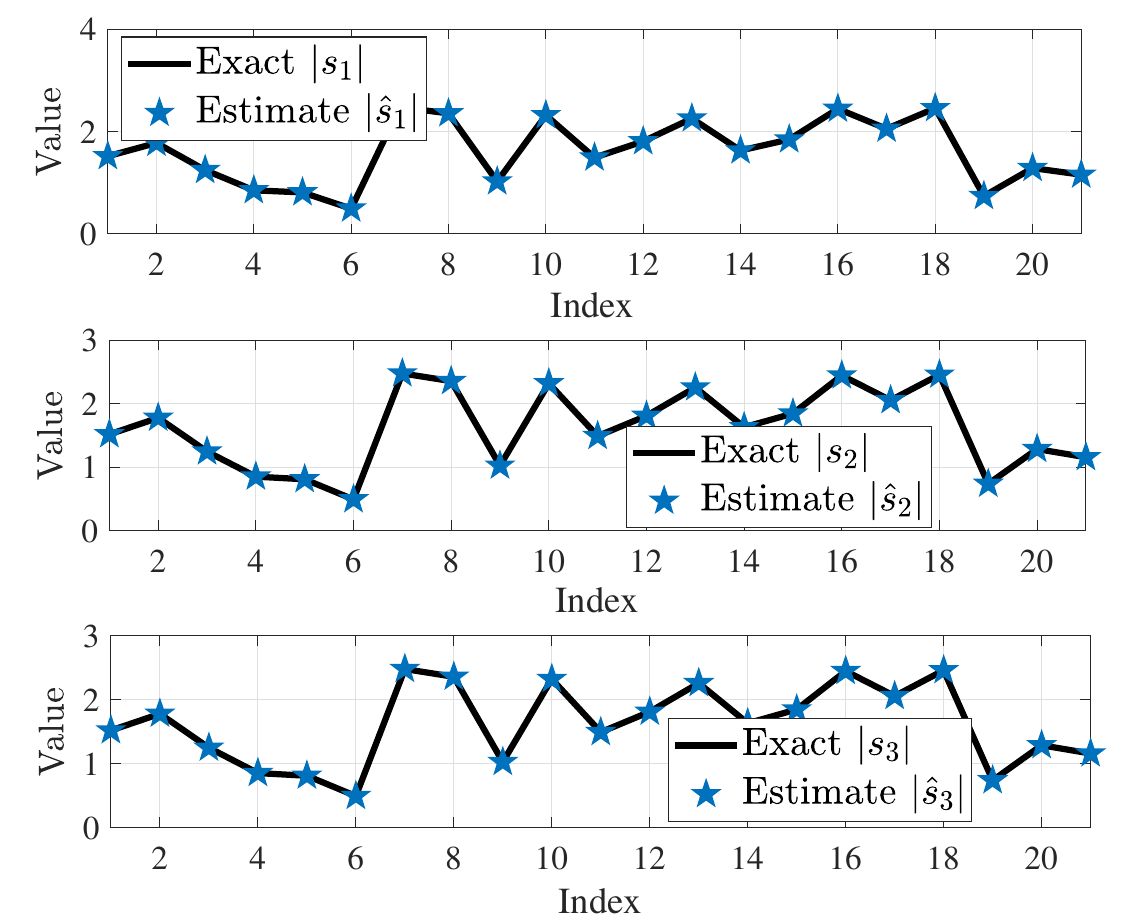}}
\caption{(a) Plot of $||\fv\left(\rv\right)||_{2}^{2}$ and the locations of the estimated shifts. (b) Comparing the estimated samples of the signals with the true ones.}
\label{fig: per 3_2 all}
\end{figure}

In the third experiment, we consider the case of $K=1$ and we set $L=21$ and $R=3$. The real and the imaginary parts of the entries of $\Dm$ are generated from a uniform distribution in $[-1,1]$ while $\hv_{j}$ are set as in the previous scenarios. Moreover, we let the real and the imaginary parts of $c_{j}$ to be fading, i.e., equal to $0.5+w^{2}$ where $w \sim \mathcal{N} \left(0,1\right)$ and we generate their signs uniformly in $[-1,1]$. Finally, the locations of the shifts are set to be $\left(0.8,0.2\right), \left(0.1,0.4\right),$ and $\left(0.7,0.6\right)$. From Fig~\ref{fig: per 3_1}, we can see that our approach recovers the shifts precisely whereas from Fig~\ref{fig: per 3_2} we can see that the estimated samples coincide with the true ones. Furthermore, we have $|\hv_{1}^{H}\hat{\hv}_{1}|\hspace{-3pt}=\hspace{-3pt}1+10^{-15}$, $\hspace{-1pt}|\hv_{2}^{H}\hat{\hv}_{2}|\hspace{-3pt}=\hspace{-3pt}1\hspace{-1pt}+2\times 10^{-15}$, $|\hv_{3}^{H}\hat{\hv}_{3}|\hspace{-3pt}=1-10^{-15}$.

Finally, we study the stability of the framework to the noise using simulation with the theoretical analysis being left to future work. Here, we set $L=15, K=3, R=1,$ and we use the same settings in the first scenario for $\Dm$ and $\hv_{j}$ and in the previous experiment for $c_{j}$. The shift pair is set at $\left(0.74,0.30\right)$. Then, a Gaussian noise vector $\tilde{\nv}$ is added to $\yv$ at $10$ dB signal-to-noise-ratio (SNR), i.e., SNR (dB) $= 10\log_{10}\left(\frac{||\yv||_{2}^{2}}{||\tilde{\nv}||_{2}^{2}}\right)$.

\begin{figure}[h!]
\vspace{-7pt}
\centering
\subfigure[]{\label{fig: per error}\includegraphics[width=1.72in]{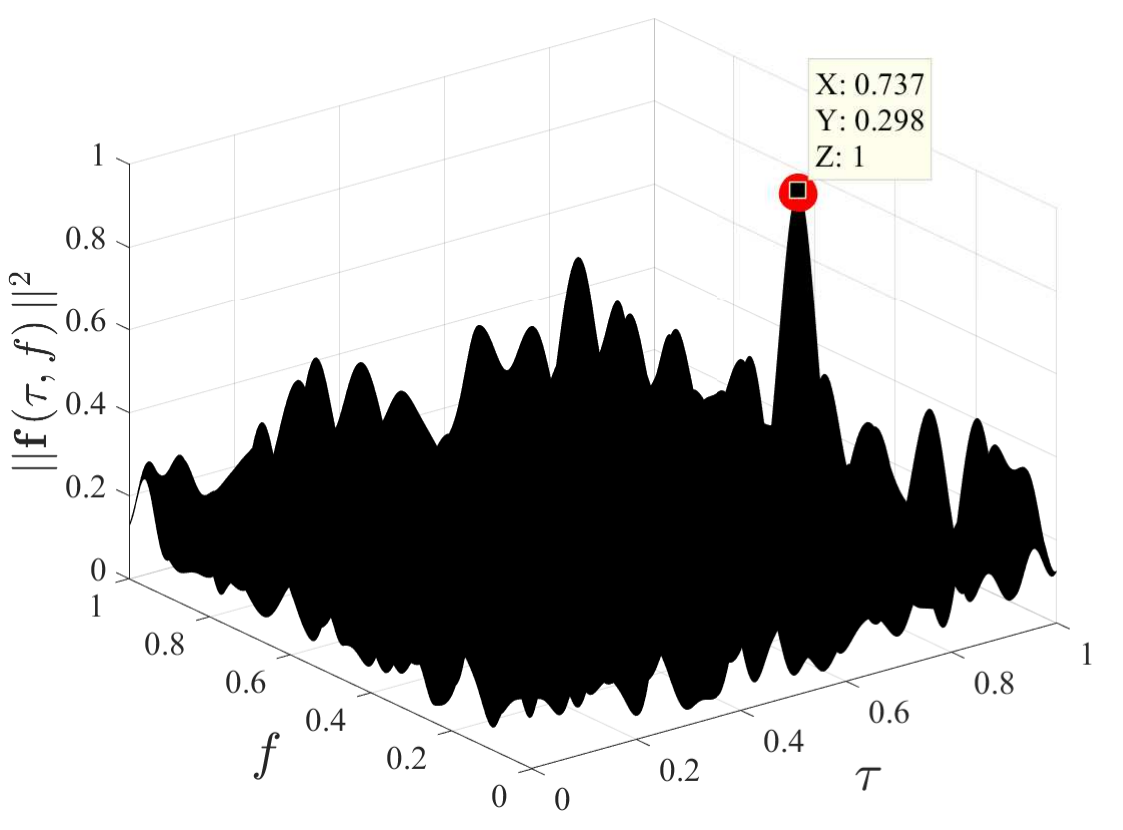}}
\subfigure[]{\label{fig: per error_2}\includegraphics[width=1.72in]{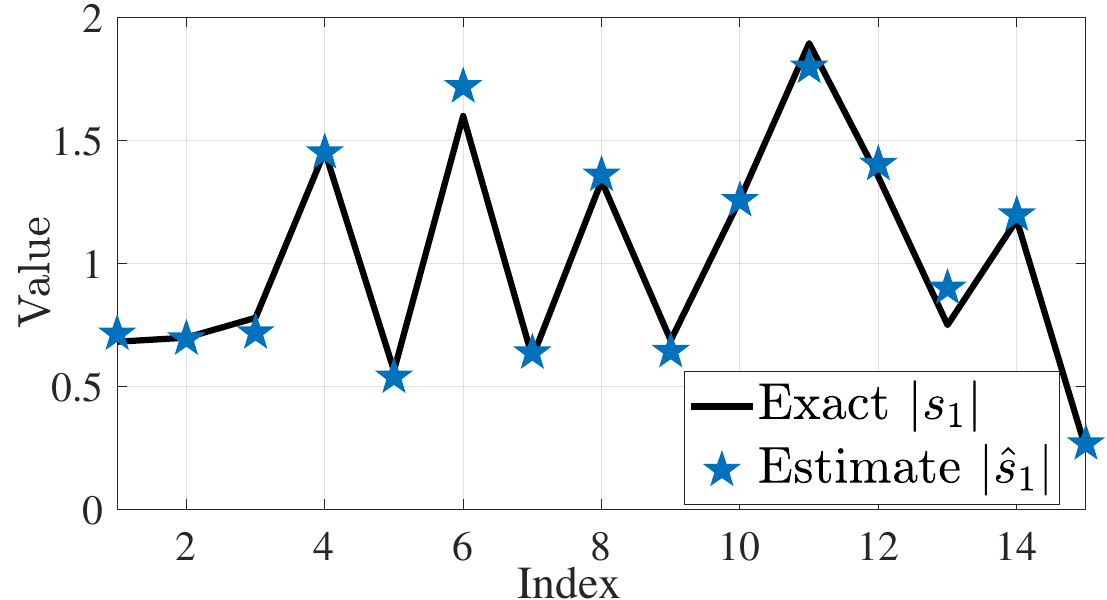}}
\caption{(a) Plot of $||\fv\left(\rv\right)||_{2}^{2}$ and the location of the estimated shift. (b) Comparing the estimated samples of the signal with the true ones.}
\label{fig: per 3_2 all}
\end{figure}

To solve (\ref{eq: model noise }), we obtain its semidefinite relaxation as
\begin{align}
\label{eq: dual of noise}
\mathcal{P}_{5}:\ \ \ &\underset{\qv, \Qm}{\text{maximize}} \ \langle\qv, \yv\rangle_{\mathbb{R}} - \zeta ||\qv||_{2} \nonumber\\
 &\text{subject to the constraints  of (\ref{eq: dual of the dual})}.
\end{align}
In Fig~\ref{fig: per error}, we plot $||\fv\left(\rv\right)||_{2}^{2}$ that is obtained by using $\qv$ upon solving (\ref{eq: dual of noise}) with  $\zeta=3$. The shift pair at which $||\fv\left(\rv\right)||_{2}^{2}=1$ is found to be $\left(0.737,0.298\right)$ which is too close to the original one. Moreover, Fig~\ref{fig: per error_2} shows that the magnitudes of the estimated signal samples are close to the original ones with a tenuous error. Finally, we find that $|\hv_{1}^{H}\hat{\hv}_{1}|= 0.9674$.

%%%%%%%%%%%%%%%%%%%%%%%%%%%%%%%%%%%%%%%%%%%%%%%%%%%%%%%%%%%%%%%%%%%%%%%%%%%%%%%%%%%%%%%%%%%%%
%%%%%%%%%%%%%%%%%%%%%%%%%%%%%%%%%%%%%%%%%%%%%%%%%%%%%%%%%%%%%%%%%%%%%%%%%%%%%%%%%%%%%%%%%%%%%
\section{Constructing the Dual Vector Polynomial: Proof of Theorem~\ref{th: main the}}
\label{sec: polynomial}
In this section, we prove Theorem~\ref{th: main the} by formulating $\fv\left(\rv\right)$ that satisfies (\ref{eq: hold assump 1}) and (\ref{eq: hold assump 2}). Obtaining such polynomial guarantees that the primal optimal solution is equal to $\Um$.

Starting from (\ref{eq: dual polynomail equa}), and based on (\ref{eq: hold assump 1}) and (\ref{eq: hold assump 2}), our goal is to acquire an expression for $\fv\left(\rv\right)$ that satisfies 
\begin{align}
\label{eq: con 1}
&\fv\left(\rv_{j}\right)= \text{sign}\left(c_{j}\right) \hv_{j} \hspace{10pt} \forall \rv_{j} \in \mathcal{R}\\
\label{eq: con 2}
-&\fv^{\left(1,0\right)}\left(\rv_{j}\right)= \bm{0}_{K\times 1} \hspace{17pt} \forall \rv_{j} \in \mathcal{R}\\
\label{eq: con 3}
-&\fv^{\left(0,1\right)}\left(\rv_{j}\right)= \bm{0}_{K\times 1}\hspace{17pt} \forall \rv_{j} \in \mathcal{R},
\end{align}
where $\fv^{(m',n')}\left(\rv\right) := \frac{\partial^{m'}}{\partial \tau^{m'}} \frac{\partial^{n'}}{\partial f^{n'}} \fv\left(\rv\right)$. Note that (\ref{eq: con 2}) and (\ref{eq: con 3}) ensure that $\fv\left(\rv\right)$ approaches a local minimum at $\rv_{j}$ which is a necessary condition for (\ref{eq: hold assump 2}) to hold. Before formulating $\fv\left(\rv\right)$, we recall some related results and definitions in the literature.

In the blind 1D super-resolution (only delay shift), the dual 1D polynomial in \cite{yang2016super} guarantees the optimality of the recovery problem. The authors in \cite{yang2016super} show that there exists a vector polynomial $\fv_{\text{1D}}\left(\tau\right)$ with: (a) $\fv_{\text{1D}}\left(\tau_{j}\right) = \text{sign}\left(c_{j}\right) \hv_{j}, \ \forall \tau_{j} \in \mathcal{R}$, \ (b) $||\fv_{\text{1D}}\left(\tau_{j}\right)||_{2} <1, \  \forall \tau_{j} \not\in \mathcal{R}$, where $\mathcal{R}$ is the set of the true shifts. This polynomial is formulated by solving a weighted least energy minimization problem and is found to be
\begin{equation}
\label{eq: tang poly}
{\fv}_{\text{1D}}\left(\tau\right) = \sum_{j=1}^{R} \Mm_{\text{1D}}\left(\tau-\tau_{j}\right) \alphav_{j} + \sum_{j=1}^{R} \Mm_{\text{1D}}^{'}\left(\tau-\tau_{j}\right) \betav_{j},
\end{equation} 
where $\Mm_{\text{1D}}\left(\tau\right) \in \mathbb{C}^{K \times K}$ is a random kernel while $\alphav_{j}, \betav_{j} \in \mathbb{C}^{K \times 1}$ are vector parameters. Finally, $\Mm_{\text{1D}}^{'}\left(\tau\right)$ is the entry-wise derivative of $\Mm_{\text{1D}}\left(\tau\right)$ with respect to (w.r.t.) $\tau$. 

To show that (\ref{eq: tang poly}) satisfies (a) and (b) above, the authors first show that under certain assumptions, the expected value of the $m$-th derivative of $\Mm_{\text{1D}}\left(\tau\right)$, i.e., $\mathbb{E}\left[\Mm_{\text{1D}}^{m}\left(\tau\right)\right]$ is $F^{m}\left(\tau\right) \Id_{\text{K}}$ where $F\left(t\right) := \left(\frac{\sin \left(T \pi t \right)}{T \sin \left(\pi t\right)}\right)^{4}; T := \frac{N}{2}+1$ 
%\begin{equation}
%%\label{eq: fejer}
%F\left(t\right) := \left(\frac{\sin \left(T \pi t \right)}{T \sin \left(\pi t\right)}\right)^{4}, \ \ T := \frac{N}{2}+1 \nonumber
%\end{equation} 
is the squared Fej\'er kernel. When $N$ is even, the Fej\'er kernel is a trigonometric polynomial of degree $(T-1)$ and can be written as 
\begin{equation}
\label{eq: another fejer}
F\left(t\right) = \frac{1}{T} \sum_{n=-N}^{N} g_{n} e^{i2 \pi n t},
\end{equation}
where 
%\begin{equation}
%\label{eq: fejer coe}
$g_{n}= \frac{1}{T} \sum_{l=\text{max}\{n-T,-T\}}^{\text{min}\{n+T,T\}} \left(1-\frac{|l|}{T}\right) \left(1-\frac{|n-l|}{T}\right).$
%\end{equation}
%Given that $F\left(0\right)=0 $, the value of $||\bar{\Mm}_{\text{1D}}\left(\tau\right)||$ will decay rapidly around $\tau=0$. 

Following that, the authors prove that there exist coefficients $\bar{\alphav}_{j}, \bar{\betav}_{j} \in \mathbb{C}^{K \times 1}, j=1,\dots, R$ such that
$\bar{\fv}_{\text{1D}}\left(\tau_{j}\right) = \text{sign}\left(c_{j}\right) \hv_{j}, \forall \tau_{j} \in \mathcal{R}$ and that $||\bar{\fv}_{\text{1D}}\left(\tau_{j}\right)||_{2} <1, \forall \tau_{j} \not\in \mathcal{R}, \nonumber
$ 
where $\bar{\fv}_{\text{1D}}\left(\tau\right) := \mathbb{E}\left[\fv_{\text{1D}}\left(\tau\right)\right]$. Finally, $\fv_{\text{1D}}\left(\tau\right)$ is shown to concentrate around $\bar{\fv}_{\text{1D}}\left(\tau\right)$ anywhere in $[0,1)$ with high probability. The fundamental idea about (\ref{eq: tang poly}) is that $\Mm_{\text{1D}}\left(\tau-\tau_{j}\right)$ interpolates $\text{sign}\left(c_{j}\right) \hv_{j}$ while $\Mm_{\text{1D}}^{'}\left(\tau-\tau_{j}\right)$ adapts this interpolation to ensure that local maxima are reached at $\tau_{j}$. This strategy is first developed in \cite{tang2013compressed} and then adapted and applied in different works in the literature, e.g., \cite{li2016off, chi2016guaranteed, heckel2016super}.

Inspired by the previous methodology and other related prior works on super-resolution, e.g., \cite{candes2014towards, heckel2016super, bendory2015super, yang2016vandermonde}, we seek to construct a 2D trigonometric vector polynomial $\fv\left(\rv\right)$ that satisfies (\ref{eq: hold assump 1}) and (\ref{eq: hold assump 2}). However, before going into in-depth technical details, it is essential to first highlight some crucial remarks. First, while (\ref{eq: tang poly}) is obtained by solving a weighted least energy minimization problem, it is impossible to generalize this problem to the 2D case upon using multiple proper weighting matrices due to nature of the problem formulation. Second, since the 2D dual vector polynomial, the interpolation functions, and the correction functions are all random, we will have to apply probabilist approaches to show that (\ref{eq: hold assump 1}) and (\ref{eq: hold assump 2}) hold true on our obtained $\fv\left(\rv\right)$. Third, given the structure of $\fv\left(\rv\right)$ in (\ref{eq: simple polynomail}), and unlike (\ref{eq: tang poly}), we cannot merely use the derivatives of the interpolating matrix as a correction function. This is due to the fact that the derivatives of a polynomial in the form as in (\ref{eq: simple polynomail}) do not necessarily have the structure in (\ref{eq: simple polynomail}). Finally, we cannot interpolate $\text{sign}\left(c_{j}\right)\hv_{j}$ using shifted versions of a single function as shifted versions of a function that represents (\ref{eq: simple polynomail}) do not necessarily have the form of (\ref{eq: simple polynomail}).  

In this paper, we set $\fv\left(\rv\right)$ using multiple random kernel matrices $\Mm_{\left(m,n\right)} \left(\rv,\rv_{j}\right) \in \mathbb{C}^{K \times K}, m,n=0,1$ in the form
\begin{eqnarray}
\label{eq: final dual}
&\fv\left(\rv\right) =\sum_{j=1}^{R} \Mm_{\left(0,0\right)}\left(\rv,\rv_{j}\right) \alphav_{j} + \Mm_{\left(1,0\right)}\left(\rv,\rv_{j}\right) \betav_{j}\nonumber\\
&+ \Mm_{\left(0,1\right)}\left(\rv,\rv_{j}\right) \gammav_{j}.
\end{eqnarray} 
The key factor of this formulation is to interpolate the vectors $\text{sign}\left(c_{j}\right) \hv_{j}$ at $\rv_{j}$ using $\Mm_{\left(0,0\right)}\left(\rv,\rv_{j}\right)$ and then to adjust this interpolation near $\rv_{j}$ by $\Mm_{\left(1,0\right)}\left(\rv,\rv_{j}\right)$ and $\Mm_{\left(0,1\right)}\left(\rv,\rv_{j}\right)$ to ensure that $\fv\left(\rv\right)$ approaches local maxima at $\rv_{j}$. The central question here is how to appropriately select the kernel matrices such that $\fv\left(\rv\right)$ satisfies (\ref{eq: hold assump 1}) and (\ref{eq: hold assump 2}). Note that it is clear based on (\ref{eq: simple polynomail}) that formulating $\fv\left(\rv\right)$ is achieved by finding the proper choice of $\qv$. With all that said, our strategy will be as follows: 
\begin{itemize}
\item We obtain an initial expression for ${\fv}\left(\rv\right)$, i.e., $\hat{\fv}\left(\rv\right)$, with
\begin{eqnarray}
\vspace{-6pt}
\label{eq: inti for}
\hat{\fv}\left(\rv\right) = \sum_{p=-N}^{N} [\hat{\qv}]_{p} \widetilde{\Dm}_{p}^{H} \av\left(\rv\right), \ \ \hat{\qv}\in \mathbb{C}^{K \times 1}; \ \hat{\qv}\neq \qv
\vspace{-6pt}
\end{eqnarray}
where $\hat{\qv}$ has unconstrained coefficients which obtained by solving unweighted least energy minimization problem. 
\item Then, we adapt this formulation by using multiple weighting functions to obtain $\Mm_{\left(m,n\right)}\left(\rv,\rv_{j}\right)$ and $\fv\left(\rv\right)$. 
\item Finally, we dedicate the remaining parts of this section to show that the obtained $\fv\left(\rv\right)$ satisfies (\ref{eq: hold assump 1}) and (\ref{eq: hold assump 2}).
\end{itemize}
To start with, consider the following linear systems $\forall \rv_{j} \in \mathcal{R}$
\begin{eqnarray}
\label{eq: equations for temp}
\hat{\fv}\left(\rv_{j}\right)= \text{sign}\left(c_{j}\right)\hv_{j}, -\hat{\fv}^{\left(1,0\right)}\left(\rv_{j}\right)= -\hat{\fv}^{\left(0,1\right)}\left(\rv_{j}\right)= \bm{0}_{K\times 1}.
\end{eqnarray}
Then, we consider solving the following problem
\begin{align}
\label{eq: poly optimzation}
  \mathcal{P}_{6}: \ \ \ &\underset{\hat{\qv}}{\text{minimize}} \ ||\hat{\qv}||_{2}^{2} \ \  \text{subject to} : (\ref{eq: equations for temp})
\end{align}
By using (\ref{eq: inti for}) we can rewrite (\ref{eq: poly optimzation}) as
\begin{eqnarray}
\label{eq: poly optimzation sim}
\mathcal{P}_{7}: \ \ \ &\underset{\hat{\qv}}{\text{minimize}} ||\hat{\qv}||_{2}^{2}\ \ \ \text{subject to}: \ \ \Fm \hat{\qv} = \gv,
\end{eqnarray}
where $\Fm \in \mathbb{C}^{3RK \times L}$ is given by
\begingroup
\fontsize{9.pt}{9.5pt}
\begin{equation}
%\label{eq:matrix h}
\Fm = {\begin{bmatrix}  \widetilde{\Dm}_{-N}^{H} \av\left(\rv_{1}\right) &\hdots  & \widetilde{\Dm}_{N}^{H} \av\left(\rv_{1}\right) \\ \vdots  & \vdots \\ \widetilde{\Dm}_{-N}^{H} \av\left(\rv_{R}\right) &\hdots  & \widetilde{\Dm}_{N}^{H} \av\left(\rv_{R}\right) \\ -\widetilde{\Dm}_{-N}^{H} \av^{\left(1,0\right)}\left(\rv\right)|_{\rv=\rv_{1}} &\hdots  & -\widetilde{\Dm}_{N}^{H} \av^{\left(1,0\right)}\left(\rv\right)|_{\rv=\rv_{1}} \\  \vdots  & \vdots \\
-\widetilde{\Dm}_{-N}^{H} \av^{\left(1,0\right)}\left(\rv\right)|_{\rv=\rv_{R}} &\hdots  & -\widetilde{\Dm}_{N}^{H} \av^{\left(1,0\right)}\left(\rv\right)|_{\rv=\rv_{R}} \\-\widetilde{\Dm}_{-N}^{H} \av^{\left(0,1\right)}\left(\rv\right)|_{\rv=\rv_{1}} &\hdots  & -\widetilde{\Dm}_{N}^{H} \av^{\left(0,1\right)}\left(\rv\right)|_{\rv=\rv_{1}} \\  \vdots  & \vdots \\
-\widetilde{\Dm}_{-N}^{H} \av^{\left(0,1\right)}\left(\rv\right)|_{\rv=\rv_{R}} &\hdots  & -\widetilde{\Dm}_{N}^{H} \av^{\left(0,1\right)}\left(\rv\right)|_{\rv=\rv_{R}}
  \end{bmatrix}} \nonumber
\end{equation}
\endgroup
while $\gv = \begin{bmatrix} \text{sign}\left(c_{1}\right)\hv_{1}^{T}, \hdots, \text{sign}\left(c_{R}\right)\hv_{R}^{T}, \bm{0}_{K \times 1}^{T}, \hdots, \bm{0}_{K \times 1}^{T} \end{bmatrix}^{T}$ $\in \mathbb{C}^{3RK \times 1}$. Using the KKT optimality conditions \cite[Section~5.5.3]{boyd2004convex}, we can show that the solution of (\ref{eq: poly optimzation sim}) is
\begin{equation}
\label{eq: opti solution}
\hat{\qv} = \Fm^{H} \vv,
\end{equation}
where $\vv  = \begin{bmatrix} \alphav^{T}, \betav^{T} , \gammav^{T} \end{bmatrix}^{T}$ with $\alphav  = \begin{bmatrix} \alphav_{1}^{T}, \dots , \alphav_{R}^{T} \end{bmatrix}^{T}$, $\betav  = \begin{bmatrix} \betav_{1}^{T}, \dots ,\betav_{R}^{T} \end{bmatrix}^{T}$, $\gammav  = \begin{bmatrix}\gammav_{1}^{T}, \dots , \gammav_{R}^{T} \end{bmatrix}^{T}$ such that $\alphav_{j}, \betav_{j}, \gammav_{j} \in \mathbb{C}^{K \times 1}$. By substituting for $\Fm$ and $\vv$ in (\ref{eq: opti solution}) we obtain 
\begingroup
\fontsize{9.3pt}{9.5pt}
\begin{align}
\label{eq: q formulation}
\hat{\qv}&= \sum_{j=1}^{R} \left( \begin{bmatrix} \av\left(\rv_{j}\right)^{H} \widetilde{\Dm}_{-N}\\ \vdots \\ \av\left(\rv_{j}\right)^{H} \widetilde{\Dm}_{N}  \end{bmatrix} \alphav_{j}  -  \begin{bmatrix}  \av^{\left(1,0\right)}\left(\rv\right)^{H}|_{\rv=\rv_{j}} \widetilde{\Dm}_{-N} \\ \vdots  \\ \av^{\left(1,0\right)}\left(\rv\right)^{H}|_{\rv=\rv_{j}} \widetilde{\Dm}_{N} \end{bmatrix} \betav_{j}  \right.  \nonumber\\
&\left. - \begin{bmatrix} \av^{\left(0,1\right)}\left(\rv\right)^{H}|_{\rv=\rv_{j}} \widetilde{\Dm}_{-N} \\ \vdots  \\  \av^{\left(0,1\right)}\left(\rv\right)^{H}|_{\rv=\rv_{j}} \widetilde{\Dm}_{N} \end{bmatrix} \gammav_{j} \right).
\end{align}
\endgroup
Now we substitute (\ref{eq: q formulation}) in (\ref{eq: inti for}) and manipulate to obtain
%\begin{align}
%\label{eq: another dual poly}
%\hat{\fv}\left(\rv\right) &=\hspace{-2pt}  \sum_{j=1}^{R} \sum_{p=-N}^{N} \hspace{-3pt} \bigg(\hspace{-1pt}  \av\left(\rv_{j}\right)^{H}\hspace{-1pt}  \widetilde{\Dm}_{p} \alphav_{j} \widetilde{\Dm}_{p}^{H} \av\left(\rv\right)-\av^{\left(1,0\right)}\left(\rv\right)^{H}|_{\rv=\rv_{j}} \times \nonumber\\
%& \widetilde{\Dm}_{p} \betav_{j} \widetilde{\Dm}_{p}^{H} \av\left(\rv\right)-\av^{\left(0,1\right)}\left(\rv\right)^{H}|_{\rv=\rv_{j}} \widetilde{\Dm}_{p} \gammav_{j} \widetilde{\Dm}_{p}^{H} \av\left(\rv\right) \bigg) \nonumber\\
%&= \sum_{j=1}^{R}\left[\left( \sum_{p=-N}^{N} \widetilde{\Dm}_{p}^{H} \av\left(\rv\right) \av^{(0,0)}\left(\rv\right)^{H}|_{\rv=\rv_{j}} \widetilde{\Dm}_{p}\right) \alphav_{j} \right.\nonumber\\
%& + \left.\left(- \sum_{p=-N}^{N}\widetilde{\Dm}_{p}^{H} \av\left(\rv\right)\av^{\left(1,0\right)}\left(\rv\right)^{H}|_{\rv=\rv_{j}} \widetilde{\Dm}_{p}\right) \betav_{j}\right.\nonumber\\
%&\left. +\left( -\sum_{p=-N}^{N}\widetilde{\Dm}_{p}^{H} \av\left(\rv\right)\av^{\left(0,1\right)}\left(\rv\right)^{H}|_{\rv=\rv_{j}} \widetilde{\Dm}_{p} \right) \gammav_{j}\right].
%\end{align}
\begingroup
\fontsize{9.3pt}{9.5pt}
\begin{align}
\label{eq: another dual poly}
\hat{\fv}\left(\rv\right) &= \sum_{j=1}^{R}\left[\left( \sum_{p=-N}^{N} \widetilde{\Dm}_{p}^{H} \av\left(\rv\right) \av^{(0,0)}\left(\rv\right)^{H}|_{\rv=\rv_{j}} \widetilde{\Dm}_{p}\right) \alphav_{j} \right.\nonumber\\
& + \left.\left(- \sum_{p=-N}^{N}\widetilde{\Dm}_{p}^{H} \av\left(\rv\right)\av^{\left(1,0\right)}\left(\rv\right)^{H}|_{\rv=\rv_{j}} \widetilde{\Dm}_{p}\right) \betav_{j}\right.\nonumber\\
&\left. +\left( -\sum_{p=-N}^{N}\widetilde{\Dm}_{p}^{H} \av\left(\rv\right)\av^{\left(0,1\right)}\left(\rv\right)^{H}|_{\rv=\rv_{j}} \widetilde{\Dm}_{p} \right) \gammav_{j}\right].
\end{align}
\endgroup
Upon defining the matrix $\widehat{\Mm}^{(m',n')}\left(\rv,\rv_{j}\right) \in \mathbb{C}^{K\times K}$ such that 
\vspace{-11pt}
%\begin{eqnarray}
%\label{eq: temp G matrix}
%\widehat{\Mm}^{(m',n')}\hspace{-2pt}\left(\rv,\rv_{j}\right) =\hspace{-2pt}-1^{m'+n'}\hspace{-5pt}\sum_{p=-N}^{N} \hspace{-5pt}\widetilde{\Dm}_{p}^{H} \av\left(\rv\right) \av^{(m',n')}\hspace{-2pt}\left(\rv\right)^{H}\hspace{-2pt}|_{\rv=\rv_{j}} \hspace{-2pt}\widetilde{\Dm}_{p}
%\end{eqnarray}
%\begingroup
%\fontsize{9pt}{9.5pt}\hspace{-1pt}
\begin{equation}
\label{eq: temp G matrix}
\hspace{-10pt}\widehat{\Mm}^{(m',n')}\hspace{-2pt}\left(\hspace{-1pt}\rv,\rv_{j}\hspace{-1pt}\right)\hspace{-2pt} =\hspace{-2pt}-1^{m'+n'}\hspace{-5pt}\sum_{p=-N}^{N}\hspace{-2pt} \widetilde{\Dm}_{p}^{H} \av\left(\rv\right) \av^{(m',n')}\hspace{-2pt}\left(\rv_{j}\right)^{H}\hspace{-2pt} \widetilde{\Dm}_{p}
\end{equation}
%\endgroup
%\begingroup
%\fontsize{9pt}{9.5pt}
%\begin{align}
%\label{eq: temp G matrix}
%&\widehat{\Mm}^{(m',n')}\left(\rv,\rv_{j}\right) =\nonumber\\
%& \left(-1\right)^{m'+n'}\sum_{p=-N}^{N} \widetilde{\Dm}_{p}^{H} \av\left(\rv\right) \av^{(m',n')}\left(\rv\right)^{H}|_{\rv=\rv_{j}} \widetilde{\Dm}_{p}
%\end{align}
%\endgroup
\vspace{-2pt}
we can rewrite (\ref{eq: another dual poly}) as
\vspace{-3pt}
\begin{eqnarray}
\label{eq: temp f matrix G}
\hat{\fv}\left(\rv\right)=  \sum_{j=1}^{R} &\widehat{\Mm}^{(0,0)}\left(\rv,\rv_{j}\right) \alphav_{j} +\widehat{\Mm}^{(1,0)}\left(\rv,\rv_{j}\right) \betav_{j} \nonumber\\
&+ \widehat{\Mm}^{(0,1)}\left(\rv,\rv_{j}\right) \gammav_{j}.
\vspace{-8pt}
\end{eqnarray}
which provides the initial formulation for $\fv\left(\rv\right)$ as in (\ref{eq: inti for}). Now, we can turn our attention into obtaining $\Mm_{\left(m,n\right)}\left(\rv,\rv_{j}\right)$ and as a result $\fv\left(\rv\right)$ by adapting (\ref{eq: temp G matrix}). Following that, we will provide our justifications for this proposed adaptation.

To start with, consider $\zv_{p}\left(\rv\right)_{\left(m,n\right)}\in \mathbb{C}^{L^{2} \times 1}$ such that
\begin{align}
\label{eq: z definition}
&[\zv_{p}\left(\rv_{j}\right)_{\left(m,n\right)}]_{(\left(k,l\right),1)} :=  g_{k}\left(i2\pi k \right)^{m} g_{p}\left(i2\pi p \right)^{n}  e^{\frac{i2\pi k \left(p+l\right)}{L}} \times \nonumber\\
&e^{-i 2 \pi \left(k\tau_{j}+pf_{j}\right)}; \ \ \ p,k,l=-N,\dots, N; \ \ m,n=0,1.
\end{align}
%where $g_{k}$ and $g_{p}$ are given by (\ref{eq: fejer coe}) and $m,n=0,1$. Based on (\ref{eq: z definition}), we propose formulating our random kernel matrix as
Based on (\ref{eq: z definition}), we propose formulating our kernel matrix as
\begin{equation}
\label{eq: G matrix}
\Mm_{\left(m,n\right)}\left(\rv,\rv_{j}\right) := \frac{1}{T^{2}} \sum_{p=-N}^{N} \widetilde{\Dm}_{p}^{H}\av\left(\rv\right) \zv_{p}\left(\rv_{j}\right)_{\left(m,n\right)}^{H} \widetilde{\Dm}_{p}.
\end{equation}
By using (\ref{eq: z definition}) and (\ref{eq: matrix D}) we can show that
\begin{align}
\label{eq: z d mult}
\zv_{p}\left(\rv_{j}\right)_{\left(m,n\right)}^{H} \widetilde{\Dm}_{p} = \hspace{-4pt}\sum_{l,k=-N}^{N}& g_{k}\left(-i2\pi k \right)^{m} g_{p}\left(-i2\pi p \right)^{n}  e^{\frac{-i2\pi k l}{L}} \times \nonumber\\
& e^{i 2 \pi \left(k\tau_{j}+pf_{j}\right)} \dv_{\left(p-l\right)}^{H}.
\end{align}
Moreover, we can also deduce that 
\begin{align}
\label{eq: int 1}
\av\left(\rv\right)^{H} \widetilde{\Dm}_{p} = \frac{1}{L}  \sum_{l,k=-N}^{N} e^{\frac{-i2\pi k l}{L}}  e^{i 2 \pi \left( k \tau +p f \right)}\dv^{H}_{\left(p-l\right)} .
\end{align}
Now from (\ref{eq: z d mult}) and (\ref{eq: int 1}) we can write
\begin{align}
\label{eq: G simpl}
&\Mm_{\left(m,n\right)}\left(\rv,\rv_{j}\right) = \nonumber\\
& \frac{1}{L} \sum_{p=-N}^{N} \frac{1}{T^{2}}\sum_{l,l',k,k'=-N}^{N} g_{k'}\left(-i2\pi k' \right)^{m} g_{p} \left(-i2\pi p \right)^{n}  \times \nonumber\\
& e^{i 2 \pi \frac{\left(kl-k'l'\right)}{L}} e^{-i2\pi \left(k\tau-k'\tau_{j}\right)} e^{-i2\pi p \left(f-f_{j}\right)}\dv_{\left(p-l\right)} \dv_{\left(p-l'\right)}^{H}.  
\end{align}
On the other hand, we can also show that
\begin{align}
\label{eq: int 2}
\widehat{\Mm}^{(0,0)}\left(\rv,\rv_{j}\right) &= 
\sum_{p=-N}^{N}  \frac{1}{L^{2}} \sum_{l,l',k,k'=-N}^{N}  e^{i 2 \pi \frac{\left(kl-k'l'\right)}{L}} \times  \nonumber\\
&\hspace{-5pt} e^{-i2\pi \left(k\tau-k'\tau_{j}\right)} e^{-i2\pi p \left(f-f_{j}\right)} \dv_{\left(p-l\right)}  \dv^{H}_{\left(p-l'\right)}.
\end{align}
Since $\fv\left(\rv\right)$ is a linear combinations of $\Mm_{\left(m,n\right)}\left(\rv,\rv_{j}\right)$, it is easy to show that it has the form in (\ref{eq: simple polynomail}) as required. Comparing $\Mm_{\left(m,n\right)}\left(\rv,\rv_{j}\right)$ with $\widehat{\Mm}^{(0,0)}\left(\rv,\rv_{j}\right)$, we can see that $\Mm_{\left(m,n\right)}\left(\rv,\rv_{j}\right)$ is a scaled version of $\widehat{\Mm}^{(0,0)}\left(\rv,\rv_{j}\right)$. The appointed choice of $\Mm_{\left(m,n\right)}\left(\rv,\rv_{j}\right)$, as we will show in Section~\ref{sec: proof small}, is motivated by the fact that it concentrates around its average deterministic version $\mathbb{E}[{\Mm}_{\left(m,n\right)}\left(\rv,\rv_{j}\right)]$ in Euclidean norm measure with high probability. This fact is crucial in showing that (\ref{eq: hold assump 1}) is satisfied (as will be shown in Section~\ref{sec: first cond}) and is also found to facilitate the proofs and to yield nicely constants. More importantly, the expression of $\Mm_{\left(m,n\right)}\left(\rv,\rv_{j}\right)$, as we will show in the remaining parts of this section, provides $\fv\left(\rv\right)$ that satisfies (\ref{eq: hold assump 1}) and (\ref{eq: hold assump 2}), which then guarantees the existence of the dual optimal solution and thus our required primal optimal solution $\Um$. We point out that anyone might suggest and use different formulation for the kernel matrices as long as they provide $\fv\left(\rv\right)$ that satisfies (\ref{eq: hold assump 1}) and (\ref{eq: hold assump 2}) and follow the same proof techniques that will be provided in this paper. Finally, by substituting (\ref{eq: G matrix}) in (\ref{eq: final dual}) we can formulate our dual trigonometric vector polynomial $\fv\left(\rv\right)$. 

Before closing this part, we will express the derivatives of $\Mm_{\left(m,n\right)}\left(\rv,\rv_{j}\right)$, i.e., $\Mm^{(m',n')}_{\left(m,n\right)}\left(\rv,\rv_{j}\right)$, in a matrix-vector form that involves $\Dm$ to facilitate our proofs later. For that, let us first define a modified version of (\ref{eq: vec 1}) as 
\begin{align}
\label{eq: matrix equi 1}
\left[\av_{p}\left(\rv\right)\right]_{(\left(k,l\right),1)} =  & D_{N}\left(\frac{k}{L}-f\right) D_{N}\left(\frac{ p-l}{L}-\tau\right),
\end{align}
with $\hspace{-1pt}p= \hspace{-2pt}-N, \dots, N$. From the periodicity property we write
\begin{equation}
\label{eq: matrix equi 2}
\widetilde{\Dm}_{p}^{H}\av\left(\rv\right) = \widehat{\Dm}_{p}^{H}\av_{p}\left(\rv\right),
\end{equation}
where $[\widehat{\Dm}_{p}]_{(\left(k,l\right),1 \to K)} = \dv_{l}^{H} e^{\frac{i 2 \pi p k}{L}}$.
%\begin{equation}
%\label{eq:tilde matrix D}
%[\widehat{\Dm}_{p}]_{(\left(k,l\right),1 \to K)} = \dv_{l}^{H} e^{\frac{i 2 \pi p k}{L}}, \ \ \ p, k, l = -N, \dots, N.
%\end{equation}
%where $\widehat{\Dm}_{p} \in \mathbb{C}^{L^{2}\times K}$ is given by
%\begin{equation}
%\label{eq:tilde matrix D}
%[\widehat{\Dm}_{p}]_{(\left(k,l\right),1 \to K)} = \dv_{l}^{H} e^{\frac{i 2 \pi p k}{L}}, \ \ \ p, k, l = -N, \dots, N.
%\end{equation}
Moreover, define the block diagonal matrix $\Jm_{p} \in \mathbb{C}^{L^{2} \times L^{2}}$ as
\begin{equation}
\label{eq: matrix E}
\Jm_{p} =\text{diag}\left(\left(\Jm_{p}^{-N},\dots, \Jm_{p}^{N}\right)\right), 
\end{equation}
where $\Jm_{p}^{k} := e^{\frac{-i2\pi p k}{L}} \Id_{\text{L}}, \ \ k=-N,\dots,N.
$
%\begin{equation}
%\label{eq: matrix Ek}
%\Jm_{p}^{k} := e^{\frac{-i2\pi p k}{L}} \Id_{\text{L}}, \ \ k=-N,\dots,N.
%\end{equation}
Finally, let
\begin{equation}
\label{eq: matrix O}
\Om := \left[\Id_{\text{L}}^{-N}, \dots, \Id_{\text{L}}^{N} \right] \in \mathbb{R}^{L \times L^{2}}.
\end{equation}
Based on (\ref{eq:matrix formulation 2}), (\ref{eq: matrix equi 2}), (\ref{eq: matrix E}), and (\ref{eq: matrix O}), we can write
\begin{equation}
\label{eq: matrix equi 3}
\widetilde{\Dm}_{p}^{H}\av\left(\rv\right) = \widehat{\Dm}_{p}^{H}\av_{p}\left(\rv\right)= \Dm^{H} \Om \Jm_{p} \av_{p}\left(\rv\right).
\end{equation}
Now, we can rewrite the derivatives of (\ref{eq: G matrix}) using (\ref{eq: matrix equi 3}) as 
\begin{align}
\label{eq: G in matrix multi 1}
&\Mm^{(m',n')}_{\left(m,n\right)}\left(\rv,\rv_{j}\right)= \nonumber\\
&\Dm^{H} \hspace{-2pt} \left[\frac{1}{T^{2}}\hspace{-2pt}\sum_{p=-N}^{N} \hspace{-2pt}\Om \Jm_{p} \av_{p}^{(m',n')}\left(\rv\right) \tilde{\zv}_{p}\left(\rv_{j}\right)_{\left(m,n\right)}^{H} \Jm_{p}^{H} \Om^{H} \right] \Dm \\
\label{eq: G in matrix multi 2}
& =: \Dm^{H} \Rm_{(m,n)}^{(m',n')}\left(\rv,\rv_{j}\right) \Dm,
\end{align}
where $\tilde{\zv}_{p}\left(\rv_{j}\right)_{\left(m,n\right)}$ is obtained by replacing $l$ with $p-l$ in (\ref{eq: z definition}) while the matrix $\Rm_{(m,n)}^{(m',n')}\left(\rv,\rv_{j}\right) \in \mathbb{C}^{L\times L}$ refers to the terms between the square brackets in (\ref{eq: G in matrix multi 1}) with $m',n'=0,1$.

\subsection{Showing that $\Big|\Big|\Mm^{(m',n')}_{\left(m,n\right)}\left(\rv,\rv_{j}\right)-\mathbb{E}\left[\Mm^{(m',n')}_{\left(m,n\right)}\left(\rv,\rv_{j}\right)\right]\Big|\Big|_{2}$ is small}
\label{sec: proof small}

In this section, we show that the our kernel matrix concentrates around its mean with high probability under certain conditions. For that, we show in Appendix \ref{A4} that
\begin{align}
\label{eq: det G}
\mathbb{E}\left[\Mm_{\left(m,n\right)}^{(m',n')}\left(\rv,\rv_{j}\right)\right]&=F^{(m+m')}\left(\tau-\tau_{j}\right)  F^{(n+n')}\left(f-f_{j}\right) \Id_{\text{K}}\nonumber\\
&=:\widebar{M}^{\left(m+m',n+n'\right)}\left(\rv-\rv_{j}\right)\Id_{\text{K}}, 
\end{align}
where $F\left(t\right)$ is given by (\ref{eq: another fejer}). Now, if we recall that the $i$-th column of $\Dm$ is denoted by $\hat{\dv}_{i} \in \mathbb{C}^{L \times 1}$, we can express the element at $(i',j')$ location in $\Mm^{(m',n')}_{\left(m,n\right)}\left(\rv,\rv_{j}\right)$ using (\ref{eq: G in matrix multi 2}) by
\begin{equation}
\label{eq: entry i j}
\left[\Mm^{(m',n')}_{\left(m,n\right)}\left(\rv,\rv_{j}\right)\right]_{(i',j')} = \hat{\dv}_{i'}^{H} \Rm_{(m,n)}^{(m',n')}\left(\rv,\rv_{j}\right) \hat{\dv}_{j'}.
\end{equation}
Moreover, we can conclude based on (\ref{eq: det G}) and (\ref{eq: entry i j}) that
\begin{eqnarray}
%\label{eq: expec realtions}
\mathbb{E}\left[\hat{\dv}_{i'}^{H} \Rm_{(m,n)}^{(m',n')}\hspace{-2pt}\left(\rv,\rv_{j}\right) \hat{\dv}_{i'}\right] \hspace{-3pt}=\hspace{-2pt}F^{(m+m')}\hspace{-2pt}\left(\tau-\tau_{j}\right) F^{(n+n')}\hspace{-2pt}\left(f-f_{j}\right) \nonumber
\end{eqnarray}
and 
\begin{eqnarray}
%\label{eq: expec realtions 2}
\mathbb{E}\left[\hat{\dv}_{i'}^{H} \Rm_{(m,n)}^{(m',n')}\left(\rv,\rv_{j}\right) \hat{\dv}_{j'}\right] \hspace{-3pt} =0,  \forall i', j'\hspace{-3pt} =\hspace{-2pt} 1,\dots, K\hspace{-3pt} : i'\hspace{-2pt} \neq j'. \nonumber
\end{eqnarray}
\begin{lemma} \normalfont
\label{th1}
Let $\rv,\rv_{j} \in [0,1]^{2}, j=1,\dots, R$ and recall $\Mm_{(m,n)}^{(m',n')}\left(\rv,\rv_{j}\right)$ in (\ref{eq: G in matrix multi 2}) with $m,m',n,n'=0,1$ and $m+m'+n+n' \leq 2$. Then, for every real $\epsilon_{1} > 0, \delta > 0$ the event $\mathcal{E}_{1}= 
\left\lbrace \frac{1}{\mu^{m+m'+n+n'}} \bigg|\bigg| \Mm_{(m,n)}^{(m',n')}\left(\rv,\rv_{j}\right)  \mathbb{E}\left[\Mm_{(m,n)}^{(m',n')}\left(\rv,\rv_{j}\right)\right]\bigg|\bigg|_{2} \leq \epsilon_{1}\right\rbrace$
%\begin{align}
%%\label{eq: convergance}
%\mathcal{E}_{1}= 
%&\left\lbrace \frac{1}{\mu^{m+m'+n+n'}} \bigg|\bigg| \Mm_{(m,n)}^{(m',n')}\left(\rv,\rv_{j}\right) \right.  \nonumber\\
%&- \left. \mathbb{E}\left[\Mm_{(m,n)}^{(m',n')}\left(\rv,\rv_{j}\right)\right]\bigg|\bigg|_{2} \leq \epsilon_{1}\right\rbrace \nonumber
%\end{align}
occurs with probability $\text{Pr}\left[\mathcal{E}_{1}\right]\geq 1-\delta/2R^{2}$
%\begin{equation}
%\text{Pr}\left[\mathcal{E}_{1}\right]\geq 1-\delta/2R^{2},  \nonumber
%\end{equation}
provided that
\begin{equation}
\label{eq: con L 1}
L \geq  \frac{ C_{1}^{2}}{\epsilon_{1}^{2}} R K \widetilde{K}^{4} \log^{2}\left(\frac{4 R^{2} K^{2}}{\delta}\right),
\end{equation}
where $\mu := \sqrt{|F''\left(0\right)|}$ and $C_{1}$ is a numerical constant. 
\end{lemma}
The proof of Lemma~\ref{th1} relies on Lemma~\ref{l2} which is built on top of Lemmas~\ref{l1} and \ref{l3} below.
\begin{lemma} \label{l1}\normalfont \cite[Theorem 1.1]{rudelson2013hanson}, \cite[Theorem 2.3]{adamczak2015note} Let $\uv \in \mathbb{C}^{N_{1} \times 1}$ be a random vector satisfying (\ref{eq: G assumption 1}), (\ref{eq: G assumption 2}) with $\Id_{\text{N}_{1}}$, and (\ref{eq: concetration x}). Then, for any $N_{1} \times N_{1}$ matrix $\Am$ and $t > 0$ we have
\begin{align}
\label{eq: h w ineq}
&\text{Pr}\left[ \left|\uv^{H}\Am \uv - \mathbb{E}\left[\uv^{H} \Am \uv\right]\right|\geq t\right] \leq \nonumber\\
&2 \exp \left(-\frac{1}{C} \min \left(\frac{t^{2}}{2 \widetilde{K}^{4}||\Am||_{F}^{2}}, \frac{t}{\widetilde{K}^{2}||\Am||_{2}}\right)\right),
\end{align}
where $C$ is a constant. Furthermore, let $\vv \in \mathbb{C}^{N_{1}\times 1}$ be another random vector that is independent of $\uv$ and satisfies (\ref{eq: G assumption 1}), (\ref{eq: G assumption 2}) with $\Id_{\text{N}_{1}}$, and (\ref{eq: concetration x}). Then, the following inequality holds true (adapted from \cite[Theorem 1.1]{rudelson2013hanson} and  \cite[Theorem 2.1]{li2014sketching})
\begin{align}
\label{eq: h w ineq 2}
&\text{Pr}\left[ \left|\uv^{H}\Am \vv - \mathbb{E}\left[\uv^{H} \Am \vv\right]\right|\geq t\right]  \leq \nonumber\\
&2 \exp \left(-\frac{1}{C} \min \left(\frac{t^{2}}{2 ||\Am||_{F}^{2}}, \frac{t}{||\Am||_{2}}\right)\right).
\end{align}
\end{lemma}
Note that the results in \cite{rudelson2013hanson, adamczak2015note, li2014sketching} are originally obtained for real random vectors. However, using standard complexification tricks, we can easily obtain their complex versions as in (\ref{eq: h w ineq}) and (\ref{eq: h w ineq 2}) (see the proof of \cite[Theorem 1.1]{rudelson2013hanson} for more details).

\begin{lemma} \normalfont
\label{l3}
Recall (\ref{eq: G in matrix multi 2}) with $j=1,\dots, R$, then
\begin{equation}
\label{eq:for norm }
\big|\big|\Rm_{(m,n)}^{(m',n')}\left(\rv,\rv_{j}\right)\big|\big|_{F} \leq \frac{C_{2}}{\sqrt{L}} \left(2 \pi N\right)^{(m+m'+n+n')},
\end{equation}
where $m,m',n,n'=0,1$ and $C_{2}$ is a numerical constant.
\end{lemma}
The proof of Lemma~\ref{l3} follows that in \cite[Lemma~3]{heckel2016super} and is provided in Appendix \ref{A5}.

In Lemma~\ref{l2} below, we apply Lemmas~\ref{l1} and \ref{l3} to show that each element in $\Mm^{(m',n')}_{\left(m,n\right)}\left(\rv,\rv_{j}\right)$ is close to its corresponding one in $\mathbb{E}[\Mm_{\left(m,n\right)}^{(m',n')}\left(\rv,\rv_{j}\right)]$ with very probability. Then, we use matrix inequalities and the union-bound along with Lemma~\ref{l2} to prove Lemma~\ref{th1}. Here, we point out that based on Assumption~\ref{as 1}, $\hat{\dv}_{i}$ also satisfies (\ref{eq: G assumption 1}) and (\ref{eq: G assumption 2}) with $\Id_{\text{L}}$.
\begin{lemma}\normalfont
\label{l2}
Let $\rv,\rv_{j} \in [0,1]^{2}$ and recall (\ref{eq: entry i j}) with $m$, $m'$, $n$, $n'$= 0, 1 and $m+m'+n+n' \leq 2$. Then, for any real $\alpha > 0$, the following two probability measures hold true
\begin{eqnarray}
\label{eq: diagonal bound}
&\hspace{-50pt}\text{Pr}\left[\frac{1}{\mu^{m+m'+n+n'}} \Big| \hat{\dv}_{i'}^{H} \Rm_{(m,n)}^{(m',n')}\left(\rv,\rv_{j}\right) \hat{\dv}_{i'}-\right. \nonumber\\
&\left.\mathbb{E}\left[\hat{\dv}_{i'}^{H} \Rm_{(m,n)}^{(m',n')}\left(\rv,\rv_{j}\right) \hat{\dv}_{i'}\right]\Big| \geq C_{2} 12^{\frac{m+m'+n+n'}{2}} \frac{\alpha}{\sqrt{L}}\right] \nonumber\\
& \leq  2 \exp\left(-\frac{1}{C} \min \left(\frac{\alpha^{2}}{2 \widetilde{K}^{4}}, \frac{\alpha}{\widetilde{K}^{2}}\right)\right), \forall i'=1,\dots, K
\end{eqnarray}
\begin{eqnarray}
\label{eq: diagonal bound 2}
&\hspace{-60pt}\text{Pr}\left[\frac{1}{\mu^{m+m'+n+n'}} \Big| \hat{\dv}_{i'}^{H} \Rm_{(m,n)}^{(m',n')}\left(\rv,\rv_{j}\right) \hat{\dv}_{j'}-\right.\nonumber\\
&\left.\mathbb{E}\left[\hat{\dv}_{i'}^{H} \Rm_{(m,n)}^{(m',n')}\left(\rv,\rv_{j}\right) \hat{\dv}_{j'}\right]\Big| \geq C_{2} 12^{\frac{m+m'+n+n'}{2}} \frac{\alpha}{\sqrt{L}}\right] \leq  \nonumber\\
& 2\exp{\left(\hspace{-3pt}-\frac{1}{C} \min \left(\hspace{-3pt}\frac{\alpha^{2}}{2}, \alpha\right)\hspace{-3pt}\right)},  \forall i',j'\hspace{-3pt}=1,\dots, K\hspace{-3pt}: i'\neq j'.
\end{eqnarray}

\end{lemma} 
The proof of Lemma~\ref{l2} is given in Appendix \ref{proof of l2}.

Now we can prove Lemma~\ref{th1} as in Appendix \ref{proof of th1}.

%Next, we show in Section~\ref{sec: first cond} that our $\fv\left(\rv\right)$ satisfies (\ref{eq: hold assump 1}) with high probability. Then, we dedicate the remaining parts of this section to prove (\ref{eq: hold assump 2}).
\subsection{Showing that $\fv\left(\rv\right)$ satisfies (\ref{eq: hold assump 1}): Obtaining $\alphav_{j}, \betav_{j}$, and $\gammav_{j}$}
\label{sec: first cond}
To prove that (\ref{eq: hold assump 1}) holds, it is enough to show that there exists $\alphav_{j}, \betav_{j}$, $\gammav_{j}$ such that $\fv\left(\rv\right)$ in (\ref{eq: final dual}) satisfies (\ref{eq: hold assump 1}) with high probability. For that, we first write
%let us first write the general expression of the derivatives of $\fv\left(\rv\right)$ as
\begin{align}
\label{eq: poly derivative}
\fv^{(m',n')}\left(\rv\right) =\sum_{j=1}^{R} &\Mm^{(m',n')}_{\left(0,0\right)}\left(\rv,\rv_{j}\right) \alphav_{j} + \Mm^{(m',n')}_{\left(1,0\right)}\left(\rv,\rv_{j}\right) \betav_{j} \nonumber\\
&+ \Mm^{(m',n')}_{\left(0,1\right)}\left(\rv,\rv_{j}\right) \gammav_{j},
\end{align}
where $m',n'=0,1$. Moreover, we can write based on (\ref{eq: det G})
\begin{align}
\label{eq: poly deriv expected}
&\bar{\fv}^{(m',n')}\left(\rv\right):=\mathbb{E}\left[\fv^{(m',n')}\left(\rv\right)\right]  =\sum_{j=1}^{R} \widebar{M}^{(m',n')}\left(\rv-\rv_{j}\right) \bar{\alphav}_{j}\nonumber\\  
&+ \widebar{M}^{(m'+1,n')}\left(\rv-\rv_{j}\right) \bar{\betav}_{j} + \widebar{M}^{(m',n'+1)}\left(\rv-\rv_{j}\right) \bar{\gammav}_{j},
\end{align}
where $ \bar{\alphav}_{j}, \bar{\betav}_{j}, \bar{\gammav}_{j} \in \mathbb{C}^{K \times 1}$ are the solutions of the equations
%\begin{align}
%\label{eq: ave con 1}
%&\bar{\fv}\left(\rv_{j}\right)= \text{sign}\left(c_{j}\right) \hv_{j} \hspace{10pt} \forall \rv_{j} \in \mathcal{R} \\
%\label{eq: ave con 2}
%-&\bar{\fv}^{\left(1,0\right)}\left(\rv_{j}\right)= \bm{0}_{K\times 1} \hspace{17pt} \forall \rv_{j} \in \mathcal{R} \\
%\label{eq: ave con 3}
%-&\bar{\fv}^{\left(0,1\right)}\left(\rv_{j}\right)= \bm{0}_{K\times 1}, \hspace{13pt} \forall \rv_{j} \in \mathcal{R}.
%\end{align}
\begin{eqnarray}
\label{eq: ave con 1}
\bar{\fv}\left(\rv_{j}\right)= \hspace{-2pt}\text{sign}\left(c_{j}\right) \hv_{j}, -\bar{\fv}^{\left(1,0\right)}\left(\rv_{j}\right)= -\bar{\fv}^{\left(0,1\right)}\left(\rv_{j}\right)= \bm{0}_{K\times 1},
\end{eqnarray}
$\forall \rv_{j} \in \mathcal{R}$. Starting from (\ref{eq: poly derivative}), we write (\ref{eq: con 1}), (\ref{eq: con 2}), and (\ref{eq: con 3}) as
\begin{eqnarray}
\label{eq: matrix poly}
\underbrace{{\begin{bmatrix} \Em^{(0,0)}_{(0,0)} & \frac{1}{\mu}\Em^{(0,0)}_{(1,0)}   & \frac{1}{\mu}\Em^{(0,0)}_{(0,1)}  \\ -\frac{1}{\mu}\Em^{(1,0)}_{(0,0)}   & -\frac{1}{\mu^2}\Em^{(1,0)}_{(1,0)} &-\frac{1}{\mu^2}\Em^{(1,0)}_{(0,1)}  \\ -\frac{1}{\mu}\Em^{(0,1)} _{(0,0)} &-\frac{1}{\mu^2}\Em^{(0,1)}_{(1,0)}   & -\frac{1}{\mu^2}\Em^{(0,1)}_{(0,1)} 
  \end{bmatrix}}}_{\Em} \hspace{-3pt} \begin{bmatrix} {\alphav} \\ \mu {\betav}  \\ \mu {\gammav} \end{bmatrix} \hspace{-3pt} = \hspace{-3pt} \begin{bmatrix} \hv \\ \bm{0}_{RK \times 1} \\  \bm{0}_{RK \times 1} \end{bmatrix}
\end{eqnarray}  
where $\Em^{(m',n')}_{(m,n)} \in \mathbb{C}^{RK \times RK}$ consists of $R \times R$ block matrices of size $K \times K$ with the one at the $\left(l,k\right)$ location being given by $[\Em^{(m',n')}_{(m,n)}]_{(l,k)} := \Mm^{(m',n')}_{(m,n)}\left(\rv_{l},\rv_{k}\right)$ (see (\ref{eq:matrix D bloack}) in Appendix \ref{proof of le: matrix inverse}) while $\hv := \left[ \text{sign}\left(c_{1}\right)\hv_{1}^{T}, \hdots, \text{sign}\left(c_{R}\right)\hv_{R}^{T}\right]^{T} \in \mathbb{C}^{RK \times 1}$. 

Moreover, we can express (\ref{eq: ave con 1}) using (\ref{eq: poly deriv expected}) as
\begin{equation}
\label{eq: matrix poly ave}
\left(\widebar{\Em} \otimes \Id_{\text{K}} \right)  \begin{bmatrix} \bar{\alphav} \\ \mu \bar{\betav}  \\ \mu \bar{\gammav} \end{bmatrix}  =  \begin{bmatrix} \hv \\ \bm{0}_{RK \times 1} \\  \bm{0}_{RK \times 1} \end{bmatrix}, 
\end{equation}
where $\widebar{\Em} \in \mathbb{C}^{3R \times 3R}$ is given by 
\begin{equation}
\label{eq:matrix D ave}
\widebar{\Em}= {\begin{bmatrix} \widebar{\Em}^{(0,0)} & \frac{1}{\mu}\widebar{\Em}^{(1,0)}   & \frac{1}{\mu}\widebar{\Em}^{(0,1)}  \\ -\frac{1}{\mu}\widebar{\Em}^{(1,0)}   & -\frac{1}{\mu^2}\widebar{\Em}^{(2,0)} &-\frac{1}{\mu^2}\widebar{\Em}^{(1,1)}  \\ -\frac{1}{\mu}\widebar{\Em}^{(0,1)} &-\frac{1}{\mu^2}\widebar{\Em}^{(1,1)}   & -\frac{1}{\mu^2}\widebar{\Em}^{(0,2)} 
  \end{bmatrix}}
\end{equation}
with $[\widebar{\Em}^{(m',n')}]_{(l,k)} := \widebar{M}^{(m',n')}\left(\rv_l - \rv_k\right)$ while $\bar{\alphav}  := [ \bar{\alphav}_{1}^{T}, \dots , \bar{\alphav}_{R}^{T} ]^{T}, \bar{\betav}  := [ \bar{\betav}_{1}^{T}, \dots ,\bar{\betav}_{R}^{T} ]^{T}, \bar{\gammav} := [\bar{\gammav}_{1}^{T}, \dots , \bar{\gammav}_{R}^{T} ]^{T}$. Note that based on (\ref{eq: det G}) and (\ref{eq: matrix poly}), $\mathbb{E}\left[\Em\right] = \left(\widebar{\Em} \otimes \Id_{\text{K}} \right)$, and that the scaling of the sub-matrices in  (\ref{eq: matrix poly}) and (\ref{eq:matrix D ave}) with $\frac{1}{\mu^k}, k=0,1,2$ is meant to make the diagonal entries of $\Em$ and $\widebar{\Em}$ equal to one which will facilitate our proofs later.

From (\ref{eq: matrix poly}) we can see that for $\alphav, \betav, \gammav$ to be well defined, $\Em$ must be invertible. To manifest that, we first show in Proposition~\ref{pro: inver} that $\mathbb{E}\left[\Em\right]$ is invertible and that $\bar{\alphav}, \bar{\betav}, \bar{\gammav}$ are well defined. Then, we prove in Lemma~\ref{le: matrix inverse} that $\Em$ is close to $\mathbb{E}\left[\Em\right] $ in Euclidean norm measure with high probability. Finally, we show in Lemma~\ref{le: inver cond l} that $\Em$ is invertible with high probability.
\begin{proposition}\normalfont
\label{pro: inver}
Under Assumption~\ref{as 4} $\mathbb{E}\left[\Em\right]$ is invertible and
\begin{align}
\label{eq: matrix in cond 1}
||\mathbb{E}\left[\Em\right]||_{2} \leq 1.19808\\
\label{eq: matrix in cond 2}
||\Id_{\text{3RK}}-\mathbb{E}\left[\Em\right]||_{2} \leq 0.19808\\
\label{eq: matrix in cond 3}
||\left(\mathbb{E}\left[\Em\right] \right)^{-1}||_{2} \leq 1.2470.
\end{align}
The proof of Proposition~\ref{pro: inver} is provided in Appendix \ref{A6}.
\end{proposition}
\begin{lemma}
\label{le: matrix inverse}
\normalfont
Consider the event $\mathcal{E}_{2} = \left\lbrace||\Em- \mathbb{E}\left[{\Em}\right]||_{2} \leq \epsilon_{1}\right\rbrace$
%\begin{equation}
%%\label{eq: event matrix}
%\mathcal{E}_{2} = \left\lbrace||\Em- \mathbb{E}\left[{\Em}\right]||_{2} \leq \epsilon_{1}\right\rbrace \nonumber
%\end{equation}
for every real $\epsilon_{1}> 0$. Then, $\mathcal{E}_{2}$ occurs with probability at least $1-\delta/2$ for every $\delta >0$ provided that (\ref{eq: con L 1}) is satisfied.
\end{lemma}
The proof of Lemma~\ref{le: matrix inverse} is appended to Appendix \ref{proof of le: matrix inverse}.
\begin{lemma}
\label{le: inver cond l}
\normalfont
The matrix $\Em$ is invertible on $\mathcal{E}_{2}$ for all $\epsilon_{1} \in (0,\frac{2}{5}]$ with probability at least $1-\delta/2$ and 
\begin{eqnarray}
\label{eq: conditon inverstion ma}
||\Id_{\text{3RK}} -\Em||_{2}  \leq  0.5981 \\
\label{eq: conditon inverstion ma 2}
||\Em^{-1}||_{2}  \leq  2.50.
\end{eqnarray}
\end{lemma}
The proof of Lemma~\ref{le: inver cond l} is provided in Appendix \ref{proof of le: inver cond l}.

%Based on Lemma \ref{le: inver cond l}, the coefficients of $\fv\left(\rv\right)$ are all well defined and can be obtained as

Since $\Em$ is invertible on $\mathcal{E}_{2}$ for $\epsilon_{1} \in (0,\frac{2}{5}]$, the coefficients of $\fv\left(\rv\right)$ are all well defined and can be obtained as
\begin{eqnarray}
\label{eq: obtianing variables}
\begin{bmatrix} {\alphav} \\ \mu {\betav}  \\ \mu {\gammav} \end{bmatrix}  = \Em^{-1} \begin{bmatrix} \hv \\ \bm{0}_{RK \times 1} \\  \bm{0}_{RK \times 1} \end{bmatrix}  = \Lm \hv, 
\end{eqnarray}
where we write $\Em^{-1} = \begin{bmatrix} \Lm \ \Gm \end{bmatrix}, \Lm \in \mathbb{C}^{3RK \times RK}, \ \Gm \in \mathbb{C}^{3RK \times 2RK}$. Finally, since $\Lm$ is a sub-matrix of $\Em^{-1}$, we can deduce that conditioned on $\mathcal{E}_{2}$ with $\epsilon_{1} \in (0,\frac{2}{5}]$ we have
\begin{equation}
\label{eq: bound L}
||\Lm||_{2} \leq ||\Em^{-1} ||_{2} \leq 2.5.
\end{equation} 

What remains now is to show that with $\alphav_{j}, \betav_{j}, \gammav_{j}$ obtained as in (\ref{eq: obtianing variables}), $\fv\left(\rv\right)$ obeys (\ref{eq: hold assump 2}) also with high probability. For that, we will pursue the following steps:
\begin{enumerate}
\item We show in Section~\ref{sec: less than 1 1} that $\fv\left(\rv\right)$, $\bar{\fv}\left(\rv\right)$, and their partial derivatives are close in Euclidean norm measure with high probability on a finite set of grid points $\Omega_{\text{S}} \subset [0,1]^{2}$.
\item Then, we prove in Section~\ref{sec: less than 1 2} that the statement in (1) holds with high probability everywhere in $[0,1]^{2}$.
\item Finally, and with the help of statements (1) and (2), we show in Section~\ref{sec: less than 1 3} that $||\fv\left(\rv\right)||_{2} < 1, \forall \rv \in [0,1]^{2} \setminus\mathcal{R}$.
\end{enumerate}
\subsection{Showing that $\fv^{(m',n')}\left(\rv\right)$ is close to $\mathbb{E}[\fv^{(m',n')}\left(\rv\right)]$ on a finite grid of points}
\label{sec: less than 1 1}
%The main result of this section is given in Lemma~\ref{th: off grid theorem}; however, we will first need to obtain some relevant results.
%
%Consider a normalized version of $\fv^{(m',n')}\left(\rv\right)$ in the form

The main result of this section is in Lemma~\ref{th: off grid theorem}; however, we first need to obtain some results. Consider the following
%Consider a normalized version of $\fv^{(m',n')}\left(\rv\right)$ in the form
\begin{eqnarray}
\label{eq: normalized dual}
& \frac{1}{\mu^{m'+n'}}\fv^{(m',n')}\left(\rv\right)=  \frac{1}{\mu^{m'+n'}} \sum_{j=1}^{R}\left(\Mm^{(m',n')}_{(0,0)}\left(\rv,\rv_{j}\right) \alphav_{j} \right. \nonumber\\
& \left.+\frac{1}{\mu}{\Mm}^{(m',n')}_{(1,0)}\left(\rv,\rv_{j}\right) \mu \betav_{j} + \frac{1}{\mu}{\Mm}^{(m',n')}_{(0,1)}\left(\rv,\rv_{j}\right) \mu \gammav_{j} \right).
\end{eqnarray}
Equation (\ref{eq: normalized dual}) can be written using matrix-vector form as
\begin{equation}
\label{eq: matrix vector dual}
\frac{1}{\mu^{m'+n'}}\fv^{(m',n')}\left(\rv\right)=  \left(\Tm^{(m',n')}\left(\rv\right)\right)^{H} \Lm \hv,
\end{equation}
where $\Tm^{(m',n')}\left(\rv\right) \in \mathbb{C}^{3RK \times K}$ is given by
%\begingroup
%\fontsize{9.4pt}{9.9pt}
%\begin{align}
%\label{eq: matrix T}
% &\left(\Tm^{(m',n')}\left(\rv\right)\right)^{H}  :=  \frac{1}{\mu^{m'+n'}} \times \nonumber\\
% &\left[ \Mm^{(m',n')}_{(0,0)}\left(\rv,\rv_{1}\right), \hdots, \Mm^{(m',n')}_{(0,0)}\left(\rv,\rv_{R}\right), \frac{1}{\mu}\Mm^{(m',n')}_{(1,0)}\left(\rv,\rv_{1}\right), \hdots,  \right.\nonumber\\ &\left.\frac{1}{\mu}\Mm^{(m',n')}_{(1,0)}\left(\rv,\rv_{R}\right), \frac{1}{\mu}\Mm^{(m',n')}_{(0,1)}\left(\rv,\rv_{1}\right), \hdots,  \frac{1}{\mu}\Mm^{(m',n')}_{(0,1)}\left(\rv,\rv_{R}\right) \right].
%\end{align}
%\endgroup
\begingroup
\fontsize{9.4pt}{9.9pt}
\begin{align}
\label{eq: matrix T}
 &\left(\Tm^{(m',n')}\left(\rv\right)\right)^{H}  :=  \frac{1}{\mu^{m'+n'}} \left[ \Mm^{(m',n')}_{(0,0)}\left(\rv,\rv_{1}\right), \hdots,\right.\nonumber\\
 &\left.  \Mm^{(m',n')}_{(0,0)}\left(\rv,\rv_{R}\right), \frac{1}{\mu}\Mm^{(m',n')}_{(1,0)}\left(\rv,\rv_{1}\right), \hdots,  \frac{1}{\mu}\Mm^{(m',n')}_{(1,0)}\left(\rv,\rv_{R}\right),\right.\nonumber\\ &\left. \frac{1}{\mu}\Mm^{(m',n')}_{(0,1)}\left(\rv,\rv_{1}\right), \hdots,  \frac{1}{\mu}\Mm^{(m',n')}_{(0,1)}\left(\rv,\rv_{R}\right) \right].
\end{align}
\endgroup
Starting from (\ref{eq: matrix vector dual}), we can show after some manipulations that
%\begin{align}
%\label{eq: T trick}
%&\frac{1}{\mu^{m'+n'}}\fv^{(m',n')}\left(\rv\right)=  \left( \Tm^{(m',n')}\left(\rv\right) -\widebar{\Tm}^{(m',n')}\left(\rv\right)\right.\nonumber\\
%& \left.+\widebar{\Tm}^{(m',n')}\left(\rv\right) \right)^{H} \hspace{-3pt} \left(\Lm  - \bar{\Lm}\otimes \Id_{\text{K}}  +\bar{\Lm}\otimes\Id_{\text{K}}\right)\hv= \hspace{-3pt}\left(\widebar{\Tm}^{(m',n')}\left(\rv\right)\right)^{H} \nonumber\\
%& \times \left(\bar{\Lm}\otimes\Id_{\text{K}}\right)\hv + \left( \Tm^{(m',n')}\left(\rv\right) -\widebar{\Tm}^{(m',n')}\left(\rv\right)\right)^{H} \Lm \hv \nonumber\\
%& +\left(\widebar{\Tm}^{(m',n')}\left(\rv\right) \right)^{H} \left(\Lm  - \bar{\Lm}\otimes \Id_{\text{K}} \right)\hv,
%\end{align}
\begin{align}
\label{eq: T trick}
&\hspace{-3pt}\frac{1}{\mu^{m'+n'}}\fv^{(m',n')}\hspace{-2pt}\left(\rv\right)\hspace{-2pt}=   \hspace{-3pt}\left(\widebar{\Tm}^{(m',n')}\left(\rv\right)\hspace{-2pt}\right)^{H}\hspace{-4pt} \left(\bar{\Lm}\otimes\Id_{\text{K}}\right)\hspace{-2pt}\hv +\hspace{-2pt} \left(\hspace{-2pt} \Tm^{(m',n')}\left(\rv\right)\right.\nonumber\\
& \left.\hspace{-2pt}-\widebar{\Tm}^{(m',n')}\left(\rv\right)\hspace{-1pt}\right)^{H} \hspace{-3pt}\Lm \hv +\hspace{-2pt}\left(\widebar{\Tm}^{(m',n')}\left(\rv\right) \hspace{-1pt}\right)^{H} \hspace{-3pt}\left(\Lm  - \bar{\Lm}\otimes \Id_{\text{K}} \right)\hv
\end{align}
where $\widebar{\Tm}^{(m',n')}\left(\rv\right):=\mathbb{E}\left[\Tm^{(m',n')}\left(\rv\right) \right]$ and $\bar{\Lm}$ is $3R \times R$ sub-matrix of $\widebar{\Em}^{-1}$ consisting of the first $R$ columns of $\widebar{\Em}^{-1}$. To simplify (\ref{eq: T trick}) note that based on (\ref{eq: det G}) and (\ref{eq: matrix T}) we can write
\begin{equation}
\label{eq: sub T }
\widebar{\Tm}^{(m',n')}\left(\rv\right) = \bar{\tv}^{(m',n')}\left(\rv\right) \otimes \Id_{\text{K}},
\end{equation}
where $\bar{\tv}^{(m',n')}\left(\rv\right) \in \mathbb{C}^{3R\times 1}$ is formed by taking the expectation for each matrix entry in (\ref{eq: matrix T}).
%given by
%\begin{eqnarray}
%\label{eq: matrix T average} 
% &\hspace{-10pt}\bar{\tv}^{(m',n')}\left(\rv\right) = \frac{1}{\mu^{m'+n'}} \left[ \widebar{M}^{(m',n')}\left(\rv-\rv_{1}\right), \hdots, \widebar{M}^{(m',n')}\left(\rv-\rv_{R}\right), \right. \nonumber\\
% &\hspace{-40pt}\left. \frac{1}{\mu}\widebar{M}^{(m'+1,n')}\left(\rv-\rv_{1}\right), \hdots, \frac{1}{\mu}M^{(m'+1,n')}\left(\rv-\rv_{R}\right),\right.\nonumber\\
% &\hspace{-40pt}\left.\frac{1}{\mu}\widebar{M}^{(m',n'+1)}\left(\rv-\rv_{1}\right), \hdots, \frac{1}{\mu}\widebar{M}^{(m',n'+1)}\left(\rv-\rv_{R}\right) \right]^{H}
%\end{eqnarray}
%\begingroup
%\fontsize{9.4pt}{9.9pt}
%\begin{align}
%\label{eq: matrix T average} 
% &\bar{\tv}^{(m',n')}\hspace{-2pt}\left(\rv\right)\hspace{-2pt} = \hspace{-2pt}\frac{1}{\mu^{m'+n'}} \hspace{-2pt}\left[ \widebar{M}^{(m',n')}\hspace{-2pt}\left(\rv-\rv_{1}\right), \hdots, \widebar{M}^{(m',n')}\hspace{-2pt}\left(\rv-\rv_{R}\right), \right. \nonumber\\
% &\left. \frac{1}{\mu}\widebar{M}^{(m'+1,n')}\left(\rv-\rv_{1}\right), \hdots, \frac{1}{\mu}M^{(m'+1,n')}\left(\rv-\rv_{R}\right),\right.\nonumber\\
% &\left.\frac{1}{\mu}\widebar{M}^{(m',n'+1)}\left(\rv-\rv_{1}\right), \hdots, \frac{1}{\mu}\widebar{M}^{(m',n'+1)}\left(\rv-\rv_{R}\right) \right]^{H}
%\end{align}
%\endgroup
Moreover
\begin{align}
\label{eq: averg h }
\left(\bar{\Lm}\otimes\Id_{\text{K}}\right)\hv=\begin{bmatrix} \bar{\alphav} \\ \mu \bar{\betav}  \\ \mu \bar{\gammav} \end{bmatrix}.
\end{align}
By using (\ref{eq: poly deriv expected}), (\ref{eq: sub T }), and (\ref{eq: averg h }), we can conclude that
\begin{align}
\label{eq: averg poly ex}
\left(\widebar{\Tm}^{(m',n')}\left(\rv\right)\right)^{H} \left(\bar{\Lm}\otimes\Id_{\text{K}}\right)\hv = \frac{1}{\mu^{m'+n'}}\bar{\fv}^{(m',n')}\left(\rv\right).
\end{align}
Substituting (\ref{eq: averg poly ex}) in (\ref{eq: T trick}) results in
\begin{align}
\label{eq: T trick 2}
&\frac{1}{\mu^{m'+n'}}\fv^{(m',n')}\left(\rv\right)=\nonumber\\
&\frac{1}{\mu^{m'+n'}}\bar{\fv}^{(m',n')}\left(\rv\right)+ \vv_{1}^{(m',n')}\left(\rv\right) + \vv_{2}^{(m',n')}\left(\rv\right),
\end{align}
where \hspace{-1pt}$\vv_{1}^{(m',n')}\hspace{-2pt}\left(\rv\right) \hspace{-2pt}\in \hspace{-2pt}\mathbb{C}^{K \times 1}$ \hspace{-2pt}and $\vv_{2}^{(m',n')}\hspace{-2pt}\left(\rv\right) \hspace{-2pt}\in \hspace{-2pt}\mathbb{C}^{K \times 1}$ \hspace{-2pt}are given by 
\begin{equation}
\label{eq: v matrix 1}
\vv_{1}^{(m',n')}\left(\rv\right):=\left( \Tm^{(m',n')}\left(\rv\right) -\widebar{\Tm}^{(m',n')}\left(\rv\right)\right)^{H} \Lm \hv 
\end{equation}
%while $\vv_{2}^{(m',n')}\left(\rv\right) \in \mathbb{C}^{K \times 1}$ is given by
\begin{equation}
\label{eq: v matrix 2}
\vv_{2}^{(m',n')}\left(\rv\right) := \left(\widebar{\Tm}^{(m',n')}\left(\rv\right) \right)^{H} \left(\Lm  - \bar{\Lm}\otimes \Id_{\text{K}} \right)\hv.
\end{equation}
Looking at (\ref{eq: T trick 2}), we can predict our steps. First, we prove in Lemmas~\ref{le: v bounds } and \ref{lemma: v2 bounds } that both $||\vv_{1}^{(m',n')}\left(\rv\right)||_{2}$ and $||\vv_{2}^{(m',n')}\left(\rv\right)||_{2}$ are small, respectively, on a finite set of grid points $\Omega_{\text{S}} \subset [0,1]^{2}$ with high probability. Then, we use these results in Lemma~\ref{th: off grid theorem} to show that $\fv^{(m',n')}\left(\rv\right)$ is close to $\bar{\fv}^{(m',n')}\left(\rv\right)$ in Euclidean norm measure on the same set. Before that, we obtain some important results. For that, we define
% and to facilitate our proofs later, we first obtain in Lemmas~\ref{lemma: T matrix} and \ref{th: pro of T} the probability bounds on $||\Delta \Tm^{(m',n')}\left(\rv\right)||_{2}$ and $||(\Delta \Tm^{(m',n')}\left(\rv\right))^{H}\Lm||_{2}$, respectively. For that, we define
\begin{equation}
\label{eq: delta T definition}
\Delta \Tm^{(m',n')}\left(\rv\right) := \Tm^{(m',n')}\left(\rv\right) -\widebar{\Tm}^{(m',n')}\left(\rv\right).
\end{equation}
\begin{lemma}\normalfont
\label{lemma: T matrix}
Let $\rv \in \Omega_{\text{S}} \subset [0,1]^{2}$ be a finite set of points and assume that $\epsilon_{2} > 0$, $\delta >0$, and $m',n'=0,1$. Then, the event $\mathcal{E}_{3}=  \left\lbrace\left| \left|\Delta \Tm^{(m',n')}\left(\rv\right)\right|\right|_{2} \leq \epsilon_{2}\right\rbrace$
%\begin{equation}
%%\label{eq: event w}
%\mathcal{E}_{3}=  \left\lbrace\left| \left|\Delta \Tm^{(m',n')}\left(\rv\right)\right|\right|_{2} \leq \epsilon_{2}\right\rbrace \nonumber
%\end{equation}
holds with probability at least $\left(1-\frac{\delta}{2|\Omega_{\text{S}}|}\right)$ provided that
\begin{equation}
\label{eq: cond L for T}
L \geq  \frac{ C_{3}^{2}}{\epsilon_{2}^{2}} RK \widetilde{K}^{4} \log^2 \left(\frac{4RK^{2} |\Omega_{\text{S}}|}{\delta}\right),
\end{equation}
where the cardinality $|\Omega_{\text{S}}|$ is to be determined in Section~\ref{sec: less than 1 2}.
\end{lemma}
The proof of Lemma~\ref{lemma: T matrix} follows the same steps of the proofs of Lemmas~\ref{th1} and \ref{le: matrix inverse}.
\begin{lemma}\normalfont
\label{th: pro of T}
Let $\epsilon_{1} \in (0,\frac{2}{5}]$, $\rv \in \Omega_{\text{S}} \subset [0,1]^{2}$. Then, the event $\mathcal{E}_{4} = \left\lbrace \max _{\rv \in \Omega_{\text{S}}} \Big|\Big|\left(\Delta \Tm^{(m',n')}\left(\rv\right)\right)^{H} \Lm \Big|\Big|_{2} \leq  2.5 \epsilon_{2}\right\rbrace$
%\begin{equation}
%%\label{eq: event e 4}
%\mathcal{E}_{4} = \left\lbrace \max _{\rv \in \Omega_{\text{S}}} \Big|\Big|\left(\Delta \Tm^{(m',n')}\left(\rv\right)\right)^{H} \Lm \Big|\Big|_{2} \leq  2.5 \epsilon_{2}\right\rbrace \nonumber
%\end{equation}
occurs with probability $\text{Pr} \left[\mathcal{E}_{4} \right] \geq  1 - \left(\delta/2 + \text{Pr}\left[\mathcal{E}_{2}^{c}\right] \right)$
%\begin{equation}
%\label{eg pro evetn e4}
%\text{Pr} \left[\mathcal{E}_{4} \right] \geq  1 - \left(\delta/2 + \text{Pr}\left[\mathcal{E}_{2}^{c}\right] \right) \nonumber
%\end{equation}
given that (\ref{eq: cond L for T}) holds where $\mathcal{E}_{2}^{c}$ is the complement of $\mathcal{E}_{2}$. 
\end{lemma}
The proof of Lemma~\ref{th: pro of T} is presented in Appendix \ref{proof of le: inver cond l}.

%Now, we are ready to derive the probability bounds on $\big|\big|\vv_{1}^{(m',n')}\left(\rv\right)\big|\big|_{2}$ and $\big|\big|\vv_{2}^{(m',n')}\left(\rv\right)\big|\big|_{2}$ as follows:
\begin{lemma}\normalfont
\label{le: v bounds }
Recall $\vv_{1}^{(m',n')}\left(\rv\right)$ in (\ref{eq: v matrix 1}) with $m',n'=0,1$ and let $\hv_{j}$ to have i.i.d. random entries on the complex unit sphere. Then, for $0 < \epsilon_{3} \leq 1$, $\delta >0,$ and $\rv \in \Omega_{\text{S}} \subset [0,1]^{2}$, we have
%\begin{align}
%\label{eq: bound v the}
$\text{Pr}\left[ \max_{\rv \in \Omega_{\text{S}}} \big|\big|\vv_{1}^{(m',n')}\left(\rv\right)\big|\big|_{2} \geq \epsilon_{3}\right] \leq 3 \delta/2$
%\end{align}
provided that 
\begin{eqnarray}
\label{eq: cond l for v}
L \geq \hspace{-1pt} \frac{\bar{C}}{\epsilon_{3}^{2}} RK\widetilde{K}^{4}\log^2 \hspace{-2pt} \left(\hspace{-2pt}\frac{4R^{2}K^{2} |\Omega_{\text{S}}|}{\delta}\hspace{-1pt}\right)\hspace{-1pt}\log^{2}\hspace{-2pt} \left(\hspace{-1pt}\frac{2\left(K+1\right)|\Omega_{\text{S}}|}{\delta}\hspace{-1pt}\right)
\end{eqnarray}
where we set $\bar{C} = \bar{C}' C_{3}^{2}$ and we assume that $\epsilon_{3} \leq \frac{0.28\sqrt{\bar{C}}}{C_{1}}$.  
\end{lemma}
\begin{lemma}\normalfont
\label{lemma: v2 bounds }
Recall $\vv_{2}^{(m',n')}\left(\rv\right)$ in (\ref{eq: v matrix 2}) with $m',n'=0,1$ and let $\hv_{j}$ to have i.i.d. random entries on the complex unit sphere. Then, for $0 < \epsilon_{3} \leq 1$, $\delta >0$, and $\rv \in \Omega_{\text{S}} \subset [0,1]^{2}$ we have
%\begin{align}
%\label{eq: bound v2 the}
$\text{Pr}\left[ \max_{\rv \in \Omega_{\text{S}}} \big|\big|\vv_{2}^{(m',n')}\left(\rv\right)\big|\big|_{2} \geq \epsilon_{3}\Big| \mathcal{E}_{2}\right] \leq \delta/2$
%\end{align}
provided that
\begin{equation}
\label{eq: expre for eps1}
\epsilon_{1} \leq C_{4}\epsilon_{3}{\log^{-1}\left(\frac{2\left(K+1\right) |\Omega_{\text{S}}|}{\delta}\right)},
\end{equation}
where $C_{4} \leq 0.55$ is a numerical constant.
\end{lemma}
The proofs of Lemmas~\ref{le: v bounds } and \ref{lemma: v2 bounds } are in Appendix \ref{proof of le: v bounds }.

%Now, we are ready to provide Lemma~\ref{th: off grid theorem} as follows:
%which shows that $\fv^{(m',n')}\left(\rv\right)$ is close to $\bar{\fv}^{(m',n')}\left(\rv\right)$ 
%in Euclidean norm distance measure on a finite set of grid points $\Omega_{\text{S}} \subset [0,1]^{2}$.
\begin{lemma}\normalfont
\label{th: off grid theorem}
Let $ \rv \in \Omega_{\text{S}} \subset [0,1]^{2}$, $\delta >0,$ and define $\mathcal{E}_{5} = \left\lbrace \max_{\rv \in \Omega_{\text{S}}} \frac{1}{\mu^{m'+n'}} \Big|\Big|\fv^{(m',n')}\left(\rv\right) - \bar{\fv}^{(m',n')}\left(\rv\right) \Big|\Big|_{2} \leq 2 \epsilon_{3} \right\rbrace$ with $0 < \epsilon_{3} \leq 1$. Then, when (\ref{eq: cond l for v}) is satisfied, $\text{Pr}\left[\mathcal{E}_{5}\right] \geq 1- 2.5\delta. \nonumber$
\end{lemma}
The proof of Lemma~\ref{th: off grid theorem} is provided in Appendix \ref{proof of th: off grid theorem}.
\vspace{-3pt}
\subsection{Showing that $\fv^{(m',n')}\left(\rv\right)$ is close to $\bar{\fv}^{(m',n')}\left(\rv\right)$ almost everywhere in $[0,1]^{2}$}
\label{sec: less than 1 2}
%Our aim in this section is to prove the following lemma:
\begin{lemma}\normalfont
\label{th: f is less 1}
Let $\rv \in [0,1]^{2}$ and assume that
\begin{equation}
\label{eq: cond L final}
L \geq  \hspace{-2pt}\frac{\bar{C}}{\epsilon_{3}^{2}} RK\widetilde{K}^{4}\log^{2}\left(\frac{12\tilde{C}_{3}^{2} R^{2} K^{2}  L^{3} }{\delta^{*} \epsilon_{3}^{2}}\right)\hspace{-2pt} \log^{2}\left(\frac{6\tilde{C}_{3}^{2} (K+1)  L^{3} }{\delta^{*} \epsilon_{3}^{2}}\right) 
\end{equation}
where $\tilde{C}_{3}$ is a numerical constant, $\delta^{*}>0$, and $0<\epsilon_{3}\leq1$. Then, it holds with probability at least $1-\delta^{*}$ that
\begin{align}
\label{eq: f to f bar al}
\max_{\substack{\rv \in [0,1]^{2}, \\ m'+n'\leq 2}} \frac{1}{\mu^{m'+n'}} \big|\big|\fv^{(m',n')}\left(\rv\right)- \bar{\fv}^{(m',n')}\left(\rv\right)\big|\big|_{2} \leq \epsilon_{3}.
\end{align}
\end{lemma}
%Appendix~\ref{proof of th: f is less 1}
The proof of Lemma~\ref{th: f is less 1} is in Appendix \ref{proof of th: f is less 1}, and is based on Lemma~\ref{lemma: close in all r}, whose proof is in Appendix \ref{A15}.
%To prove Lemma~\ref{th: f is less 1}, we need Lemma~\ref{lemma: close in all r} below whose proof is given in Appendix~\ref{A15}.
\begin{lemma}\normalfont
\label{lemma: close in all r}
Recall (\ref{eq: poly derivative}) with $m',n'=0,1$ and let $ \rv \in [0,1]^{2}$. Then, conditioned on $\mathcal{E}_{2}$ with $\epsilon_{1} \in (0,\frac{2}{5}]$ the event $\mathcal{E}_{6} = \left\lbrace \max_{\substack{\rv \in [0,1]^{2}, m'+n'\leq 2}} \frac{1}{\mu^{m'+n'}}||\fv^{(m',n')}\left(\rv\right)||_{2} \leq  \frac{\tilde{C}_{2}}{4} \sqrt{L} \right\rbrace$
holds with probability at least $1- \frac{\delta}{2}$ given that (\ref{eq: con L 1}) is satisfied where $\tilde{C}_{2}$ is a numerical constant.
\end{lemma}
%Now, we are ready to prove Lemma~\ref{th: f is less 1} as in Appendix~\ref{proof of th: f is less 1}.
\subsection{Showing that $||\fv\left(\rv\right)||_{2} < 1, \forall \rv \in [0,1]^{2} \setminus\mathcal{R}$}
\label{sec: less than 1 3}
To start with, consider the definitions of the following sets
\begin{align}
\label{eq: close set}
&\Omega_{\text{far}}=  \forall \rv \in [0,1]^{2} :  \min_{\rv_{j} \in \mathcal{R}} |\rv-\rv_{j}| \geq 0.2447/N \\
\label{eq: far set}
&\Omega_{\text{close}}=  \forall \rv \notin \mathcal{R},  \rv_{j} \in \mathcal{R}:0 < |\rv-\rv_{j}| \leq 0.2447/N
\end{align}
where $\Omega_{\text{close}}$ has the points in $[0,1]^{2}$ that are close to $\rv_{j} \in \mathcal{R}$ while $\Omega_{\text{far}}$ has the points that are far away from it. To show that $\fv\left(\rv\right)$ in (\ref{eq: final dual}), with its coefficients being obtained as in (\ref{eq: obtianing variables}), satisfies (\ref{eq: hold assump 2}), it is enough to show that $||\fv\left(\rv\right)||_{2} < 1$ $ \forall \rv \in \Omega_{\text{far}}$ and $\forall  \rv \in \Omega_{\text{close}}$. For that, we first rewrite (\ref{eq: cond L final}) as
\begin{eqnarray}
\label{eq: L cond close and far}
L \geq \hspace{-2pt} C_{1}^{*}RK\widetilde{K}^{4}\hspace{-2pt}\log^{2}\hspace{-2pt}\left(\hspace{-2pt}\frac{{C}_{2}^{*} R^{2}  K^{2}  L^{3} }{\delta^{*}}\hspace{-1pt}\right)\hspace{-2pt}\log^{2}\hspace{-2pt}\left(\hspace{-2pt}\frac{{C}_{2}^{*} (K+1)  L^{3} }{\delta^{*}}\hspace{-1pt}\right)
\end{eqnarray}
\begin{lemma}\normalfont
\label{lemma: close}
Let (\ref{eq: seperation condition}) and (\ref{eq: L cond close and far}) be satisfied, then with probability $1 - \delta^{*}$, each of the following bounds holds
\begin{align}
\label{eq: Q less 0.9}
&||\fv\left(\rv\right)||_{2} < 1,    \  \forall \rv \in \Omega_{\text{far}}. \\
\label{eq: Q2 less 0.9}
&||\fv\left(\rv\right)||_{2} < 1,   \  \forall \rv \in \Omega_{\text{close}}.
\end{align}
\end{lemma}
The proof of Lemma~\ref{lemma: close} is given in Appendix \ref{proof of lemma: far}.

Using Lemma~\ref{lemma: close}, we get $||\fv\left(\rv\right)||_{2} < 1, \forall \rv \in [0,1]^{2} \setminus\mathcal{R}$.

\section{Conclusions and Future Work Directions}
\label{sec: conc}
In this work, we developed a general framework for blind 2D super-resolution. We showed that given the response of a linear system to multiple unknown time-delayed and frequency-shifted waveforms, we could recover, with infinite precision, the locations of the shifts upon applying the atomic norm. To convert the problem into a well-posed one, we assumed that the unknown waveforms lie in a common low-dimensional subspace. The exact recovery holds provided that a bound on the number of the observed samples is satisfied.  

We conclude by pointing-out possible future extensions. First, it is of interest to study the stability of the framework to noise. In this case, the exact recovery for the unknowns does not exist; however, given the stability that we experienced in our simulations, we do hope that a theoretical stability result exists and easy to derive. Second, we encountered a significant computational complexity issue throughout our simulations while solving (\ref{eq: dual of the dual}). Thus, it is of interest to investigate alternative ways to formulate and solve (\ref{eq: optimzation dual}). Finally, a promising path is to consider developing a general framework for MD blind super-resolution to cover various applications.

\bibliographystyle{IEEEbib}
\bibliography{refs2}

\appendices
\numberwithin{equation}{section}

\section{Equivalence Between (\ref{eq: model 2}) and (\ref{eq: model 3})}
\label{sec:proof A}
Starting from the expression in (\ref{eq: model 3}), we can write based on (\ref{eq: diric kernel})
\begin{align}
\label{eq: model 3 proof1}
&\sum_{k=-N}^{N} D_{N}\left(\frac{k}{L}-f_{j}\right) e^{\frac{i 2 \pi p k}{L} } = \frac{1}{L} \sum_{r=-N}^{N} \sum_{k=-N}^{N} e^{i2\pi \left(\frac{k}{L}-f_{j}\right) r} e^{\frac{i 2 \pi p k}{L} } \nonumber\\
&= \sum_{r=-N}^{N} e^{-i2\pi f_{j} r} \frac{1}{L} \sum_{k=-N}^{N} e^{ \frac{i2\pi k \left( r+p\right)}{L} }  =   e^{i2\pi p f_{j}},
\end{align}
where the last equality follows from
\begin{align}
\label{eq: A3 3}
     \sum_{k=-N}^{N} e^{\frac{i2\pi k \left(r+p\right)}{L} } =\left\{
                \begin{array}{ll}
                  L &\text{if} \ \ r = -p\\
                  0 &\text{if} \ \ r \neq -p\\
                \end{array}
              \right.
\end{align}
Now, by substituting (\ref{eq: model 3 proof1}) in (\ref{eq: model 3}) we obtain
%\begin{align}
%\label{eq: model 3 proof2 temp1 }
%y\left(p\right)&= \sum_{j=1}^{R} c_{j} e^{i2\pi p f_{j}} \sum_{l=-N}^{N} D_{N}\left(\frac{l}{L}-\tau_{j}\right) \dv^{H}_{\left(p-l\right)} \hv_{j}  =  \sum_{j=1}^{R} c_{j} e^{i2\pi p f_{j}} \sum_{l=-N}^{N} \frac{1}{L} \sum_{r=-N}^{N} e^{i2\pi \left(\frac{l}{L}-\tau_{j}\right) r}  \dv^{H}_{\left(p-l\right)} \hv_{j} \\
%\label{eq: model 3 proof2 temp2 }
%&= \sum_{j=1}^{R} c_{j} e^{i2\pi p f_{j}} \sum_{l=-N}^{N} \frac{1}{L} \sum_{r=-N}^{N} e^{i2\pi \left(\frac{p-l}{L}-\tau_{j}\right) r}  \dv^{H}_{l} \hv_{j},  
%\end{align} 

\begin{align}
y\left(p\right)&= \sum_{j=1}^{R} c_{j} e^{i2\pi p f_{j}} \sum_{l=-N}^{N} D_{N}\left(\frac{l}{L}-\tau_{j}\right) \dv^{H}_{\left(p-l\right)} \hv_{j}  \nonumber \\
\label{eq: model 3 proof2 temp1 }
&=  \sum_{j=1}^{R} c_{j} e^{i2\pi p f_{j}} \sum_{l=-N}^{N} \frac{1}{L} \sum_{r=-N}^{N} e^{i2\pi \left(\frac{l}{L}-\tau_{j}\right) r}  \dv^{H}_{\left(p-l\right)} \hv_{j} \\
\label{eq: model 3 proof2 temp2 }
&= \sum_{j=1}^{R} c_{j} e^{i2\pi p f_{j}} \sum_{l=-N}^{N} \frac{1}{L} \sum_{r=-N}^{N} e^{i2\pi \left(\frac{p-l}{L}-\tau_{j}\right) r}  \dv^{H}_{l} \hv_{j},  
\end{align} 
where (\ref{eq: model 3 proof2 temp1 }) is based on (\ref{eq: diric kernel}) while (\ref{eq: model 3 proof2 temp2 }) is a consequence of the periodicity property of $s_{j}\left(l\right)$. Finally, by rearranging the terms in (\ref{eq: model 3 proof2 temp2 }), we can obtain (\ref{eq: model 3}). 

\section{Proof of Proposition~\ref{pro: main pro}}
\label{A2}
First, the variable $\qv$ that satisfies (\ref{eq: hold assump 1}) and (\ref{eq: hold assump 2}) is dual feasible. To show that, we have
\begin{align}
&||\mathcal{X}^{*}\left(\qv\right) ||_{\mathcal{A}}^{*}= \sup_{||\Um ||_{\mathcal{A}}\leq 1} \big\langle\mathcal{X}^{*}\left(\qv\right),\Um \big\rangle_{\mathbb{R}}  \nonumber\\
&= \sup_{\rv \in [0,1]^{2}, ||\hv||_{2}=1} \big\langle\mathcal{X}^{*}\left(\qv\right),\hv \av\left(\rv \right)^{H}\big\rangle_{\mathbb{R}}  \nonumber \\
&= \sup_{\rv \in [0,1]^{2}, ||\hv||_{2}=1} \big| \big\langle \hv, \mathcal{X}^{*}\left(\qv\right)  \av\left(\rv\right) \big\rangle\big| \nonumber\\
\label{eq: app dual of atomic 3}
&\leq \sup_{\rv \in [0,1]^{2}} || \mathcal{X}^{*}\left(\qv\right)  \av\left(\rv\right)||_{2} = \sup_{\rv \in [0,1]^{2}} || \fv\left(\rv\right)||_{2} \leq 1,
\end{align}
where the last inequality is based on (\ref{eq: hold assump 1}) and (\ref{eq: hold assump 2}).

Next, we show that $\Um$ is a primal optimal solution for (\ref{eq: optimization atomic}) and $\qv$ is a dual optimal solution for (\ref{eq: optimzation dual}) when $\qv$ satisfies (\ref{eq: hold assump 1}) and (\ref{eq: hold assump 2}). For that, we can write based on (\ref{eq: sensing to vector})
%Next, we show that any vector $\qv$ that satisfies (\ref{eq: hold assump 1}) and (\ref{eq: hold assump 2}) implies that $\langle\qv, \yv\rangle_{\mathbb{R}} = ||\Um||_{\mathcal{A}}$. For that we can write based on (\ref{eq: sensing to vector})
\begin{align}
&\langle\qv, \yv\rangle_{\mathbb{R}}= \big\langle\qv, \mathcal{X}\left(\Um\right)\big\rangle_{\mathbb{R}} = \big\langle\mathcal{X}^{*}\left(\qv\right), \Um\big\rangle_{\mathbb{R}}\nonumber\\
& =\bigg\langle\mathcal{X}^{*}\left(\qv\right), \sum_{j=1}^{R} c_{j} \hv_{j} \av\left(\rv_{j}\right)^{H}\bigg\rangle_{\mathbb{R}}  \nonumber\\
&=  \sum_{j=1}^{R} c_{j}^{*} \big\langle\mathcal{X}^{*}\left(\qv\right), \hv_{j} \av\left(\rv_{j}\right)^{H}\big\rangle_{\mathbb{R}}=  \sum_{j=1}^{R} c_{j}^{*} \big\langle \hv_{j}, \fv\left(\rv_{j}\right)\big\rangle_{\mathbb{R}}  \nonumber\\
\label{eq: app prop proof}
& =  \sum_{j=1}^{R} \text{Re}\left[ c_{j}^{*} \text{sign}\left(c_{j}\right)\right] = \sum_{j=1}^{R}|c_{j}| \\
\label{eq: app prop proof 2}
& \geq ||\Um||_{\mathcal{A}},
\end{align}
where the first equality in (\ref{eq: app prop proof}) is based on (\ref{eq: hold assump 1}) and $||\hv_{j}||_{2}=1$ while (\ref{eq: app prop proof 2}) is from the atomic norm definition. On the other hand, we have based on H\"{o}lder inequality
\begin{align}
\label{eq: app prop proof 2 sub}
\langle\qv, \yv\rangle_{\mathbb{R}}= \big\langle\mathcal{X}^{*}\left(\qv\right), \Um\big\rangle_{\mathbb{R}} \leq ||\mathcal{X}^{*}\left(\qv\right)||_{\mathcal{A}}^{*}||\Um||_{\mathcal{A}} \leq ||\Um||_{\mathcal{A}}
\end{align}
where the last inequality is based on (\ref{eq: hold assump 1}) and (\ref{eq: hold assump 2}). Thus, we conclude from (\ref{eq: app prop proof 2}) and (\ref{eq: app prop proof 2 sub}) that $\langle\qv, \yv\rangle_{\mathbb{R}} = ||\Um||_{\mathcal{A}}$ when $\qv$ satisfies (\ref{eq: hold assump 1}) and (\ref{eq: hold assump 2}). Now, since the pair $\left(\Um,\qv\right)$ is primal-dual feasible to (\ref{eq: optimization atomic}) and (\ref{eq: optimzation dual}), it means that $\Um$ is the primal optimal and $\qv$ is the dual optimal based on strong duality. 

What remains now is to show that $\Um$ is the unique optimal solution to (\ref{eq: optimization atomic}). To this end, let us assume that there exists another solution $\widebar{\Um}:= \sum_{\bar{\rv}_{j} \in \bar{\mathcal{R}}} \bar{c}_{j} \bar{\hv}_{j} \av\left(\bar{\rv}_{j}\right)^{H}$ such that $||\widebar{\Um}||_{\mathcal{A}} =\sum_{\bar{\rv}_{j} \in \bar{\mathcal{R}}} |\bar{c}_{j}|$ where $\mathbf{\bar{\mathcal{R}}}\neq\mathbf{{\mathcal{R}}}$. Since the set of atoms with its shifts in $\mathcal{R}$ are linearly independent, it will be enough for us to prove that $\Um$ and $\widebar{\Um}$ have the same support if we would like to show that they match. Starting from the definition of $\widebar{\Um}$ above we can write
\begin{align}
&\langle\qv, \yv\rangle_{\mathbb{R}} = \big\langle\mathcal{X}^{*}\left(\qv\right), \widebar{\Um}\big\rangle_{\mathbb{R}} \nonumber\\
&= \sum_{\bar{\rv}_{j} \in \mathcal{{R}}} \bar{c}_{j}^{*} \big\langle \bar{\hv}_{j}, \fv\left(\bar{\rv}_{j}\right)\big\rangle_{\mathbb{R}} + \sum_{\bar{\rv}_{j} \in \mathcal{\bar{R}} \backslash \mathcal{{R}}} \bar{c}_{j}^{*} \big\langle \bar{\hv}_{j}, \fv\left(\bar{\rv}_{j}\right)\big\rangle_{\mathbb{R}} \nonumber\\
& < \sum_{\bar{\rv}_{j} \in \mathcal{R}}|\bar{c}_{j}| +  \sum_{\bar{\rv}_{j} \in \mathcal{\bar{R}} \backslash \mathcal{{R}}}|\bar{c}_{j}| = ||\widebar{\Um}||_{\mathcal{A}}, \nonumber
\end{align}
where the strict inequality is based on (\ref{eq: hold assump 2}). However, this contradicts with strong duality and, therefore, we can conclude that all shifts are supported on $\mathcal{R}$.

On the other hand, if we refer to the estimate of $\rv_{j}$ by $\hat{\rv}_{j}$, then, condition (2) in Proposition~\ref{pro: main pro} ensures that estimating $c_{j}\hv_{j}$ by solving the following linear system 
\begin{equation}
%\label{eq: app matrix formulation 2}
{\begin{bmatrix}  \av\left(\hat{\rv}_{1}\right)^{H} \widetilde{\Dm}_{-N} &\dots & \av\left(\hat{\rv}_{R}\right)^{H} \widetilde{\Dm}_{-N} \\ \vdots  & \ddots & \vdots \\ \av\left(\hat{\rv}_{1}\right)^{H} \widetilde{\Dm}_{N} &\dots & \av\left(\hat{\rv}_{R}\right)^{H} \widetilde{\Dm}_{N} \end{bmatrix}}  \begin{bmatrix} c_{1}\hv_{1} \\ \vdots \\ c_{R}\hv_{R} \end{bmatrix}  = \begin{bmatrix} y\left(-N\right) \\ \vdots \\ y\left(N\right) \end{bmatrix} \nonumber
\end{equation}
which is based on (\ref{eq: final model}) provides a unique solution. Therefore, we can conclude that $\Um$ is the unique optimal solution to (\ref{eq: optimization atomic}) if Proposition~\ref{pro: main pro} conditions are satisfied.

\section{Proof of (\ref{eq: simple polynomail})}
\label{A3}
By substituting (\ref{eq: vec 1}) and (\ref{eq: matrix D}) into (\ref{eq: dual polynomail equa}), we obtain
\begin{align}
\label{eq: A3 1}
\fv\left(\rv\right) &= \sum_{p=-N}^{N} [\qv]_{p} \sum_{l=-N}^{N} D_{N}\left(\frac{l}{L}-\tau\right)\dv_{\left(p-l\right)}\times \nonumber\\
&  \sum_{k=-N}^{N} D_{N}\left(\frac{k}{L}-f\right)e^{\frac{-i2\pi p k}{L}}. 
\end{align}
The last summation in (\ref{eq: A3 1}) can be written using (\ref{eq: diric kernel}) as
%\begin{align}
%\label{eq: A3 2}
%\sum_{k=-N}^{N} D_{N}\left(\frac{k}{L}-f\right) e^{\frac{-i2\pi p k}{L}} &=  \sum_{k=-N}^{N}  \frac{1}{L} \sum_{r=-N}^{N} e^{i2\pi \left(\frac{k}{L}-f\right) r} e^{\frac{-i2\pi p k}{L}} = \sum_{r=-N}^{N} e^{-i2\pi f r}  \frac{1}{L} \sum_{k=-N}^{N} e^{\frac{i2\pi k \left(r-p\right)}{L} } =e^{-i2\pi p f}, 
%\end{align}
\begin{align}
\label{eq: A3 2}
&\sum_{k=-N}^{N} D_{N}\left(\frac{k}{L}-f\right) e^{\frac{-i2\pi p k}{L}} =  \sum_{k=-N}^{N}  \frac{1}{L} \sum_{r=-N}^{N} e^{i2\pi \left(\frac{k}{L}-f\right) r} \times \nonumber\\
&e^{\frac{-i2\pi p k}{L}}= \sum_{r=-N}^{N} e^{-i2\pi f r}  \frac{1}{L} \sum_{k=-N}^{N} e^{\frac{i2\pi k \left(r-p\right)}{L} } =e^{-i2\pi p f}, 
\end{align}
where the last equality follows from (\ref{eq: A3 3}). Now, by substituting (\ref{eq: A3 2}) into (\ref{eq: A3 1}) we obtain
%\begin{align}
%\label{eq: A3 6}
%\fv\left(\rv\right) &= \sum_{p=-N}^{N} [\qv]_{p} \sum_{l=-N}^{N} D_{N}\left(\frac{l}{L}-\tau\right)  \dv_{\left(p-l\right)} e^{-i2\pi p f}= \sum_{p=-N}^{N} [\qv]_{p} \sum_{l=-N}^{N} \frac{1}{L} \sum_{r=-N}^{N} e^{i2\pi \left(\frac{l}{L}-\tau\right) r}\dv_{\left(p-l\right)} e^{-i2\pi p f}  \nonumber\\
%&=\sum_{p=-N}^{N} [\qv]_{p}\frac{1}{L} \sum_{l,r=-N}^{N} e^{ \frac{i2\pi l r}{L} }\dv_{\left(p-l\right)} e^{-i2\pi \left(r \tau+p f \right)} =\sum_{p=-N}^{N} [\qv]_{p}\frac{1}{L} \sum_{l,r=-N}^{N} e^{\frac{i2\pi r \left(p-l\right)}{L} } e^{-i2\pi \left(r \tau+p f \right)}\dv_{l},
%\end{align}
\begin{align}
\label{eq: A3 6}
\fv\left(\rv\right) &= \sum_{p=-N}^{N} [\qv]_{p} \sum_{l=-N}^{N} D_{N}\left(\frac{l}{L}-\tau\right)  \dv_{\left(p-l\right)} e^{-i2\pi p f}\nonumber\\
&= \sum_{p=-N}^{N} [\qv]_{p} \sum_{l=-N}^{N} \frac{1}{L} \sum_{r=-N}^{N} e^{i2\pi \left(\frac{l}{L}-\tau\right) r}\dv_{\left(p-l\right)} e^{-i2\pi p f}  \nonumber\\
&=\sum_{p=-N}^{N} [\qv]_{p}\frac{1}{L} \sum_{l,r=-N}^{N} e^{ \frac{i2\pi l r}{L} }\dv_{\left(p-l\right)} e^{-i2\pi \left(r \tau+p f \right)} \nonumber\\
&=\sum_{p=-N}^{N} [\qv]_{p}\frac{1}{L} \sum_{l,r=-N}^{N} e^{\frac{i2\pi r \left(p-l\right)}{L} } e^{-i2\pi \left(r \tau+p f \right)}\dv_{l},
\end{align}
where the last equality is from the periodicity property.

\section{Proof of (\ref{eq: det G})}
\label{A4}
Starting from the left-hand side of (\ref{eq: det G}), and by using (\ref{eq: G simpl}), we can write
%\begin{align}
%\label{eq: A4 1}
%\mathbb{E}\left[\Mm^{(m',n')}_{\left(m,n\right)}\left(\rv,\rv_{j}\right)\right]  &= \frac{1}{L} \sum_{p=-N}^{N} \frac{1}{T^{2}}\sum_{l,l',k,k'=-N}^{N} g_{k'}\left(-i2\pi k' \right)^{m} \left(-i2\pi k \right)^{m'}g_{p} \left(-i2\pi p \right)^{(n+n')}\times \nonumber\\
%& e^{i 2 \pi \frac{\left(kl-k'l'\right)}{L}} e^{-i2\pi \left(k\tau-k'\tau_{j}\right)} e^{-i2\pi p \left(f-f_{j}\right)} \mathbb{E}\left[\dv_{\left(p-l\right)} \dv_{\left(p-l'\right)}^{H}\right]. 
%\end{align}
\begin{align}
\label{eq: A4 1}
&\mathbb{E}\left[\Mm^{(m',n')}_{\left(m,n\right)}\left(\rv,\rv_{j}\right)\right]  =\nonumber\\
& \frac{1}{L} \sum_{p=-N}^{N} \frac{1}{T^{2}}\sum_{l,l',k,k'=-N}^{N} g_{k'}\left(-i2\pi k' \right)^{m} \left(-i2\pi k \right)^{m'}g_{p} \times \nonumber\\
&   \left(-i2\pi p \right)^{(n+n')} e^{i 2 \pi \frac{\left(kl-k'l'\right)}{L}} e^{-i2\pi \left(k\tau-k'\tau_{j}\right)} e^{-i2\pi p \left(f-f_{j}\right)}\times \nonumber\\
& \mathbb{E}\left[\dv_{\left(p-l\right)} \dv_{\left(p-l'\right)}^{H}\right]. 
\end{align}

Based on Assumption~\ref{as 1}, we have $\mathbb{E}[\dv_{\left(p-l\right)} \dv_{\left(p-l'\right)}^{H}] =\Id_{\text{K}}$ for $l=l'$ and $\bm{0}$ for $l \neq l'$. Substituting this in (\ref{eq: A4 1}) we obtain
%\begin{align}
%\label{eq: A4 2}
%&\mathbb{E}\left[\Mm^{(m',n')}_{\left(m,n\right)}\left(\rv,\rv_{j}\right)\right] \nonumber\\
%  &=\frac{1}{L} \sum_{p=-N}^{N} \frac{1}{T^{2}}\sum_{l,k,k'=-N}^{N} g_{k'}\left(-i2\pi k' \right)^{m}\left(-i2\pi k \right)^{m'} g_{p} \left(-i2\pi p \right)^{(n+n')}  e^{ \frac{i 2 \pi l\left(k-k'\right)}{L}}e^{-i2\pi \left(k\tau-k'\tau_{j}\right)} e^{-i2\pi p \left(f-f_{j}\right)}\Id_{\text{K}} \nonumber\\
%  &= \sum_{p=-N}^{N} \frac{1}{T^{2}}\sum_{k,k'=-N}^{N} g_{k'}\left(-i2\pi k' \right)^{m}  \left(-i2\pi k \right)^{m'}g_{p} \left(-i2\pi p \right)^{(n+n')} e^{-i2\pi \left(k\tau-k'\tau_{j}\right)} e^{-i2\pi p \left(f-f_{j}\right)}\frac{1}{L}  \sum_{l=-N}^{N} e^{ \frac{i 2 \pi l\left(k-k'\right)}{L}} \Id_{\text{K}} \nonumber\\
%  &=\sum_{p=-N}^{N} \frac{1}{T^{2}}\sum_{k=-N}^{N} g_{k}\left(-i2\pi k \right)^{(m+m')}  g_{p} \left(-i2\pi p \right)^{(n+n')}e^{-i2\pi k \left(\tau-\tau_{j}\right)} e^{-i2\pi p \left(f-f_{j}\right)}   \Id_{\text{K}},
%\end{align}
\begin{align}
\label{eq: A4 2}
&\mathbb{E}\left[\Mm^{(m',n')}_{\left(m,n\right)}\left(\rv,\rv_{j}\right)\right]=
  \frac{1}{L} \sum_{p=-N}^{N} \frac{1}{T^{2}}\sum_{l,k,k'=-N}^{N} g_{k'}\left(-i2\pi k' \right)^{m} \times \nonumber\\
  &\left(-i2\pi k \right)^{m'} g_{p} \left(-i2\pi p \right)^{(n+n')}  e^{ \frac{i 2 \pi l\left(k-k'\right)}{L}}e^{-i2\pi \left(k\tau-k'\tau_{j}\right)} \times \nonumber\\
 &e^{-i2\pi p \left(f-f_{j}\right)}\Id_{\text{K}} = \sum_{p=-N}^{N} \frac{1}{T^{2}}\sum_{k,k'=-N}^{N} g_{k'}\left(-i2\pi k' \right)^{m} \times \nonumber\\
&  \left(-i2\pi k \right)^{m'}g_{p} \left(-i2\pi p \right)^{(n+n')} e^{-i2\pi \left(k\tau-k'\tau_{j}\right)} e^{-i2\pi p \left(f-f_{j}\right)} \times  \nonumber\\
&\frac{1}{L}  \sum_{l=-N}^{N} e^{ \frac{i 2 \pi l\left(k-k'\right)}{L}} \Id_{\text{K}} =\sum_{p=-N}^{N} \frac{1}{T^{2}}\sum_{k=-N}^{N} g_{k}\left(-i2\pi k \right)^{(m+m')}  \times \nonumber\\
&  g_{p} \left(-i2\pi p \right)^{(n+n')}e^{-i2\pi k \left(\tau-\tau_{j}\right)} e^{-i2\pi p \left(f-f_{j}\right)}   \Id_{\text{K}},
\end{align}
where the last equality is based on (\ref{eq: A3 3}).

Now, given the fact that $g_{k}$ and $g_{p}$ are even functions, we can simplify (\ref{eq: A4 2}) as
%\begin{align}
%&\mathbb{E}\left[\Mm^{(m',n')}_{\left(m,n\right)}\left(\rv,\rv_{j}\right)\right]  =  \left(\frac{1}{T}\sum_{k=-N}^{N} g_{k}\left(i2\pi k \right)^{(m+m')}  e^{i2\pi k \left(\tau-\tau_{j}\right)} \right)  \left( \frac{1}{T} \sum_{p=-N}^{N} g_{p} \left(i2\pi p \right)^{(n+n')} e^{i2\pi p \left(f-f_{j}\right)} \right)   \Id_{\text{K}} \nonumber
%\end{align}
\begin{align}
%\label{eq: A4 4}
&\mathbb{E}\left[\Mm^{(m',n')}_{\left(m,n\right)}\left(\rv,\rv_{j}\right)\right]  =  \frac{1}{T}\sum_{k=-N}^{N} g_{k}\left(i2\pi k \right)^{(m+m')}  e^{i2\pi k \left(\tau-\tau_{j}\right)}   \nonumber\\
&  \frac{1}{T} \sum_{p=-N}^{N} g_{p} \left(i2\pi p \right)^{(n+n')} e^{i2\pi p \left(f-f_{j}\right)}  \Id_{\text{K}} \nonumber
\end{align}
which leads to (\ref{eq: det G}) upon using the definition in (\ref{eq: another fejer}).
\section{Proof of Lemma~\ref{l3}}
\label{A5}
Based on (\ref{eq: G in matrix multi 2}) we can write the entry at $\left(l,l'\right)$ location in $\Rm_{(m,n)}^{(m',n')}\left(\rv,\rv_{j}\right)$ as
\begin{align}
\label{eq: a5}
&\left[\Rm_{(m,n)}^{(m',n')}\left(\rv,\rv_{j}\right)\right]_{(l,l')} \hspace{-3pt} = \frac{1}{L}\sum_{p=-N}^{N} \frac{1}{T^{2}} \sum_{k,k'=-N}^{N} \hspace{-3pt} g_{k'}\left(-i2\pi k' \right)^{m}\times \nonumber\\
&  \left(-i2\pi k \right)^{m'}  g_{p} \left(-i2\pi p \right)^{(n+n')} e^{i 2 \pi \frac{k\left(p-l\right)}{L}}  e^{-i 2 \pi \frac{k'\left(p- l'\right)}{L}} \times \nonumber\\
& e^{-i2\pi \left(k\tau-k'\tau_{j}\right)} e^{-i2\pi p \left(f-f_{j}\right)} \nonumber\\
&= \frac{1}{L} \hspace{-2pt}\sum_{p=-N}^{N} \left(\frac{1}{T}\sum_{k'=-N}^{N}\hspace{-3pt} g_{k'}\left(-i2\pi k' \right)^{m}   e^{-i 2 \pi \frac{k'\left(p- l'\right)}{L}} e^{i2\pi k'\tau_{j}} \right) \times  \nonumber\\
& \left( \frac{1}{T}\sum_{k=-N}^{N}\left(-i2\pi k \right)^{m'}   e^{i 2 \pi \frac{k\left(p-l\right)}{L}}  e^{-i2\pi k\tau} \right) \times \nonumber\\
& g_{p} \left(-i2\pi p \right)^{(n+n')}e^{-i2\pi p \left(f-f_{j}\right)}.
\end{align}
Since $g_{k}$ is an even function, we can write based on (\ref{eq: another fejer})
\begin{align}
\label{eq: a5 2}
&\frac{1}{T}\sum_{k'=-N}^{N} g_{k'}\left(-i2\pi k' \right)^{m}   e^{-i 2 \pi \frac{k'\left(p- l'\right)}{L}} e^{i2\pi k'\tau_{j}} \nonumber\\
& = \frac{1}{T}\sum_{k'=-N}^{N} g_{k'}\left(i2\pi k' \right)^{m}   e^{i 2 \pi \frac{k'\left(p- l'\right)}{L}} e^{-i2\pi k'\tau_{j}}\nonumber\\
& = F^{m}\left(\frac{p-l'}{L} -\tau_{j}\right).
\end{align}

Substituting (\ref{eq: a5 2}) in (\ref{eq: a5}) we obtain
\begin{align}
\label{eq: a5 3}
&\left[\Rm_{(m,n)}^{(m',n')}\left(\rv,\rv_{j}\right)\right]_{(l,l')}  =\frac{1}{L} \sum_{p=-N}^{N} F^{m}\left(\frac{p-l'}{L} -\tau_{j}\right) \times \nonumber\\
& \left( \frac{1}{T}\sum_{k=-N}^{N}\left(-i2\pi k \right)^{m'}   e^{i 2 \pi \frac{k\left(p-l\right)}{L}}  e^{-i2\pi k\tau} \right)\times \nonumber\\
&g_{p} \left(-i2\pi p \right)^{(n+n')}e^{-i2\pi p \left(f-f_{j}\right)}.
\end{align}
Now, given that $|\left(-i2\pi p \right)^{(n+n')} e^{-i 2 \pi p \left(f-f_{j}\right)}| \leq \left(2\pi N\right)^{(n+n')}$ and that $|g_{p}| \leq 1$, we can bound the absolute value of (\ref{eq: a5 3}) as
\begin{align}
\label{eq: a5 4}
&\Big|\left[\Rm_{(m,n)}^{(m',n')}\left(\rv,\rv_{j}\right)\right]_{(l,l')}\Big| \leq \frac{1}{L} \sum_{p=-N}^{N}\Bigg| F^{m}\left(\frac{p-l'}{L} -\tau_{j}\right)\Bigg| \times \nonumber\\
&\Bigg|\frac{1}{T}\sum_{k=-N}^{N} \left(-i2\pi k \right)^{m'}   e^{i 2 \pi \frac{k\left(p-l\right)}{L}}  e^{-i2\pi k\tau} \Bigg|\left(2\pi N\right)^{(n+n')}
\end{align}
Based on the result obtained in \cite[Lemma~3]{heckel2016super} we have
\begin{align}
\label{eq: a5 5}
\Bigg| F^{m}\left(\frac{p-l'}{L} -\tau_{j}\right)\Bigg| \leq \bar{C}^{*} \left(2 \pi N \right)^{m} \min \left(1, \frac{1}{p^4}\right),
\end{align}
where $\bar{C}^{*}$ is a numerical constant. Substituting this in (\ref{eq: a5 4}) we obtain
\begingroup
\fontsize{9.4pt}{9.9pt}
\begin{align}
\label{eq: a5 6}
&\Big|\left[\Rm_{(m,n)}^{(m',n')}\left(\rv,\rv_{j}\right)\right]_{(l,l')}\Big| \leq \frac{1}{L}  \left(2 \pi N \right)^{m+m'+n+n'} \times \nonumber\\
& \underbrace{\bar{C}^{*}\left(2 \pi N \right)^{-m'} \hspace{-6pt}\sum_{p=-N}^{N}\hspace{-6pt}\min \left(1, \frac{1}{p^4}\right)\Bigg|\frac{1}{T} \hspace{-3pt}\sum_{k=-N}^{N} \hspace{-6pt} \left(-i2\pi k \right)^{m'}   \hspace{-3pt} e^{i 2 \pi k \frac{\left(p-l\right)-L\tau}{L} } \Bigg|}_{U\left(\tau + \frac{l}{L}\right)}
\end{align}
\endgroup
Furthermore, it is shown in \cite[Appendix~F]{heckel2016super} that $U\left(t\right)$ as defined in (\ref{eq: a5 6}) is a 1-periodic function that satisfies
\begin{equation}
\label{eq: a5 7}
U\left(t\right) \leq \hat{C}^{*} \min \left(1 , \frac{1}{L |t|}\right),
\end{equation}
where $\hat{C}^{*}$ is a constant. Finally, we can conclude based on (\ref{eq: a5 6}) and (\ref{eq: a5 7}) that
\begin{align}
\label{eq: a5 8}
&\big|\big|\Rm_{(m,n)}^{(m',n')}\left(\rv,\rv_{j}\right)\big|\big|^{2}_{F} = \sum_{l,l'=-N}^{N} \Big| \left[\Rm_{(m,n)}^{(m',n')}\left(\rv,\rv_{j}\right)\right]_{(l,l')}\Big|^{2}\nonumber\\
&\leq \frac{1}{L^{2}}  \left(2 \pi N \right)^{2(m+m'+n+n')} \sum_{l',l=-N}^{N}\left(\hat{C}^{*}\min \left(1 , \frac{1}{L |\frac{l}{L}|}\right)\right)^{2} \nonumber\\
& \leq \left(\frac{\hat{C}^{*}}{L} \right)^{2} \left(2 \pi N \right)^{2(m+m'+n+n')} \sum_{l'=-N}^{N} \left(1+2\sum_{l\geq 1} \frac{1}{l^{2}}\right) \nonumber\\
& = \frac{(\hat{C}^{*})^{2}}{L} \left(2 \pi N \right)^{2(m+m'+n+n')} \left(1+\frac{\pi^2}{3}\right),
\end{align}
which boils down to (\ref{eq:for norm }) upon setting $C_{2}=\hat{C}^{*} \sqrt{1+\frac{\pi^2}{3}}$.

\section{Proof of Lemma~\ref{l2}}
\label{proof of l2}
In the following, we will provide the proof of (\ref{eq: diagonal bound}) as that of (\ref{eq: diagonal bound 2}) follows the same steps. First, given the fact that $\mu = \sqrt{\frac{\pi^{2}}{3}\left(N^{2}+4N\right)}$ \cite{candes2014towards}, we can write
\begin{align}
\label{eq: lemma proof 1}
\frac{\left(2 \pi N\right)^{m}}{\mu^{m}} = \frac{\left(2 \pi N\right)^{m} 3^{\frac{m}{2}}}{\pi^{m}\left(N^2+4N\right)^{\frac{m}{2}}}  \leq 12^{\frac{m}{2}}.
\end{align}
Starting from the left-hand side of (\ref{eq: diagonal bound}), we can write
\begin{align}
\text{Pr} &\left[\frac{1}{\mu^{m+m'+n+n'}} \Big| \hat{\dv}_{i'}^{H} \Rm_{(m,n)}^{(m',n')}\left(\rv,\rv_{j}\right) \hat{\dv}_{i'}-\right. \nonumber\\
&\left.\mathbb{E}\left[\hat{\dv}_{i'}^{H} \Rm_{(m,n)}^{(m',n')}\left(\rv,\rv_{j}\right) \hat{\dv}_{i'}\right]\Big| \geq C_{2} 12^{\frac{m+m'+n+n'}{2}} \frac{\alpha}{\sqrt{L}}\right] \nonumber\\
\label{eq: diagonal bound proof 1_1}
& \leq  \text{Pr} \left[ \Big| \hat{\dv}_{i'}^{H} \Rm_{(m,n)}^{(m',n')}\left(\rv,\rv_{j}\right) \hat{\dv}_{i'}-\right. \nonumber\\
&\left.\mathbb{E}\left[\hat{\dv}_{i'}^{H} \Rm_{(m,n)}^{(m',n')}\left(\rv,\rv_{j}\right) \hat{\dv}_{i'}\right]\Big| \geq C_{2} \left(2 \pi N\right)^{m+m'+n+n'} \frac{\alpha}{\sqrt{L}}\right] \\
\label{eq: diagonal bound proof 1_2}
& \leq  \text{Pr} \left[ \Big| \hat{\dv}_{i'}^{H} \Rm_{(m,n)}^{(m',n')}\left(\rv,\rv_{j}\right) \hat{\dv}_{i'}-\right. \nonumber\\
&\left.\mathbb{E}\left[\hat{\dv}_{i'}^{H} \Rm_{(m,n)}^{(m',n')}\left(\rv,\rv_{j}\right) \hat{\dv}_{i'}\right]\Big| \geq \alpha ||\Rm_{(m,n)}^{(m',n')}\left(\rv,\rv_{j}\right)||_{F} \right] \\
\label{eq: diagonal bound proof 1_3}
& \leq  2 \exp {\left(-\frac{1}{C} \min \left(\frac{\alpha^{2}}{2 \widetilde{K}^{4}}, \frac{ \alpha}{\widetilde{K}^{2}}\right)\right)},
\end{align}
where (\ref{eq: diagonal bound proof 1_1}) is based on (\ref{eq: lemma proof 1}) while (\ref{eq: diagonal bound proof 1_2}) is obtained by using Lemma~\ref{l3}. To prove (\ref{eq: diagonal bound proof 1_3}), we set $\Am =\Rm_{(m,n)}^{(m',n')}\left(\rv,\rv_{j}\right)$ and $t = \alpha ||\Rm_{(m,n)}^{(m',n')}\left(\rv,\rv_{j}\right)||_{F}$ in (\ref{eq: h w ineq}), then, we use the fact that $||\Am||_{F} \geq \||\Am ||_{2}$ and that $\exp\left(-x\right)$ is a decaying function for $x \in [0,\infty)$. By following the same steps, and upon applying (\ref{eq: h w ineq 2}), we can prove (\ref{eq: diagonal bound 2}).

\section{Proof of Lemma~\ref{th1}}
\label{proof of th1}
Starting from the definition of $\mathcal{E}_{1}$ we can write
\begin{align}
\text{Pr}&\left[\frac{1}{\mu^{m+m'+n+n'}}\Big|\Big| \Mm_{(m,n)}^{(m',n')}\left(\rv,\rv_{j}\right) - \right.\nonumber\\
&\left.\mathbb{E}\left[\Mm_{(m,n)}^{(m',n')}\left(\rv,\rv_{j}\right) \right]\Big|\Big|_{2} \geq \epsilon_{1}\right]  \nonumber\\
\label{eq: conver proof 1}
&\leq \text{Pr}\left[\sqrt{K} \max_{l,k,m,m',n,n'}\frac{1}{\mu^{m+m'+n+n'}}\Big| \Mm_{(m,n)}^{(m',n')}\left(\rv,\rv_{j}\right) - \right.\nonumber\\
&\left.\mathbb{E}\left[\Mm_{(m,n)}^{(m',n')}\left(\rv,\rv_{j}\right) \right]\Big|_{(l,k)} \geq \epsilon_{1}\right]   \\
\label{eq: conver proof 2}
&\leq\sum_{l,k,m,m',n,n'}\text{Pr}\left[\frac{1}{\mu^{m+m'+n+n'}}\Big| \Mm_{(m,n)}^{(m',n')}\left(\rv,\rv_{j}\right) - \right.\nonumber\\
&\left.\mathbb{E}\left[\Mm_{(m,n)}^{(m',n')}\left(\rv,\rv_{j}\right) \right]\Big|_{(l,k)} \geq \frac{\epsilon_{1}}{\sqrt{K}}\right] \\
\label{eq: conver proof 3}
& =\sum_{l,k,m,m',n,n'}\text{Pr}\left[\frac{1}{\mu^{m+m'+n+n'}}\Big| \Mm_{(m,n)}^{(m',n')}\left(\rv,\rv_{j}\right) - \right.\nonumber\\
&\left.\mathbb{E}\left[\Mm_{(m,n)}^{(m',n')}\left(\rv,\rv_{j}\right)\right]\Big|_{(l,k)} \geq \frac{12 C_{2} \alpha }{\sqrt{L}}\right], 
\end{align}
where (\ref{eq: conver proof 1}) follows from the fact that $\forall \Am, \Bm \in \mathbb{C}^{K \times K}, ||\Am-\Bm||_{2} \leq \sqrt{K} \max_{i,j} |\Am-\Bm|_{(i,j)}$ where $|\cdot|_{(i,j)}$ refers to the absolute value of the $(i,j)$ entry. Next, (\ref{eq: conver proof 2}) is based on the union bound while (\ref{eq: conver proof 3}) is obtained by setting $\epsilon_{1} = \frac{12 \alpha C_{2} \sqrt{K}}{\sqrt{L}}$.

Now given the fact that $12^{\frac{m+m'+n+n'}{2}} \leq 12$ for $m+m'+n+n'\leq 2$ we can write starting from (\ref{eq: conver proof 3})
\begin{align}
&\sum_{l,k,m,m',n,n'}\text{Pr}\left[\frac{1}{\mu^{m+m'+n+n'}}\bigg| \Mm_{(m,n)}^{(m',n')}\left(\rv,\rv_{j}\right) - \right.\nonumber\\
&\left.\mathbb{E}\left[\Mm_{(m,n)}^{(m',n')}\left(\rv,\rv_{j}\right)\right]\bigg|_{(l,k)} \geq \frac{12 C_{2} \alpha }{\sqrt{L}}\right] \nonumber\\
\label{eq: conver proof 4}
& \leq \sum_{l,k,m,m',n,n'}\text{Pr}\left[\frac{1}{\mu^{m+m'+n+n'}}\bigg| \Mm_{(m,n)}^{(m',n')}\left(\rv,\rv_{j}\right) -\right.\nonumber\\
&\left.\mathbb{E}\left[\Mm_{(m,n)}^{(m',n')}\left(\rv,\rv_{j}\right) \right]\bigg|_{(l,k)} \geq 12^{\frac{m+m'+n+n'}{2}}\frac{ C_{2} \alpha }{\sqrt{L}}\right]  \\
\label{eq: conver proof 5}
& \leq \sum_{l,m,m',n,n'}K \ \text{Pr}\left[\frac{1}{\mu^{m+m'+n+n'}}\bigg| \Mm_{(m,n)}^{(m',n')}\left(\rv,\rv_{j}\right) - \right.\nonumber\\
&\left.\mathbb{E}\left[\Mm_{(m,n)}^{(m',n')}\left(\rv,\rv_{j}\right) \right]\bigg|_{(l,l)} \geq 12^{\frac{m+m'+n+n'}{2}}\frac{ C_{2} \alpha }{\sqrt{L}}\right] \\
\label{eq: conver proof 6}
&  \leq 2 K^{2} \exp \left(-\frac{1}{C} \min \left( \frac{\alpha^{2}}{2 \widetilde{K}^{4}}, \frac{\alpha}{\widetilde{K}^{2}}\right)\right).
\end{align}
To show (\ref{eq: conver proof 5}), we know that since $\widetilde{K} \geq 1$ we have (\ref{eq: diagonal bound}) $\geq$ (\ref{eq: diagonal bound 2}). Hence, we can upper bound (\ref{eq: conver proof 4}) by replacing the sum over all the matrix entries by a sum over the diagonal entries only multiplied by $K$. Finally, (\ref{eq: conver proof 6}) is obtained by using Lemma~\ref{l2}.

Now, by substituting $\alpha = \frac{\epsilon_{1}\sqrt{L}}{12 C_{2} \sqrt{K}}$ in (\ref{eq: conver proof 6}) we obtain
\begin{align}
\label{eq: conver proof 7}
\text{Pr}&\left[\frac{1}{\mu^{m+m'+n+n'}}\Big|\Big| \Mm_{(m,n)}^{(m',n')}\left(\rv,\rv_{j}\right) - \right.\nonumber\\
&\left.\mathbb{E}\left[\Mm_{(m,n)}^{(m',n')}\left(\rv,\rv_{j}\right) \right]\Big|\Big|_{2} \geq \epsilon_{1}\right] \leq 2 K^{2} \times \nonumber\\
&\exp \left(-\frac{1}{C} \min \left(\frac{\epsilon_{1}^{2} L}{2 (12)^{2} K \widetilde{K}^{4}C_{2}^{2}}, \frac{\epsilon_{1} \sqrt{L}}{(12)\sqrt{K}\widetilde{K}^{2} C_{2}}\right)\right)
\end{align}
which can be easily shown to be $\leq \delta/2R^{2}$ provided that (\ref{eq: con L 1}) is satisfied with $C_{1} = C' C C_{2}$ where $C'$ is a constant.

\section{Proof of Proposition~\ref{pro: inver}}
\label{A6}
First, note that $\widebar{\Em}^{(0,0)},\widebar{\Em}^{(1,1)}, \widebar{\Em}^{(2,0)},$ and $\widebar{\Em}^{(0,2)}$ are symmetric matrices while $\widebar{\Em}^{(1,0)}$ and $\widebar{\Em}^{(0,1)}$ are antisymmetric matrices. Therefore, $\widebar{\Em}$ and $\widebar{\Em} \otimes \Id_{\text{K}}$ are symmetric matrices. 

To show that any symmetric matrix $\Sm$ with unit diagonal entries is invertible, it is enough to prove that \cite[Theorem~6.1.1]{horn2013matrix}
\begin{equation}
||\Id -\Sm||_{\infty} < 1. \nonumber
\end{equation}
Now, based on the result obtained in \cite[Proposition~5]{heckel2016super}, and given that (\ref{eq: seperation condition}) is satisfied, the matrix $\widebar{\Em}$ is invertible and satisfies 
\begin{align}
\label{eq: supper invert result}
&||\Id_{\text{3R}} - \widebar{\Em}||_{\infty} \leq 0.19808 \nonumber\\
&||\widebar{\Em}||_{2} \leq 1.19808 \nonumber\\
&|| \widebar{\Em}^{-1}||_{2} \leq 1.2470. \nonumber
\end{align}
Furthermore, for any two matrices $\Am$ and $\Bm$ and any $\ell_{p}$ norm function $||\cdot||_{p}$ we have
% \cite[Theorem~8]{lancaster1972norms} 
\begin{equation}
\label{eq: a5 proof 1}
||\Am\otimes\Bm||_{p} = ||\Am||_{p} \ ||\Bm||_{p}.
\end{equation}
Starting from (\ref{eq: a5 proof 1}) we can deduce that
\begin{align}
\label{eq: a5 proof 3}
%\label{eq: ap6 proof 2}
 \left|\left|\mathbb{E}\left[\Em\right]\right|\right|_{2}  = \left|\left|\widebar{\Em} \otimes \Id_{\text{K}} \right|\right|_{2} = ||\widebar{\Em}||_{2} \leq 1.19808. \nonumber 
\end{align}
On the other hand, we can also write
\begin{align}
&||\Id_{\text{3RK}} - \mathbb{E}\left[\Em\right]||_{\infty} = ||\Id_{\text{3RK}} - \left(\widebar{\Em} \otimes \Id_{\text{K}} \right)||_{\infty} \nonumber\\
&= ||\left(\Id_{\text{3R}} - \widebar{\Em} \right)\otimes \Id_{\text{K}} ||_{\infty} = ||\Id_{\text{3R}} - \widebar{\Em}||_{\infty} \leq 0.19808. 
\end{align}
Now since $\Id_{\text{3RK}} - \mathbb{E}\left[\Em\right]$ is a symmetric matrix with zero diagonals we have \cite[Theorem~6.1.1]{horn2013matrix}
\begin{equation}
||\Id_{\text{3RK}} - \mathbb{E}\left[\Em\right]||_{2} \leq ||\Id_{\text{3RK}} - \mathbb{E}\left[\Em\right]||_{\infty}  \nonumber
\end{equation}
which leads to (\ref{eq: matrix in cond 2}) upon using (\ref{eq: a5 proof 3}). 

Finally, to prove (\ref{eq: matrix in cond 3}) we write
\begin{align}
||\left(\mathbb{E}\left[\Em\right] \right)^{-1}||_{2} = || \widebar{\Em}^{-1} \otimes \Id_{\text{K}}||_{2} \leq 1.2470. \nonumber
\end{align}

\section{Proof of Lemma \ref{le: matrix inverse}}
\label{proof of le: matrix inverse} eq:matrix D bloack
First, note that $\Em^{(m',n')}_{(m,n)}$ is given by
\begin{align}
\label{eq:matrix D bloack}
&\Em^{(m',n')}_{(m,n)} = {\begin{bmatrix} \Mm_{(m,n)}^{(m',n')}\left(\rv_{1},\rv_{1}\right) & \hdots & \Mm_{(m,n)}^{(m',n')}\left(\rv_{1},\rv_{R}\right)   \\\vdots & \ddots & \vdots \\ \Mm_{(m,n)}^{(m',n')}\left(\rv_{R},\rv_{1}\right) & \hdots & \Mm_{(m,n)}^{(m',n')}\left(\rv_{R},\rv_{R}\right)  
  \end{bmatrix}}.
\end{align}
Starting from the definitions of $\mathcal{E}_{2}$ and $\Em$ we can write
\begin{align}
&\text{Pr}\left[||\Em - \mathbb{E}\left[{\Em}\right]||_{2} \geq \epsilon_{1}\right] \leq  \text{Pr}\left[ \sqrt{3} \max_{m,m',n,n'} \frac{1}{\mu^{m+m'+n+n'}}\times \right. \nonumber\\
%\label{eq: proof inverse 1}
& \left.\Big|\Big|\Em^{(m',n')}_{(m,n)} - \mathbb{E}\left[\Em^{(m',n')}_{(m,n)}\right]\Big|\Big|_{2} \geq \epsilon_{1}\right] \nonumber \\
\label{eq: proof inverse 2}
&  \leq  \text{Pr}\left[\sqrt{3RK} \max_{q,r,j,k,m,m',n,n'} \frac{1}{\mu^{m+m'+n+n'}} \times \right. \nonumber\\
& \left. \Big| \Mm_{(m,n)}^{(m',n')}\left(\rv_{j},\rv_{k}\right) - \mathbb{E}\left[\Mm_{(m,n)}^{(m',n')}\left(\rv_{j},\rv_{k}\right) \right] \Big|_{(q,r)} \geq \epsilon_{1} \right]
\end{align}
where (\ref{eq: proof inverse 2}) is obtained by using (\ref{eq:matrix D bloack}) with the matrix norm bound. Now, we can apply the union bound to (\ref{eq: proof inverse 2}) in order to obtain 
\begin{align}
&\text{Pr}\left[||\Em - \mathbb{E}\left[{\Em}\right]||_{2} \geq \epsilon_{1}\right] \leq \sum _{q,r,j,k,m,m',n,n'} \text{Pr}\left[ \frac{1}{\mu^{m+m'+n+n'}}\times \right.\nonumber\\
&\left.   \Big| \Mm_{(m,n)}^{(m',n')}\left(\rv_{j},\rv_{k}\right) - \mathbb{E}\left[\Mm_{(m,n)}^{(m',n')}\left(\rv_{j},\rv_{k}\right) \right]\Big|_{(q,r)} \right.    \left. \geq \frac{\epsilon_{1}}{\sqrt{3RK}} \right] \nonumber\\
&  \leq \sum _{q,j,k,m,m',n,n'} K \ \text{Pr}\left[ \frac{1}{\mu^{m+m'+n+n'}} \times \right. \nonumber\\
\label{eq: proof inverse 4}
&\left.   \Big| \Mm_{(m,n)}^{(m',n')}\left(\rv_{j},\rv_{k}\right) - \mathbb{E}\left[\Mm_{(m,n)}^{(m',n')}\left(\rv_{j},\rv_{k}\right) \right] \Big|_{(q,q)}  \geq \frac{\epsilon_{1}}{\sqrt{3RK}} \right]\\
&\leq 2 R ^2 K^{2}  \times  \nonumber\\
\label{eq: proof inverse 5}
&\exp \left(\left(-\frac{1}{C} \min \left(\frac{\epsilon_{1}^{2} L}{(12^{2}\cdot 6) RK \widetilde{K}^{4}C_{2}^{2}}, \frac{\epsilon_{1} \sqrt{L}}{12\sqrt{3RK}\widetilde{K}^{2}C_{2}}\right)\right)\right)
\end{align}
where (\ref{eq: proof inverse 4}) is obtained by using the same justification led to (\ref{eq: conver proof 5}) while (\ref{eq: proof inverse 5}) is based on using (\ref{eq: h w ineq}) followed by the same steps that led to (\ref{eq: conver proof 7}). Based on (\ref{eq: proof inverse 5}), it is easy to show that when (\ref{eq: con L 1}) is satisfied, $\text{Pr}\left[\mathcal{E}_{2}\right] \geq 1-\delta/2$.

\section{Proofs of Lemmas~\ref{le: inver cond l} and \ref{th: pro of T}}
\label{proof of le: inver cond l}

\subsection{Proof of Lemma~\ref{le: inver cond l}}
By using triangular inequality we can write
\begin{align}
||\Id_{\text{3RK}} -\Em||_{2} &= ||\left(\Id_{\text{3RK}} -\mathbb{E}\left[\Em\right] \right)+ \left(\mathbb{E}\left[\Em\right]  -\Em \right)||_{2} \nonumber \\
%\label{eq: proof inver ineq 1}
& \leq ||\Id_{\text{3RK}} -\mathbb{E}\left[\Em\right]||_{2}+ ||\mathbb{E}\left[\Em\right]  -\Em||_{2} \nonumber \\
\label{eq: proof inver ineq 2}
& \leq 0.19808 + \epsilon_{1} \leq  0.5981,
\end{align}
where (\ref{eq: proof inver ineq 2}) is obtained by using (\ref{eq: matrix in cond 2}), Lemma~\ref{le: matrix inverse}, and the fact that $\epsilon_{1} \in (0,\frac{2}{5}]$. Moreover, we show in Appendix~\ref{A8} that
\begin{equation}
\label{eq: tem proof gor}
||\Em^{-1}||_{2} \leq 2||\left(\mathbb{E}\left[\Em\right]\right)^{-1}||_{2}
\end{equation}
which leads to (\ref{eq: conditon inverstion ma 2}) based on (\ref{eq: matrix in cond 3}). 

\subsection{Proof of Lemma~\ref{th: pro of T}}
Starting from the definition of $\mathcal{E}_{4}$ we can write
\begin{align}
&\text{Pr} \left[ \max _{\rv \in \Omega_{\text{S}}} \Big|\Big|\left(\Delta \Tm^{(m',n')}\left(\rv\right)\right)^{H} \Lm\Big|\Big|_{2} \geq  2.5 \epsilon_{2} \right] \leq \nonumber \\
&\text{Pr} \left[ \left\lbrace \max _{\rv \in \Omega_{\text{S}}} \Big|\Big|\hspace{-3pt}\left(\Delta \Tm^{(m',n')}\left(\rv\right)\right)^{H}\hspace{-3pt} \Lm\Big|\Big|_{2} \hspace{-3pt} \geq ||\Lm||_{2} \epsilon_{2} \right\rbrace  \hspace{-2pt} \cup \left\lbrace ||\Lm||_{2} \hspace{-2.5pt} \geq 2.5\right\rbrace  \right]  \nonumber \\
\label{eq: pr of T prof 2}
& \leq \sum_{\rv \in \Omega_{\text{S}}} \text{Pr} \left[ \Big|\Big|\left(\Delta \Tm^{(m',n')}\left(\rv\right)\right)^{H} \Lm\Big|\Big|_{2}\geq   ||\Lm||_{2} \epsilon_{2} \right] \nonumber\\
&+ \text{Pr} \left[||\Lm||_{2} \geq 2.5 \right] \leq \delta/2 +  \text{Pr}\left[\mathcal{E}_{2}^{c}\right],
\end{align}
where the first and the second inequalities are based on the union bound while the last inequality follows from the triangular inequality, Lemma~\ref{lemma: T matrix}, and the fact that $\left\lbrace ||\Lm||_{2} \geq 2.5\right\rbrace \subseteq \mathcal{E}_{2}$ when $\epsilon_{1} \in (0,\frac{2}{5}]$ as in (\ref{eq: bound L}).

\section{Proof of Lemmas~\ref{le: v bounds } and \ref{lemma: v2 bounds }}
\label{proof of le: v bounds }
The proofs of Lemmas~\ref{le: v bounds } and \ref{lemma: v2 bounds } are based on Matrix Bernstein inequality which is given by the following lemma:
\begin{lemma}
\label{le: matrix berns}
\normalfont(Matrix Bernstein inequality)\cite[Theorem 1.6.2]{tropp2015introduction}
Let $\Sm_{1},\dots,\Sm_{n}$ be $N_{1} \times N_{2}$ independent, centred random matrices that are uniformly bounded, i.e.,
\begin{equation}
%\label{eq: con matrix ber}
\mathbb{E}\left[\Sm_{k}\right]=\bm{0}, \ \ \ \ ||\Sm_{k}||_{2} \leq q, \ \ \ \ k= 1, \dots, n.  \nonumber
\end{equation}
Moreover, define the sum
\begin{equation}
\Zm = \sum_{k=1}^{n} \Sm_{k} \nonumber \nonumber
\end{equation}
and let $\nu(\Zm)$ to denote the matrix variance statistic of the sum, i.e.,
\begin{equation}
\nu(\Zm) := \max \left\lbrace\big|\big|\mathbb{E}\left[\Zm^{H}\Zm\right]\big|\big|_{2}, \big|\big|\mathbb{E}\left[\Zm\Zm^{H}\right]\big|\big|_{2} \right\rbrace. \nonumber
\end{equation}
Then, for every $t \geq 0$ we have
\begin{align}
\text{Pr}\left[||\Zm||_{2} \geq t\right] \leq \left(N_{1} +N_{2}\right) \exp \left( \frac{-t^{2}/2}{\nu(\Zm)+ q t/3}\right). \nonumber
\end{align}
\end{lemma}
Now, we are ready to prove Lemma~\ref{le: v bounds } as follows

\subsection{Proof of Lemma~\ref{le: v bounds }}
First, let us consider the following matrix definition
\begin{align}
\label{eq: new matrix W}
\Wm^{(m',n')}\left(\rv\right) &:=\left(\Delta \Tm^{(m',n')}\left(\rv\right)\right)^{H} \Lm  \nonumber\\
&=\begin{bmatrix} \Wm_{1}^{(m',n')}\left(\rv\right), \dots, \Wm_{R}^{(m',n')}\left(\rv\right)\end{bmatrix}  \in \mathbb{C}^{K \times RK},
\end{align}
where $\Wm_{j}\left(\rv\right)^{(m',n')} \in \mathbb{C}^{K \times K}$. Upon using the definition of $\hv$ in (\ref{eq: matrix poly}) and based on (\ref{eq: new matrix W}) we can rewrite $\vv_{1}^{(m',n')}\left(\rv\right)$ as
\begin{equation}
\label{eq: redefine v vector}
\vv_{1}^{(m',n')}\left(\rv\right) =  \sum_{j=1}^{R} \Wm_{j}^{(m',n')}\left(\rv\right) \ \text{sign}\left(c_{j}\right) \hv_{j} =: \sum_{j=1}^{R} \wv_{j}^{(m',n')}.
\end{equation}
From (\ref{eq: redefine v vector}), it is easy to show that $\vv_{1}^{(m',n')}\left(\rv\right)$ is a sum of independent zero-mean vectors based on Assumptions~\ref{as 1} and \ref{as 3}. Therefore, we can apply the Matrix Bernstein inequality in Lemma~\ref{le: matrix berns} to obtain a probability measure on the bound of $\big|\big|\vv_{1}^{(m',n')}\left(\rv\right)\big|\big|_{2}$. However, we first need to calculate the values of $q$ and $\nu\left(\vv^{(m',n')}_{1}\right)$ as in Lemma~\ref{le: matrix berns}. 

Starting with $q$ we can write conditioned on $\mathcal{E}_{4}$
\begin{align}
&\big|\big|\wv_{j}^{(m',n')}\big|\big|_{2}  = \big|\big|\Wm_{j}^{(m',n')}\left(\rv\right) \text{sign}\left(c_{j}\right) \hv_{j}\big|\big|_{2}  \nonumber\\
\label{eq: s vector bound 2}
& \leq \big|\big|\Wm_{j}^{(m',n')}\left(\rv\right)\big|\big|_{2}  \leq \big|\big|\Wm^{(m',n')}\left(\rv\right)\big|\big|_{2}  \leq  2.5 \epsilon_{2}=: q,
\end{align}
where the first inequality follows from triangular inequality and Assumption~\ref{as 3} while the second inequality is based on the fact that $\Wm_{j}^{(m',n')}\left(\rv\right)$ is a sub-matrix of $\Wm^{(m',n')}\left(\rv\right)$. Finally, the last inequality follows from Lemma~\ref{th: pro of T}.

On the other hand, we prove in Appendix~\ref{A7} that, conditioned on $\mathcal{E}_{4}$, we have
\begin{equation}
\label{eq: varince v vector}
\nu\left(\vv_{1}^{(m',n')}\left(\rv\right)\right) = 6.25 \epsilon_{2}^{2}.
\end{equation}
%Starting from the left-hand side of (\ref{eq: bound v the}) we can write
Now we can write
\begin{align}
&\text{Pr}\left[ \max_{\rv \in \Omega_{\text{S}}} \big|\big|\vv_{1}^{(m',n')}\left(\rv\right)\big|\big|_{2} \geq \epsilon_{3}\right] \nonumber\\
\label{eq: bound v proof 1_1}
& \leq \text{Pr}\left[ \max_{\rv \in \Omega_{\text{S}}} \big|\big|\vv_{1}^{(m',n')}\left(\rv\right)\big|\big|_{2} \geq \epsilon_{3} \Big| \mathcal{E}_{4} \right] + \text{Pr}\left[\mathcal{E}_{4}^{c}\right]  \\
\label{eq: bound v proof 1}
&\leq \left(K+1\right)|\Omega_{\text{S}}| \exp \left(\frac{-3\epsilon_{3}^{2}}{37.5 \epsilon_{2}^{2}+ 5 \epsilon_{2} \epsilon_{3}}\right)+ \text{Pr}\left[\mathcal{E}_{4}^{c}\right] \\
\label{eq: bound v proof 2}
&  \leq  \left\{
                \begin{array}{ll}
                  \left(K+1\right)|\Omega_{\text{S}}| \exp \left(\frac{-0.04\epsilon_{3}^{2}}{\epsilon_{2}^{2}}\right)+ \text{Pr}\left[\mathcal{E}_{4}^{c}\right]  &\text{if} \     \epsilon_{3} \leq  7.5 \epsilon_{2}\\
                  \left(K+1\right)|\Omega_{\text{S}}| \exp \left(\frac{-0.3\epsilon_{3}}{\epsilon_{2} }\right)+ \text{Pr}\left[\mathcal{E}_{4}^{c} \right] & \text{if} \  \epsilon_{3} \geq 7.5 \epsilon_{2} \
                \end{array}
              \right. \\
              \label{eq: bound v proof 3}
              & \leq 1.5 \delta,
\end{align}
where (\ref{eq: bound v proof 1_1}) is based on the fact that for any two events $A_{1}$ and $A_{2}$, $\text{Pr}\left[A_{1}\right] \leq \text{Pr}\left[A_{1}|A_{2}\right] +\text{Pr}\left[{A_{2}}^{c}\right]$ while (\ref{eq: bound v proof 1}) is obtained by using the union bound and Lemma~\ref{le: matrix berns} with (\ref{eq: s vector bound 2}) and (\ref{eq: varince v vector}). 

To show (\ref{eq: bound v proof 3}), first note that based on Lemma~\ref{th: pro of T}, $\text{Pr}\left[\mathcal{E}_{4}^{c} \right] \leq \delta/2 +\text{Pr}\left[\mathcal{E}_{2}^{c} \right]$ provided that (\ref{eq: cond L for T}) is satisfied whereas $\text{Pr}\left[\mathcal{E}_{2}^{c}\right] \leq \delta/2$ when (\ref{eq: con L 1}) is satisfied as in Lemma~\ref{le: matrix inverse}. Therefore, $\text{Pr}\left[\mathcal{E}_{4}^{c} \right] \leq \delta$ given that $\max \left\lbrace(\ref{eq: con L 1}),(\ref{eq: cond L for T})\right\rbrace$ is satisfied. 

On the other hand, in order for the first terms in (\ref{eq: bound v proof 2}) to be less than or equal $\delta/2$ we should have
\begin{align}
\label{eq: multi epsi}
 \epsilon_{2} = \left\{
                \begin{array}{ll}
                   \frac{0.2 \epsilon_{3}}{\sqrt{\log \left(\frac{2\left(K+1\right)|\Omega_{\text{S}}|}{\delta}\right)}} &\text{if} \  \epsilon_{3} \leq  7.5\epsilon_{2}  \\
                    \frac{0.3 \epsilon_{3}}{\log \left(\frac{2\left(K+1\right)|\Omega_{\text{S}}|}{\delta}\right)} & \text{if} \  \epsilon_{3} \geq   7.5\epsilon_{2}\
                \end{array}
              \right.
\end{align}
Upon substituting (\ref{eq: multi epsi}) in (\ref{eq: cond L for T}) and manipulating, we obtain the following bound for $\epsilon_{3} \leq   7.5\epsilon_{2}$
\begin{equation}
\label{eq: first L boud y}
 L \geq  25 C_{3}^{2}\frac{ RK \widetilde{K}^{4}}{\epsilon_{3}^{2}} \log^2 \left(\frac{4RK^{2} |\Omega_{\text{S}}|}{\delta}\right)\log \left(\frac{2\left(K+1\right)|\Omega_{\text{S}}|}{\delta}\right)
\end{equation}
whereas for $\epsilon_{3} \geq  7.5\epsilon_{2}$ we obtain
\begin{equation}
\label{eq: first L boud y 2}
 L \geq  \hspace{-3pt} \frac{100}{9}C_{3}^{2} \frac{ RK \widetilde{K}^{4}}{\epsilon_{3}^{2}}\hspace{-1.5pt} \log^2 \left(\frac{4RK^{2} |\Omega_{\text{S}}|}{\delta}\right)\hspace{-2pt}\log^{2} \left(\frac{2\left(K+1\right)|\Omega_{\text{S}}|}{\delta}\right)
\end{equation}
Now, based on (\ref{eq: con L 1}), (\ref{eq: first L boud y}), (\ref{eq: first L boud y 2}), and by setting $\epsilon_{1} = \frac{2}{5}$ in (\ref{eq: con L 1}), we can easily show that (\ref{eq: bound v proof 3}) is satisfied under the hypotheses of Lemma~\ref{le: v bounds }.

\subsection{Proof of Lemma~\ref{lemma: v2 bounds }}
To prove Lemma~\ref{lemma: v2 bounds } we need to obtain some results first. To begin note that
\begin{align}
\label{eq: T aver bound 1}
 \big|\big|\widebar{\Tm}^{(m',n')}\left(\rv\right)\big|\big|_{F}^{2} &= \big|\big|\bar{\tv}^{(m',n')}\left(\rv\right)\big|\big|_{F}^{2} ||\Id_{\text{K}}||_{F}^{2}  \\
 \label{eq: T aver bound 2}
& = K \big|\big|\bar{\tv}^{(m',n')}\left(\rv\right)\big|\big|_{F}^{2} \leq K\tilde{C}_{1},
\end{align}
where (\ref{eq: T aver bound 1}) is based on (\ref{eq: sub T }) and the fact that $||\Am\otimes\Bm||_{F} = ||\Am||_{F} \ ||\Bm||_{F}$ while the inequality in (\ref{eq: T aver bound 2}) follows from the fact that $ \frac{1}{\mu^{m'+n'}}\widebar{M}^{(m',n')}\left(\rv\right)$ is a bounded function where $\tilde{C}_{1}$ is a constant (see \cite[Appendix~H]{tang2013compressed}). On the other hand, we can write conditioned on $\mathcal{E}_{2}$ with $\epsilon_{1} \in (0,\frac{2}{5}]$
\begin{align}
\label{eq: bound condt v2 }
&\big|\big|\left(\widebar{\Tm}^{(m',n')}\left(\rv\right) \right)^{H} \left(\Lm  - \bar{\Lm}\otimes \Id_{\text{K}} \right)\big|\big|_{F}^{2}\leq \nonumber\\
&\big|\big|\widebar{\Tm}^{(m',n')}\left(\rv\right)\big|\big|_{F}^{2} \big|\big|\Lm  - \bar{\Lm}\otimes \Id_{\text{K}} \big|\big|_{2}^{2}  \leq \left(3.11\right)^{2}K\tilde{C}_{1} \epsilon_{1}^{2},
\end{align}
where the first inequality is based on the fact that for any two matrices $\Am$ and $\Bm$, $||\Am \Bm||_{F} \leq ||\Am||_{2} ||\Bm||_{F}$ whereas the second inequality is based on (\ref{eq: T aver bound 2}) and the fact that
\begin{equation}
\label{eq: boud of L-L bar}
\big|\big|\Lm- \bar{\Lm} \otimes \Id_{\text{K}} \big|\big|_{2}  \leq  3.11 \epsilon_{1},
\end{equation}
with its proof being provided in Appendix~\ref{A8}. 

Now, if we define  
\begin{align}
\label{eq: w bar definition}
&\widetilde{\Wm}^{(m',n')}\left(\rv\right) := \left(\widebar{\Tm}^{(m',n')}\left(\rv\right) \right)^{H} \left(\Lm  - \bar{\Lm}\otimes \Id_{\text{K}} \right) \nonumber\\
&=\begin{bmatrix} \widetilde{\Wm}_{1}^{(m',n')}\left(\rv\right), \dots, \widetilde{\Wm}_{R}^{(m',n')}\left(\rv\right)\end{bmatrix} \in \mathbb{C}^{K \times RK}
\end{align}
with $\widetilde{\Wm}_{j}\left(\rv\right)^{(m',n')} \in \mathbb{C}^{K \times K}$, we can rewrite $\vv_{2}^{(m',n')}\left(\rv\right)$ in (\ref{eq: v matrix 2}) as
\begin{equation}
\label{eq: redefine v2 vector}
\vv_{2}^{(m',n')}\left(\rv\right) =  \sum_{j=1}^{R} \widetilde{\Wm}_{j}^{(m',n')}\left(\rv\right) \ \text{sign}\left(c_{j}\right) \hv_{j} =: \sum_{j=1}^{R} \tilde{\wv}_{j}^{(m',n')}.
\end{equation}
Based on Assumption~\ref{as 3}, we can easily show that $\vv_{2}^{(m',n')}\left(\rv\right)$ is a sum of independent, centred random vectors of dimension $K\times 1$. Therefore, we can apply Lemma~\ref{le: matrix berns} to prove Lemma~\ref{lemma: v2 bounds } as follows:

Conditioned on $\mathcal{E}_{2}$ for all $\epsilon_{1} \in (0,\frac{2}{5}]$ we can write 
\begin{align}
\label{eq: mean and var of s hat}
&\big|\big|\tilde{\wv}_{j}^{(m',n')}\big|\big|_{2} = \big|\big|\widetilde{\Wm}_{j}^{(m',n')}\left(\rv\right) \text{sign}\left(c_{j}\right) \hv_{j}\big|\big|_{2} \nonumber\\
&\leq \big|\big|\widetilde{\Wm}_{j}^{(m',n')}\left(\rv\right)\big|\big|_{2} \leq \big|\big|\widebar{\Tm}^{(m',n')}\left(\rv\right) \big|\big|_{2} \ \big|\big|\Lm  - \bar{\Lm}\otimes \Id_{\text{K}}\big|\big|_{2} \nonumber\\
&  \leq  3.11 \tilde{C}_{1} \epsilon_{1}=: q.
\end{align}
On the other hand, and upon following the same steps that led to (\ref{eq: varince v vector}), we can show that
\begin{equation}
\label{eq: s hat var opt}
\nu\left(\vv_{2}^{(m',n')}\left(\rv\right)\right) = 9.672 \ \tilde{C}_{1} \epsilon_{1}^{2}.
\end{equation}
%Now, starting from the left-hand side of (\ref{eq: bound v2 the}), and upon applying the Matrix Bernstein inequality with (\ref{eq: mean and var of s hat}) and (\ref{eq: s hat var opt}), we can show that
Now by applying the Matrix Bernstein inequality with (\ref{eq: mean and var of s hat}) and (\ref{eq: s hat var opt}), we can show that
\begin{align}
&\text{Pr}\left[ \max_{\rv \in \Omega_{\text{S}}} \big|\big|\vv_{2}^{(m',n')}\left(\rv\right)\big|\big|_{2} \geq \epsilon_{3} \Big| \mathcal{E}_{2} \right] \nonumber\\
& \leq \left(K+1\right)|\Omega_{\text{S}}| \exp \left(\frac{-3\epsilon_{3}^{2}}{6 (3.11)^{2} \tilde{C}_{1} \epsilon_{1}^{2}+ 6.22 \tilde{C}_{1}\epsilon_{1}\epsilon_{3}}\right) \nonumber\\
& \leq  \left\{
                \begin{array}{ll}
                  \left(K+1\right)|\Omega_{\text{S}}| \exp \left(\frac{-\epsilon_{3}^{2}}{(6.22)^{2} \tilde{C}_{1} \epsilon_{1}^{2}}\right)  &\text{if} \   \epsilon_{1} \geq \frac{1}{9.33}\epsilon_{3}  \nonumber\\
                  \left(K+1\right)|\Omega_{\text{S}}| \exp \left(\frac{-3\epsilon_{3}}{12.44 \tilde{C}_{1}\epsilon_{1} }\right)  & \text{if} \   \epsilon_{1} \leq \frac{1}{9.33}\epsilon_{3}\
                \end{array}
              \right. \\
              \label{eq: bound v2 proof 3}
              & \leq \delta/2,
\end{align}
where the last inequality can be shown to hold true provided that (\ref{eq: expre for eps1}) is satisfied. Note that since $\epsilon_{3} \leq 1$ and $ C_{4} \leq 0.55$ we have $\epsilon_{1} \leq 2/5$ based on (\ref{eq: expre for eps1}).

\section{Proof of Lemma~\ref{th: off grid theorem}}
\label{proof of th: off grid theorem}
Starting from the definition of $\mathcal{E}_{5}$ we can write
\begin{align}
&\text{Pr}\left[ \max_{\rv \in \Omega_{\text{S}}} \frac{1}{\mu^{m'+n'}} \Big|\Big|\fv^{(m',n')}\left(\rv\right) - \bar{\fv}^{(m',n')}\left(\rv\right) \Big|\Big|_{2} \geq 2 \epsilon_{3} \right]  \leq \nonumber\\
&\text{Pr}\left[ \max_{\rv \in \Omega_{\text{S}}} \big|\big|\vv_{1}^{(m',n')}\left(\rv\right)\big|\big|_{2} \geq \epsilon_{3}\right] + \text{Pr}\left[ \mathcal{E}_{2}^{c}\right]\nonumber\\
\label{eq: final bound delta5}
&+\text{Pr}\left[ \max_{\rv \in \Omega_{\text{S}}} \big|\big|\vv_{2}^{(m',n')}\left(\rv\right)\big|\big|_{2} \geq \epsilon_{3}\Big| \mathcal{E}_{2}\right] \\ 
\label{eq: final bound delta52}
&\leq 2.5 \delta.
\end{align}
To show (\ref{eq: final bound delta52}), we choose $\epsilon_{1}$ as in (\ref{eq: expre for eps1}) and thus, the last term in (\ref{eq: final bound delta5}) is less than or equal to $\delta/2$ based on Lemma~\ref{lemma: v2 bounds }. Next, note that when (\ref{eq: con L 1}) is satisfied, $\text{Pr}\left[ \mathcal{E}_{2}^{c}\right]\leq \delta/2$ based on Lemma~\ref{le: matrix inverse}. By substituting (\ref{eq: expre for eps1}) in (\ref{eq: con L 1}) we obtain $L \geq  \frac{ C_{1}^{2}}{C^{2}_{4}\epsilon_{3}^{2}} R K \widetilde{K}^{4} \log^{2}\left(\frac{4 R^{2} K^{2}}{\delta}\right) \log^{2}\left(\frac{2\left(K+1\right) |\Omega_{\text{S}}|}{\delta}\right)$ which is strictly less than (\ref{eq: cond l for v}) upon defining $\bar{C}=\frac{C_{1}^{2}}{C^{2}_{4}}$ and given that $|\Omega_{\text{S}}| > 1$. Finally, note that the first term in (\ref{eq: final bound delta5}) is less than or equal $3\delta/2$ when (\ref{eq: cond l for v}) is satisfied.

\section{Proofs of (\ref{eq: tem proof gor}) and (\ref{eq: boud of L-L bar})}
\label{A8}
\subsubsection{Proof of (\ref{eq: tem proof gor})}
We start by stating that for any two invertible matrices $\Am$ and $\Bm$ that satisfy $||\Am-\Bm||_{2}\ ||\Bm^{-1}||_{2} \leq 0.5$, the following inequalities hold true \cite[Appendix~E]{tang2013compressed}
\begin{align}
\label{eq: app8 2_n}
&||\Am^{-1}||_{2} \leq 2||\Bm^{-1}||_{2} \\
\label{eq: app8 2}
&||\Am^{-1} - \Bm^{-1}||_{2}  \leq 2 \ ||\Bm^{-1}||_{2} ^{2} \ ||\Am-\Bm||_{2}.
\end{align}
Now, to prove (\ref{eq: tem proof gor}), we know that conditioned on $\mathcal{E}_{2}$ with $\epsilon_{1} \in (0,\frac{2}{5}]$ we can write based on (\ref{eq: matrix in cond 3}) and Lemma~\ref{le: matrix inverse}
\begin{equation}
\label{eq: app8 3}
||\Em- \mathbb{E}\left[{\Em}\right]||_{2}\ ||\left(\mathbb{E}\left[{\Em}\right] \right)^{-1}||_{2} \leq 0.4988 < 0.5
\end{equation}
which leads to (\ref{eq: tem proof gor}) based on (\ref{eq: app8 2_n}).
\subsubsection{Proof of  (\ref{eq: boud of L-L bar})}
To show (\ref{eq: boud of L-L bar}), recall first the definitions of $\Lm$ and $\bar{\Lm}$ as applied in (\ref{eq: bound L}) and (\ref{eq: T trick}), respectively. Then, conditioned on $\mathcal{E}_{2}$ with $\epsilon_{1} \in (0,\frac{2}{5}]$ we can write
\begin{align}
\label{eq: app8 1}
\big|\big|\Lm-\bar{\Lm} \otimes \Id_{\text{K}} \big|\big|_{2}   \leq \big|\big|\Em^{-1}- \left(\mathbb{E}\left[\Em\right]\right)^{-1} \big|\big|_{2} . 
\end{align}
Now, based on (\ref{eq: app8 2}), (\ref{eq: app8 3}), and (\ref{eq: app8 1}) we can conclude
\begin{equation}
||\Lm-\bar{\Lm} \otimes \Id_{\text{K}} ||_{2}   \leq   2 (1.247)^{2} \epsilon_{1} \leq  3.11 \epsilon_{1}.
\end{equation}

\section{Proof of (\ref{eq: varince v vector})}
\label{A7}
Conditioned on the event $\mathcal{E}_{4}$ we can write
\begin{align}
&\Bigg|\Bigg|\sum_{j=1}^{R}\mathbb{E}\left[\left(\wv_{j}^{(m',n')}\right)^{H}\wv_{j}^{(m',n')}\right]\Bigg|\Bigg|_{2}\hspace{-7pt} = \Bigg|\Bigg|\sum_{j=1}^{R} \mathbb{E}\left[ \left(\text{sign}\left(c_{j}\right) \hv_{j}\right)^{H} \right.\nonumber\\
&\left.\left(\Wm_{j}^{(m',n')}\left(\rv\right)\right)^{H}\Wm_{j}^{(m',n')}\left(\rv\right) \text{sign}\left(c_{j}\right) \hv_{j}\right]\Bigg|\Bigg|_{2}  \nonumber\\
& = \Bigg|\Bigg|\sum_{j=1}^{R} \mathbb{E}\left[\hv_{j}^{H}\left(\Wm_{j}^{(m',n')}\left(\rv\right)\right)^{H}\Wm_{j}^{(m',n')}\left(\rv\right)  \hv_{j}\right]\Bigg|\Bigg|_{2}  \nonumber\\
&= \sum_{j=1}^{R} \mathbb{E}\left[\text{Tr}\left(\left(\Wm_{j}^{(m',n')}\left(\rv\right)\right)^{H}\Wm_{j}^{(m',n')}\left(\rv\right)  \hv_{j}\hv_{j}^{H}\right)\right] \nonumber\\
&= \sum_{j=1}^{R} \text{Tr}\left(\left(\Wm_{j}^{(m',n')}\left(\rv\right)\right)^{H}\Wm_{j}^{(m',n')}\left(\rv\right) \mathbb{E}\left[\hv_{j}\hv_{j}^{H}\right]\right) \nonumber\\
\label{eq: app7 1}
&= \frac{1}{K}\sum_{j=1}^{R} \text{Tr}\left(\left(\Wm_{j}^{(m',n')}\left(\rv\right)\right)^{H}\Wm_{j}^{(m',n')}\left(\rv\right)\right)\\
%\label{eq: app7 2}
&= \frac{1}{K} \text{Tr}\left(\left(\Wm^{(m',n')}\left(\rv\right)\right)^{H}\Wm^{(m',n')}\left(\rv\right)\right) \nonumber\\
\label{eq: app7 3}
&= \frac{1}{K} \big|\big|\Wm^{(m',n')}\left(\rv\right)\big|\big|^{2}_{F},
\end{align}
where (\ref{eq: app7 1}) is based on Lemma~\ref{lemma: covarince of h} given below. Next, we can write conditioned on $\mathcal{E}_{4}$
\begin{align}
\label{eq: app7 4}
&\big|\big|\Wm^{(m',n')}\left(\rv\right)\big|\big|^{2}_{F} \leq ||\Lm||^{2}_{2} \big|\big|\Delta \Tm^{(m',n')}\left(\rv\right)\big|\big|_{F}^{2}  \\
\label{eq: app7 5}
& \leq (2.5)^{2}  K \big|\big|\Delta \Tm^{(m',n')}\left(\rv\right)\big|\big|_{2}^{2}
 \leq 6.25 K \epsilon_{2}^{2},
\end{align}
where (\ref{eq: app7 4}) is based on the fact that for any two matrices $\Am$ and $\Bm$, $||\Am \Bm||^{2}_{F} \leq ||\Am||^{2}_{2} ||\Bm||^{2}_{F}$  while (\ref{eq: app7 5}) follows from the fact that $||\Am||_{F} \leq \sqrt{r_{\Am}} ||\Am||_{2}$ ($r_{\Am}$ is the rank of $\Am$), (\ref{eq: bound L}), and Lemma~\ref{lemma: T matrix}. Note that the event $\mathcal{E}_{4}$ includes $\mathcal{E}_{3}$ and $\mathcal{E}_{2}$ with $\epsilon_{1} \in (0, \frac{2}{5}]$.

Finally, by substituting (\ref{eq: app7 5}) in (\ref{eq: app7 3}) we obtain (\ref{eq: varince v vector}).
\begin{lemma}\normalfont\cite[Lemma~21]{yang2016super}
\label{lemma: covarince of h}
Let $\hv_{j} \in \mathbb{C}^{K \times 1}$ have i.i.d. entries on the complex unit sphere. Then, $\mathbb{E}\left[\hv_{j}\hv_{j}^{H}\right] = \frac{1}{K} \Id_{\text{K}}.$  
\end{lemma}

\section{Proof of Lemma~\ref{th: f is less 1}}
\label{proof of th: f is less 1}

To start with, we consider a dense set of point vectors $\rv_{p}$ on $\Omega_{\text{S}}$ to be on the rectangular grid closet to $\rv$ that is defined by
\begin{equation}
\label{eq: point grid definition}
\max_{\rv \in [0,1]^{2}} \min _{\rv_{p} \in \Omega_{\text{S}}} |\rv-\rv_{p}|\leq \frac{4\epsilon_{3}}{3 \pi \tilde{C}_{2}\sqrt{K}L^{3/2}},
\end{equation}
where the cardinality of $\Omega_{\text{S}}$ is set to be
\begin{equation}
\label{eq: card omega}
|\Omega_{\text{S}}| = \left(\frac{3 \pi \tilde{C}_{2}\sqrt{K} L^{3/2}}{4\epsilon_{3}} \right)^{2}=: \frac{\tilde{C}_{3}^{2} L^{3}}{\epsilon_{3}^{2}}.
\end{equation}
Starting from the norm function in (\ref{eq: f to f bar al}), and upon letting $\rv \in [0,1]^{2}$ and considering $\rv_{p}$ to be a vector in $\Omega_{\text{S}}$ that is closest to $\rv$ as in (\ref{eq: point grid definition}), we can write
\begin{align}
\label{eq: f to f bar al 1}
&\frac{1}{\mu^{m'+n'}}  \big|\big|\fv^{(m',n')}\left(\rv\right)- \bar{\fv}^{(m',n')}\left(\rv\right)    \big|\big|_{2}  \leq  \nonumber \\
&  \frac{1}{\mu^{m'+n'}} \left[ \big|\big|\fv^{(m',n')}\left(\rv\right)- \fv^{(m',n')}\left(\rv_{p}\right)  \big|\big|_{2} \right. \nonumber \\
& + \left.\big|\big|\fv^{(m',n')}\left(\rv_{p}\right)- \bar{\fv}^{(m',n')}\left(\rv_{p}\right)    \big|\big|_{2}\right. \nonumber\\
& \left.+ \big|\big|\bar{\fv}^{(m',n')}\left(\rv_{p}\right)- \bar{\fv}^{(m',n')}\left(\rv\right)    \big|\big|_{2}\right].
\end{align}
Now, we consider each term in the left-hand side of (\ref{eq: f to f bar al 1}) separately. Starting with the first term, we can write
\begin{align}
\label{eq: f to f bar al term 1}
&\frac{1}{\mu^{m'+n'}}  \big|\big|\fv^{(m',n')}\left(\rv\right)- \fv^{(m',n')}\left(\rv_{p}\right)  \big|\big|_{2}  \leq \nonumber\\
&\frac{\sqrt{K}}{\mu^{m'+n'}}\max_{i} \big|\fv^{(m',n')}\left(\rv\right)- \fv^{(m',n')}\left(\rv_{p}\right)  \big|_{i}, 
\end{align}
where $|\cdot|_{i}$ refers to the absolute value of the $i$-th entry of the vector. The absolute value function in (\ref{eq: f to f bar al term 1}) can be upper bounded by
\begin{align}
&\big|\fv^{(m',n')}\left(\rv\right)- \fv^{(m',n')}\left(\rv_{p}\right)  \big|_{i}  \leq \nonumber\\
&\big|\fv^{(m',n')}\left(\tau, f\right)- \fv^{(m',n')}\left(\tau, f_{p}\right)  \big|_{i} \nonumber\\ 
&+\big|\fv^{(m',n')}\left(\tau, f_{p}\right) - \fv^{(m',n')}\left(\tau_{p}, f_{p}\right)   \big|_{i}  \nonumber\\
& \leq |f-f_{p}|\sup_{x} \big|\fv^{(m',n'+1)}\left(\tau, x\right) \big|_{i} \nonumber \\
\label{eq: f to f bar al term 1 1}
&+  |\tau-\tau_{p}|\sup_{x} \big|\fv^{(m'+1,n')}\left(x, f_{p}\right)\big|_{i} \\
& \leq |f-f_{p}| \left(\pi L\right) \sup_{x} \big|\fv^{(m',n')}\left(\tau, x\right) \big|_{i} \nonumber\\
\label{eq: f to f bar al term 1 2}
&+  |\tau-\tau_{p}| \left(\pi L\right) \sup_{x} \big|\fv^{(m',n')}\left(x, f_{p}\right)\big|_{i} \\
\label{eq: f to f bar al term 1 3}
& \leq |f-f_{p}| \left(\pi L\right) \sup_{x} \big|\big|\fv^{(m',n')}\left(\tau, x\right) \big|\big|_{2} \nonumber\\
&+  |\tau-\tau_{p}| \left(\pi L\right) \sup_{x} \big|\big|\fv^{(m',n')}\left(x, f_{p}\right)\big|\big|_{2},
\end{align}
where (\ref{eq: f to f bar al term 1 1}) follows from the definition of the derivative of the function while (\ref{eq: f to f bar al term 1 2}) is obtained by applying Bernstein's inequality \cite{rahman2002analytic}. Upon substituting (\ref{eq: f to f bar al term 1 3}) into (\ref{eq: f to f bar al term 1}) and then using the result in Lemma~\ref{lemma: close in all r} we can obtain 
\begin{align}
\label{eq: f to f bar al term 1 4}
&\frac{1}{\mu^{m'+n'}}  \big|\big|\fv^{(m',n')}\left(\rv\right)- \fv^{(m',n')}\left(\rv_{p}\right)  \big|\big|_{2}   \nonumber\\
&  \leq  \left(\pi L\right) \frac{\tilde{C}_{2}}{4} \sqrt{KL} |\rv-\rv_{p}| \leq \frac{\epsilon_{3}}{3},
\end{align}
where the last inequality is based on (\ref{eq: point grid definition}). Now, by following the same steps we can show that
\begin{align}
\label{eq: f to f bar al term 1 5}
&\frac{1}{\mu^{m'+n'}}  \big|\big|\bar{\fv}^{(m',n')}\left(\rv_{p}\right)- \bar{\fv}^{(m',n')}\left(\rv\right)    \big|\big|_{2} \leq \frac{\epsilon_{3}}{3}.
\end{align}
On the other hand, we can deduce based on Lemma~\ref{th: off grid theorem} that
\begin{equation}
\label{eq: f to f bar al term 1 6}
\max_{\rv_{p} \in \Omega_{\text{S}}, m'+n'\leq 2} \frac{1}{\mu^{m'+n'}} \big|\big|\fv^{(m',n')}\left(\rv_{p}\right)- \bar{\fv}^{(m',n')}\left(\rv_{p}\right)  \big|\big|_{2}  \leq \frac{\epsilon_{3}}{3}
\end{equation}
holds with probability at least $1- 2.5 \delta $ for all pairs $(m',n')$ with $m'+n'\leq 2$ provided that (\ref{eq: cond L final}) is satisfied. Note that the occurrence of (\ref{eq: cond L final}) implies that (\ref{eq: cond l for v}) and (\ref{eq: con L 1}) are satisfied. Finally, the proof of Lemma~\ref{th: f is less 1} is concluded by substituting (\ref{eq: f to f bar al term 1 4}), (\ref{eq: f to f bar al term 1 5}), and (\ref{eq: f to f bar al term 1 6}) in (\ref{eq: f to f bar al 1}) and setting $\delta^{*}= 3\delta$.

\section{Proof of Lemma~\ref{lemma: close in all r}}
\label{A15}
Starting from  (\ref{eq: matrix vector dual}) we can write
\begin{align}
&\frac{1}{\mu^{m'+n'}}\big|\big|\fv^{(m',n')}\left(\rv\right)\big|\big|_{2} = \Big|\Big|\left(\Tm^{(m',n')}\left(\rv\right)\right)^{H} \Lm \hv\Big|\Big|_{2}  \nonumber\\
& \leq \big|\big|\Tm^{(m',n')}\left(\rv\right)\big|\big|_{2}  \ ||\Lm||_{2} \ || \hv||_{2} =  \sqrt{R}  \big|\big|\Tm^{(m',n')}\left(\rv\right)\big|\big|_{2}  \ ||\Lm||_{2} \nonumber\\
\label{eq: proof f bo 2}
& \leq   \max_{\substack{j, (m,n)\in \\ \{(0,0),(1,0),(0,1)\}}} \frac{ \sqrt{3} R ||\Lm||_{2} }{\mu^{m+m'+n+n'}} \Big|\Big|  \Dm^{H} \Rm_{(m,n)}^{(m',n')}\left(\rv,\rv_{j}\right) \Dm   \Big|\Big|_{2},
\end{align}
where the last inequality follows from (\ref{eq: G in matrix multi 2}), (\ref{eq: matrix T}), and the union bound. Now, based on Lemma~\ref{l3} and (\ref{eq: lemma proof 1}), we can write
\begin{align}
\label{eq: proof f bo 3}
&\frac{1}{\mu^{m+m'+n+n'}} \Big|\Big|    \Dm^{H} \Rm_{(m,n)}^{(m',n')}\left(\rv,\rv_{j}\right) \Dm   \Big|\Big|_{2}  \nonumber\\
& \leq  \frac{1}{\mu^{m+m'+n+n'}} ||\Dm||_{2}^{2} \Big|\Big| \Rm_{(m,n)}^{(m',n')}\left(\rv,\rv_{j}\right) \Big|\Big|_{2}   \nonumber\\
&\leq  \frac{1}{\mu^{m+m'+n+n'}}\  ||    \Dm\||_{2}^{2}  \Big|\Big| \Rm_{(m,n)}^{(m',n')}\left(\rv,\rv_{j}\right) \Big|\Big| _{F} \leq  \nonumber\\
& \frac{C_{2}}{\sqrt{L}} 12^{\frac{m+m'+n+n'}{2}}   ||\Dm||_{2}^{2} \leq  C_{2} 12^{\frac{m+m'+n+n'}{2}} \sqrt{L} K \widetilde{K}^{2}.
\end{align}
%where the first inequality in (\ref{eq: proof f bo 3}) follows from Lemma~\ref{l3} and  (\ref{eq: lemma proof 1}) while the second inequality is based on (\ref{eq: concetration x}).
Upon substituting (\ref{eq: proof f bo 3}) into (\ref{eq: proof f bo 2}) and manipulating, we obtain
\begin{align}
%\label{eq: proof f bo 4}
\frac{1}{\mu^{m'+n'}}\big|\big|\fv^{(m',n')}\left(\rv\right)\big|\big|_{2}  &\leq  12^{\frac{3}{2}} C_{2}R  K \widetilde{K}^{2} \sqrt{3L} \ ||\Lm||_{2}\nonumber \\
&   = 0.1\tilde{C}_{2}\sqrt{L}||\Lm||_{2}, \nonumber
\end{align}

where we used the fact that $ m+m'+n+n' \leq 3$ $(m'+n'\leq 2)$ and we set $\tilde{C}_{2} =  10 \cdot 12^{\frac{3}{2}} \sqrt{3} C_{2} R  K \widetilde{K}^{2}$. Now, conditioned on $\mathcal{E}_{2}$ with $\epsilon_{1}\in (0,\frac{2}{5}]$ we can write
\begin{align}
&\text{Pr}\left[ \max_{\rv \in [0,1]^{2}, m'+n'\leq 2 }\frac{1}{\mu^{m'+n'}}||\fv^{(m',n')}\left(\rv\right)||_{2} \geq  \frac{\tilde{C}_{2}}{4} \sqrt{L} \right]  \leq \nonumber\\
& \text{Pr}\left[ 0.1\tilde{C}_{2}\sqrt{L} \ ||\Lm||_{2}  \geq \frac{\tilde{C}_{2}}{4} \sqrt{L} \right]  \leq  \text{Pr}\left[ ||\Lm||_{2}  \geq 2.5  \right]   \leq \frac{\delta}{2},\nonumber
\end{align}
where the last inequality holds when (\ref{eq: con L 1}) is satisfied.

\section{Upper Bound on $||\bar{\fv}\left(\rv\right)||_{2}$: Proof of (\ref{eq: temp proof asa})}
\label{A9}
Starting from (\ref{eq: averg poly ex}), and based on the definition of the Euclidean norm function, we can write

\begin{align}
||\bar{\fv}\left(\rv\right)||_{2} &=  \sup_{\xv : \ ||\xv||_{2}=1} \xv^{H} \left(\left(\widebar{\Tm}\left(\rv\right)\right)^{H} \left(\bar{\Lm}\otimes\Id_{\text{K}}\right)\hv\right) \nonumber\\
\label{eq: a9 1}
&=  \sup_{\xv : \ ||\xv||_{2}=1}   \xv^{H}\left(\left(\bar{\tv}\left(\rv\right) \otimes \Id_{\text{K}}\right)^{H} \left(\bar{\Lm}\otimes\Id_{\text{K}}\right)\hv\right)
\\
\label{eq: a9 2}
&=\sup_{\xv : \ ||\xv||_{2}=1} \sum_{j=1}^{R} \left[\bar{\tv}^{H}\left(\rv\right) \bar{\Lm}\right]_{j} \left(\xv^{H} \text{sign}\left(c_{j}\right)\hv_{j}\right) \\
\label{eq: a9 3}
& =\sup_{\xv : \ ||\xv||_{2}=1} \sum_{j=1}^{R} \left[\bar{\tv}^{H}\left(\rv\right) \bar{\Lm}\right]_{j} \left( [\cv]_{j}\xv^{H}\hv_{j}\right),
\end{align}
where (\ref{eq: a9 1}) is based on (\ref{eq: sub T }) while $\left[\bar{\tv}^{H}\left(\rv\right) \bar{\Lm}\right]_{j}$ in (\ref{eq: a9 2}) refers to the $j$-th entry of the vector. Finally, the vector $\cv$ is defined as $\cv = \left[ \text{sign}\left(c_{1}\right), \dots, \text{sign}\left(c_{R}\right) \right]^{T}$. Now, based on the result obtained in \cite[Lemma~C.4]{candes2014towards} and the fact that $\big| [\cv]_{j} \xv^{H}\hv_{j}\big| \leq 1$, we can conclude that 
\begin{equation}
\label{eq: cand re}
||\bar{\fv}\left(\rv\right)||_{2} \leq 0.9958, \ \forall \rv \in \Omega_{\text{far}}.
\end{equation}

\section{Proof of Lemma~\ref{lemma: close}}
\label{proof of lemma: far}
\subsection{Proof of (\ref{eq: Q less 0.9})}
By setting $\epsilon_{3} = 2 \times 10^{-3} $ in Lemma~\ref{th: f is less 1} we have
\begin{align}
\label{eq: f to f lemma 1}
\big|\big|\fv\left(\rv\right)- \bar{\fv}\left(\rv\right)    \big|\big|_{2} \leq 0.002
\end{align}
holds with probability at least $1-\delta^{*}$. On the other hand, we prove in Appendix~\ref{A9} that
\begin{equation}
\label{eq: temp proof asa}
||\bar{\fv}\left(\rv\right)||_{2} \leq  0.9958, \  \forall \rv \in \Omega_{\text{far}}.
\end{equation}
Finally, we can write based on (\ref{eq: f to f lemma 1}) and (\ref{eq: temp proof asa})
\begin{align}
\label{eq: proof close 1}
||\fv\left(\rv\right)||_{2}  &\leq  ||\fv\left(\rv\right)-\bar{\fv}\left(\rv\right)||_{2} +||\bar{\fv}\left(\rv\right)||_{2} \leq  0.9978. 
\end{align}

\subsection{Proof of (\ref{eq: Q2 less 0.9})}
Without loss of generality, we assume that $\bm{0} \in \mathcal{R}$ i.e., $|\rv| \leq 0.2447/N$ based on (\ref{eq: far set}), and that $N \geq 512$. Now, to prove that $||\fv\left(\rv\right)||_{2} < 1, \ \forall \rv \in \Omega_{\text{close}}$, it is enough to show that the normalized Hessian matrix of $ ||\fv\left(\rv\right)||_{2}^{2}$, i.e., 
\begin{equation}
\label{eq: matrix Q}
\frac{1}{\mu^{2}}\Hm = \begin{bmatrix} \frac{\partial^{2}}{ \partial \tau^{2}} \frac{1}{\mu^{2}}||\fv\left(\rv\right)||_{2}^{2}  & \frac{\partial}{\partial f  \partial \tau} \frac{1}{\mu^{2}}||\fv\left(\rv\right)||_{2}^{2}  \\ \frac{\partial}{\partial f  \partial \tau} \frac{1}{\mu^{2}}||\fv\left(\rv\right)||_{2}^{2}   & \frac{\partial^{2}}{ \partial f^{2}} \frac{1}{\mu^{2}}||\fv\left(\rv\right)||_{2}^{2}    \end{bmatrix}
\end{equation}
is negative definite $\forall \rv \in \Omega_{\text{close}}$. From the properties of $2 \times 2$ block matrices, we know that $\Hm$ will become a negative definite matrix if the following two conditions are satisfied: 
\begin{align}
\label{eq: negative cond}
&\frac{1}{\mu^{2}}\text{Tr}\left(\Hm\right) = \frac{\partial^{2}}{ \partial \tau^{2}} \Big|\Big|\frac{1}{\mu}\fv\left(\rv\right)\Big|\Big|_{2}^{2}+  \frac{\partial^{2}}{ \partial f^{2}} \Big|\Big|\frac{1}{\mu}\fv\left(\rv\right)\Big|\Big|_{2}^{2}  < 0
\end{align}
\begin{align}
\label{eq: negative cond 2}
\frac{1}{\mu^{2}}\text{det} \left(\Hm\right) &= \left(\frac{\partial^{2}}{ \partial \tau^{2}} \Big|\Big|\frac{1}{\mu}\fv\left(\rv\right)\Big|\Big|_{2}^{2}\right) \left(\frac{\partial^{2}}{ \partial f^{2}} \Big|\Big|\frac{1}{\mu}\fv\left(\rv\right)\Big|\Big|_{2}^{2}\right) \nonumber\\
&- \left(\frac{\partial}{\partial f  \partial \tau} \Big|\Big|\frac{1}{\mu}\fv\left(\rv\right)\Big|\Big|_{2}^{2}\right)^{2} > 0.
\end{align}
Note that (\ref{eq: negative cond}) is nothing but the normalized sum of the eigenvalues while (\ref{eq: negative cond 2}) is equal to their normalized product.

To show (\ref{eq: negative cond}) and (\ref{eq: negative cond 2}), we first derive in Appendix~\ref{A10} the following bounds $\forall \rv \in \Omega_{\text{close}}$ and $N \geq 512$
\begin{align}
\label{eq: bounds to app 1}
&\big|\big|  \bar{\fv}\left(\rv\right)\big|\big|_{2} \leq 1.1295 + 0.0475/N  \\
\label{eq: bounds to app 2}
&\big|\big|  \bar{\fv}^{(1,0)}\left(\rv\right)\big|\big|_{2} \leq  0.8874+0.2148N \\
\label{eq: bounds to app 3}
&\big|\big|  \bar{\fv}^{(1,1)}\left(\rv\right)\big|\big|_{2} \leq  0.8459N+0.2129N^{2} \\
\label{eq: bounds to app 4}
&\big|\big|  \bar{\fv}^{(2,0)}\left(\rv\right)\big|\big|_{2} \leq 0.5025N+ 3.8845 N^{2}. 
\end{align}
Note that the bounds in (\ref{eq: bounds to app 2}) and (\ref{eq: bounds to app 4}) also hold for  $\big|\big|  \bar{\fv}^{(0,1)}\left(\rv\right)\big|\big|_{2}$ and $\big|\big|  \bar{\fv}^{(0,2)}\left(\rv\right)\big|\big|_{2}$, respectively.
\subsubsection{Showing (\ref{eq: negative cond})} 
Starting from the first term in (\ref{eq: negative cond}), we can write
\begin{align}
\label{eq: proving deri}
&\frac{\partial^{2}}{ \partial \tau^{2}} \Big|\Big|\frac{1}{\mu}\fv\left(\rv\right)\Big|\Big|_{2}^{2} = \frac{\partial}{ \partial \tau} \frac{2}{\mu^{2}} \left\langle\fv^{(1,0)}\left(\rv\right),\fv\left(\rv\right)\right\rangle_{\mathbb{R}} \nonumber\\
&  = 2 \Big|\Big|\frac{1}{\mu}\fv^{(1,0)}\left(\rv\right)\Big|\Big|_{2}^{2}+\frac{2}{\mu^{2}}\text{Re}\left[\left(\fv^{(2,0)}\left(\rv\right)\right)^{H}\fv\left(\rv\right)\right].
\end{align}
Now, the first term in (\ref{eq: proving deri}) can be bounded as
\begin{align}
\label{eq: proving deri 3}
&\Big|\Big|\frac{1}{\mu}\fv^{(1,0)}\left(\rv\right)\Big|\Big|_{2}^{2} \leq \Big|\Big|\frac{1}{\mu}\left(\fv^{(1,0)}\left(\rv\right)-\bar{\fv}^{(1,0)}\left(\rv\right)\right)\Big|\Big|_{2}^{2} \nonumber\\
& + \Big|\Big|\frac{1}{\mu}\bar{\fv}^{(1,0)}\left(\rv\right)\Big|\Big|_{2}^{2}  \leq \epsilon_{3}^{2}+\frac{1}{\mu^{2}} \left(0.8874+0.2148N\right)^{2}\nonumber\\
& \leq  \epsilon_{3}^{2}+ 0.0142,
\end{align}
where the first inequality is from triangular inequality while the last inequality is based on Lemma~\ref{th: f is less 1}, (\ref{eq: bounds to app 2}), and the fact that $\mu^{2} > \frac{\pi^{2}}{3}N^{2}$.

Next, we consider obtaining an upper bound for the second term in (\ref{eq: proving deri}) as
\begin{align}
&\frac{1}{\mu^{2}}\text{Re}\left[\left(\fv^{(2,0)}\left(\rv\right)\right)^{H}\fv\left(\rv\right)\right] =  \text{Re}\bigg[\frac{1}{\mu^{2}}\left( \fv^{(2,0)}\left(\rv\right) \right.\nonumber\\
&\left.-\bar{\fv}^{(2,0)}\left(\rv\right)+ \bar{\fv}^{(2,0)}\left(\rv\right)\right)^{H} \left( \fv\left(\rv\right)+ \bar{\fv}\left(\rv\right)-\bar{\fv}\left(\rv\right) \right) \bigg] = \nonumber\\
& \text{Re}\left[\frac{1}{\mu^{2}}\left( \fv^{(2,0)}\left(\rv\right) -\bar{\fv}^{(2,0)}\left(\rv\right) \right)^{H} \left( \fv\left(\rv\right)-\bar{\fv}\left(\rv\right) \right) \right] \nonumber\\
&+ \text{Re}\left[\frac{1}{\mu^{2}}\left(\bar{\fv}^{(2,0)}\left(\rv\right) \right)^{H} \bar{\fv}\left(\rv\right) \right] \nonumber\\
&+\text{Re}\left[\frac{1}{\mu^{2}}\left( \fv^{(2,0)}\left(\rv\right) -\bar{\fv}^{(2,0)}\left(\rv\right) \right)^{H} \bar{\fv}\left(\rv\right)  \right] \nonumber\\
&+ \text{Re}\left[\frac{1}{\mu^{2}}\left( \bar{\fv}^{(2,0)}\left(\rv\right) \right)^{H} \left( \fv\left(\rv\right)-\bar{\fv}\left(\rv\right) \right) \right] \nonumber \\
\label{eq: bound the real fuc 1}
&
\leq \epsilon_{3}^{2}  + \text{Re}\left[\frac{1}{\mu^{2}}\left(\bar{\fv}^{(2,0)}\left(\rv\right)\right)^{H} \bar{\fv}\left(\rv\right) \right]+ 1.129\epsilon_{3}+ 1.181\epsilon_{3}
\\
\label{eq: bound the real fuc 2}
&\leq  \epsilon_{3}^{2} +  2.31\epsilon_{3} -0.307,
\end{align}
where the inequality in (\ref{eq: bound the real fuc 1}) is obtained by using Lemma~\ref{th: f is less 1}, (\ref{eq: bounds to app 1}), (\ref{eq: bounds to app 4}), and the fact that $\mu^{2} > \frac{\pi^{2}}{3}N^{2}$. Finally, the inequality in (\ref{eq: bound the real fuc 2}) is based on
\begin{equation}
\label{eq: term bound Re}
\text{Re}\left[\frac{1}{\mu^{2}}\left(\bar{\fv}^{(2,0)}\left(\rv\right)\right)^{H} \bar{\fv}\left(\rv\right) \right] \leq -0.307
\end{equation}
with its proof being provided in Appendix~\ref{A10}. Now, by substituting (\ref{eq: proving deri 3}) and (\ref{eq: bound the real fuc 2}) in (\ref{eq: proving deri}) and then manipulating we obtain
\begin{align}
\label{eq: negative cond veri}
&\frac{1}{\mu^{2}}\text{Tr}\left(\Hm\right)\leq  8\epsilon_{3}^{2}+9.24\epsilon_{3}-1.171.
\end{align}
The above expression can be easily shown to be strictly negative for all $\epsilon_{3}\leq 0.1$. 
\subsubsection{Showing (\ref{eq: negative cond 2})} Starting from the second term in (\ref{eq: negative cond 2}), we can write
\begin{align}
\label{eq: proving deri 5}
\frac{\partial}{\partial f  \partial \tau} \Big|\Big|\frac{1}{\mu}\fv\left(\rv\right)\Big|\Big|_{2}^{2} &= \frac{2}{\mu^{2}} \left\langle \fv^{(1,0)}\left(\rv\right),\fv^{(0,1)} \left(\rv\right)\right\rangle_{\mathbb{R}}\nonumber\\
&+ \frac{2}{\mu^{2}} \left\langle \fv^{(1,1)}\left(\rv\right),\fv\left(\rv\right) \right\rangle_{\mathbb{R}}.
\end{align}
The first term in (\ref{eq: proving deri 5}) can be upper bounded by
\begin{align}
\label{eq: proving deri 6}
&\frac{1}{\mu^{2}} \left\langle \fv^{(1,0)}\left(\rv\right),\fv^{(0,1)} \left(\rv\right)\right\rangle_{\mathbb{R}}  = \nonumber\\
&\text{Re}\left[\frac{1}{\mu^{2}}\left( \fv^{(1,0)}\left(\rv\right) -\bar{\fv}^{(1,0)}\left(\rv\right) \right)^{H} \left( \fv^{(0,1)}\left(\rv\right)-\bar{\fv}^{(0,1)}\left(\rv\right) \right) \right] \nonumber\\
&+ \text{Re}\left[\frac{1}{\mu^{2}}\left(\bar{\fv}^{(1,0)}\left(\rv\right) \right)^{H} \bar{\fv}^{(0,1)}\left(\rv\right) \right] \nonumber\\
&+\text{Re}\left[\frac{1}{\mu^{2}}\left( \fv^{(1,0)}\left(\rv\right) -\bar{\fv}^{(1,0)}\left(\rv\right) \right)^{H} \bar{\fv}^{(0,1)}\left(\rv\right)  \right] \nonumber\\
&+ \text{Re}\left[\frac{1}{\mu^{2}}\left( \bar{\fv}^{(1,0)}\left(\rv\right) \right)^{H} \left( \fv^{(0,1)}\left(\rv\right)-\bar{\fv}^{(0,1)}\left(\rv\right) \right) \right] \nonumber\\
&\leq \epsilon_{3}^{2} + 0.238\epsilon_{3}+0.0142, 
\end{align}
where the last inequality follows from Lemma~\ref{th: f is less 1}, (\ref{eq: bounds to app 2}), and the fact that $\mu^{2} > \frac{\pi^{2}}{3}N^{2}$. By following the same steps that led to (\ref{eq: proving deri 6}), we can show using (\ref{eq: bounds to app 3}) that
\begin{align}
\label{eq: proving deri 7}
&\frac{1}{\mu^{2}} \left\langle \fv^{(1,1)}\left(\rv\right),\fv \left(\rv\right)\right\rangle_{\mathbb{R}}  \leq  \epsilon_{3}^{2} + 1.1948\epsilon_{3}+0.0736.
\end{align}
Now, substituting (\ref{eq: proving deri 6}) and (\ref{eq: proving deri 7}) in (\ref{eq: proving deri 5}), then manipulating, we obtain
%\begin{align}
%\label{eq: ddddre}
%\frac{1}{2}\frac{\partial}{\partial f  \partial \tau} \Big|\Big|\frac{1}{\mu}\fv\left(\rv\right)\Big|\Big|_{2}^{2} &\leq 2\epsilon_{3}^{2}+1.4338\epsilon_{3}+0.0879.
%\end{align}
\begin{align}
\label{eq: ddddre}
\frac{\partial}{\partial f  \partial \tau} \left|\left|\frac{1}{\mu}\fv\left(\rv\right)\right|\right|_{2}^{2} &\leq 4\epsilon_{3}^{2}+2.865\epsilon_{3}+0.175.
\end{align}
Finally, by using the bound obtained for (\ref{eq: proving deri}) with that in (\ref{eq: ddddre}), we can easily show that (\ref{eq: negative cond 2}) is satisfied for all $\epsilon_{3} \leq 0.051$. 
This completes the proof of (\ref{eq: Q2 less 0.9}).

\section{Various Important Results}
\label{A10}
The proofs in this appendix are based on the assumptions that $\bm{0} \in \mathcal{R}$ and $N \geq 512$. Starting from the results obtained in \cite[Lemma~2.3 and Section~C.2]{candes2014towards}, we can show that for $|\rv|\leq 0.2447/N$ and $N \geq 512$ we have
\begin{align}
\label{eq: app candice bounds}
&\big| \widebar{M}^{(1,0)}\left(\rv\right) \big|  \leq 0.8113,  \hspace{21pt} \big| \widebar{M}^{(1,1)}\left(\rv\right) \big|  \leq 0.6531N, \nonumber\\
&{{\big| \widebar{M}^{(2,0)}\left(\rv\right) \big|  \leq 3.393N^{2}}},  \hspace{13pt} \big| \widebar{M}^{(2,1)}\left(\rv\right) \big|  \leq 2.669N^{2}, \nonumber\\
&\big| \widebar{M}^{(3,0)}\left(\rv\right) \big|  \leq 8.070N^{3},  \hspace{13pt}  \widebar{M}^{(2,0)}\left(\rv\right)   \leq -2.097N^{2}, \nonumber\\
& \widebar{M}\left(\rv\right) \geq 0.8113,     \hspace{47pt} |\widebar{M}\left(\rv\right)| \leq 1, 
 \end{align}
where $\widebar{M}^{(m',n')}\left(\rv\right)$ is as defined in (\ref{eq: det G}). Moreover, by defining 
\begin{equation}
\label{eq: temp app}
\tilde{Z}^{(m',n')} \left(\rv\right) := \sum_{\rv_{j}\in \mathcal{R} \setminus \{\bm{0}\}}\big|\widebar{M}^{(m',n')}\left(\rv-\rv_{j}\right)\big|
\end{equation}
we can obtain the following bounds based on \cite[Section~C.2]{candes2014towards}
\begin{align}
\label{eq: bounds candice 3}
&\tilde{Z}^{(0,0)} \left(\rv\right) \leq 6.405\times 10^{-2},  \hspace{10pt} \tilde{Z}^{(1,0)} \left(\rv\right) \leq 0.1047N, \nonumber\\
&\tilde{Z}^{(2,0)} \left(\rv\right) \leq 0.4019N,  \hspace{27pt} \tilde{Z}^{(1,1)} \left(\rv\right) \leq 0.1642N^{2}, \nonumber\\
&\tilde{Z}^{(2,1)} \left(\rv\right) \leq 0.6751N^{3}, \hspace{22pt} \tilde{Z}^{(3,0)} \left(\rv\right) \leq  1.574N^{3}.
\end{align}
Finally, we can also conclude based on \cite{candes2014towards} and \cite{yang2014exact}
\begin{align}
\label{eq: app bound can 2}
&||\bar{\alphav}_{j}||_{2} \leq \alpha_{\text{max}} = 1+5.577\times 10^{-2} \nonumber\\
&||\bar{\alphav}_{j}||_{2}\geq \alpha_{\text{min}} = 1-5.577\times 10^{-2} \nonumber\\
&||\bar{\betav}_{j}||_{2} \leq \beta_{\text{max}} = \frac{2.93}{N}\times 10^{-2}  \nonumber\\
&||\bar{\gammav}_{j}||_{2} \leq \gamma_{\text{max}} = \frac{2.93}{N}\times 10^{-2}.
\end{align}
\subsection{Proofs of (\ref{eq: bounds to app 1}) and (\ref{eq: bounds to app 2})}
In this section, we will provide the proofs of (\ref{eq: bounds to app 1}) and (\ref{eq: bounds to app 2}) as those of (\ref{eq: bounds to app 3}) and (\ref{eq: bounds to app 4}) follow the same steps. Starting from (\ref{eq: poly deriv expected}) we can write
%\begin{align}
%\big|\big|\bar{\fv}\left(\rv\right)\big|\big|_{2} &= \Bigg|\Bigg|\sum_{j=1}^{R} \widebar{M}^{(0,0)}\left(\rv-\rv_{j}\right) \bar{\alphav}_{j}+ \widebar{M}^{(1,0)}\left(\rv-\rv_{j}\right) \bar{\betav}_{j} + \widebar{M}^{(0,1)}\left(\rv-\rv_{j}\right) \bar{\gammav}_{j}\Bigg|\Bigg|_{2} \nonumber\\  
%& \leq  \alpha_{\text{max}} \left(\big|\widebar{M}^{(0,0)}\left(\rv\right)\big| + \tilde{Z}^{(0,0)} \left(\rv\right)\right) +2\beta_{\text{max}}\left(\big|\widebar{M}^{(1,0)}\left(\rv\right)\big| + \tilde{Z}^{(1,0)} \left(\rv\right)\right) \leq  1.1295 + 0.0475/N, \nonumber
%\end{align}
\begin{align}
&\big|\big|\bar{\fv}\left(\rv\right)\big|\big|_{2} = \Bigg|\Bigg|\sum_{j=1}^{R} \widebar{M}^{(0,0)}\left(\rv-\rv_{j}\right) \bar{\alphav}_{j}+ \widebar{M}^{(1,0)}\left(\rv-\rv_{j}\right) \bar{\betav}_{j} \nonumber\\  
&+ \widebar{M}^{(0,1)}\left(\rv-\rv_{j}\right) \bar{\gammav}_{j}\Bigg|\Bigg|_{2} \leq  \alpha_{\text{max}} \left(\big|\widebar{M}^{(0,0)}\left(\rv\right)\big| + \tilde{Z}^{(0,0)} \left(\rv\right)\right) \nonumber\\
& +2\beta_{\text{max}}\left(\big|\widebar{M}^{(1,0)}\left(\rv\right)\big| + \tilde{Z}^{(1,0)} \left(\rv\right)\right) \leq  1.1295 + 0.0475/N, \nonumber
\end{align}
where the last inequality is based on (\ref{eq: app candice bounds}), (\ref{eq: bounds candice 3}), and (\ref{eq: app bound can 2}).
On the other hand, we can also obtain
\begin{align}
&\big|\big|\bar{\fv}^{(1,0)}\left(\rv\right)\big|\big|_{2}  \leq  
\alpha_{\text{max}} \left(\big|\widebar{M}^{(1,0)}\left(\rv\right)\big| + \tilde{Z}^{(1,0)} \left(\rv\right) \right)  \nonumber\\
& +\beta_{\text{max}}\left(\big|\widebar{M}^{(2,0)}\left(\rv\right)\big| +\tilde{Z}^{(2,0)} \left(\rv\right) \right) \nonumber\\
&+\gamma_{\text{max}}\left(\big|\widebar{M}^{(1,1)}\left(\rv\right)\big| + \tilde{Z}^{(1,1)} \left(\rv\right) \right) \leq 0.8874 +0.2148N. \nonumber
\end{align}
\subsection{Proof of (\ref{eq: term bound Re})}
Starting from the expression in (\ref{eq: poly deriv expected}), we can write after some algebraic manipulations
\begin{align}
%\label{eq: a10 proof 3}
&\text{Re}\left[\frac{1}{\mu^{2}}\left(\bar{\fv}^{(2,0)}\left(\rv\right)\right)^{H} \bar{\fv}\left(\rv\right) \right] = \frac{1}{\mu^{2}}\text{Re} \left[T_{1}\left(\rv\right)+T_{2}\left(\rv\right)\right], \nonumber
\end{align}
where 
%\begin{align}
%T_{1}\left(\rv\right)&= ||\bar{\alphav}_{l}||_{2}^{2}\widebar{M}^{(2,0)}\left(\rv\right) \widebar{M}^{(0,0)}\left(\rv\right) + \bar{\alphav}_{l}^{H} \widebar{M}^{(2,0)}\left(\rv\right) \sum_{\rv_{j}\in \mathcal{R} \setminus \{\bm{0}\}} \widebar{M}^{(0,0)}\left(\rv-\rv_{j}\right)\bar{\alphav}_{j}\nonumber \\
%& + \bar{\alphav}_{l}^{H} \bar{\betav}_{l} \widebar{M}^{(2,0)}\left(\rv\right) \widebar{M}^{(1,0)}\left(\rv\right)+  \bar{\alphav}_{l}^{H} \widebar{M}^{(2,0)}\left(\rv\right) \sum_{\rv_{j}\in \mathcal{R} \setminus \{\bm{0}\}} \widebar{M}^{(1,0)}\left(\rv-\rv_{j}\right)\bar{\betav}_{j} \nonumber\\
%&+ \bar{\alphav}_{l}^{H}\bar{\gammav}_{l} \widebar{M}^{(2,0)}\left(\rv\right) \widebar{M}^{(0,1)}\left(\rv\right)   + \bar{\alphav}_{l}^{H} \widebar{M}^{(2,0)}\left(\rv\right) \sum_{\rv_{j}\in \mathcal{R} \setminus \{\bm{0}\}} \widebar{M}^{(0,1)}\left(\rv-\rv_{j}\right)\bar{\gammav}_{j}, \nonumber
%\end{align}
%\begin{align}
%T_{2}\left(\rv\right)&=\left( \sum_{\rv_{j}\in \mathcal{R} \setminus \{\bm{0}\}} \widebar{M}^{(2,0)}\left(\rv-\rv_{j}\right)\bar{\alphav}_{j}\right)^{H} \bar{\fv}\left(\rv\right) +\left( \bar{\betav}_{l} \widebar{M}^{(3,0)}\left(\rv\right)+ \sum_{\rv_{j}\in \mathcal{R} \setminus \{\bm{0}\}} \widebar{M}^{(3,0)}\left(\rv-\rv_{j}\right)\bar{\betav}_{j}\right)^{H}\bar{\fv}\left(\rv\right) \nonumber\\
%& + \left(\bar{\gammav}_{l} \widebar{M}^{(2,1)}\left(\rv\right) +  \sum_{\rv_{j}\in \mathcal{R} \setminus \{\bm{0}\}} \widebar{M}^{(2,1)}\left(\rv-\rv_{j}\right)\bar{\gammav}_{j}\right)^{H} \bar{\fv}\left(\rv\right), \nonumber
%\end{align}
\begin{align}
%\label{eq: a10 proof 4}
&T_{1}\left(\rv\right)= ||\bar{\alphav}_{l}||_{2}^{2}\widebar{M}^{(2,0)}\left(\rv\right) \widebar{M}^{(0,0)}\left(\rv\right) + \bar{\alphav}_{l}^{H} \widebar{M}^{(2,0)}\left(\rv\right) \times 
\nonumber \\
& \sum_{\rv_{j}\in \mathcal{R} \setminus \{\bm{0}\}} \widebar{M}^{(0,0)}\left(\rv-\rv_{j}\right)\bar{\alphav}_{j} + \bar{\alphav}_{l}^{H} \bar{\betav}_{l} \widebar{M}^{(2,0)}\left(\rv\right) \widebar{M}^{(1,0)}\left(\rv\right)+ \nonumber\\
& \bar{\alphav}_{l}^{H} \widebar{M}^{(2,0)}\left(\rv\right) \hspace{-5pt}\sum_{\rv_{j}\in \mathcal{R} \setminus \{\bm{0}\}} \hspace{-5pt} \widebar{M}^{(1,0)}\left(\rv-\rv_{j}\right)\bar{\betav}_{j}+ \bar{\alphav}_{l}^{H}\bar{\gammav}_{l} \widebar{M}^{(2,0)}\left(\rv\right) \times  \nonumber\\
&\widebar{M}^{(0,1)}\left(\rv\right)   + \bar{\alphav}_{l}^{H} \widebar{M}^{(2,0)}\left(\rv\right) \sum_{\rv_{j}\in \mathcal{R} \setminus \{\bm{0}\}} \widebar{M}^{(0,1)}\left(\rv-\rv_{j}\right)\bar{\gammav}_{j}, \nonumber
\end{align}
\begin{align}
%\label{eq: a10 proof 5}
&T_{2}\left(\rv\right)=\left( \sum_{\rv_{j}\in \mathcal{R} \setminus \{\bm{0}\}} \widebar{M}^{(2,0)}\left(\rv-\rv_{j}\right)\bar{\alphav}_{j}\right)^{H} \bar{\fv}\left(\rv\right) \nonumber\\
&+\left( \bar{\betav}_{l} \widebar{M}^{(3,0)}\left(\rv\right)+ \sum_{\rv_{j}\in \mathcal{R} \setminus \{\bm{0}\}} \widebar{M}^{(3,0)}\left(\rv-\rv_{j}\right)\bar{\betav}_{j}\right)^{H}\bar{\fv}\left(\rv\right) \nonumber\\
& + \left(\bar{\gammav}_{l} \widebar{M}^{(2,1)}\left(\rv\right) +  \sum_{\rv_{j}\in \mathcal{R} \setminus \{\bm{0}\}} \widebar{M}^{(2,1)}\left(\rv-\rv_{j}\right)\bar{\gammav}_{j}\right)^{H} \bar{\fv}\left(\rv\right), \nonumber
\end{align}

while $l$ is the index at which $\rv_{l}=\bm{0}$. Now, by using the bounds in (\ref{eq: app candice bounds}), (\ref{eq: bounds candice 3}), and (\ref{eq: app bound can 2}), and after some algebraic manipulations, we can show that
\begin{align}
%\label{eq: T1 bound }
\text{Re} \left[T_{1}\left(\rv\right)\right] &\leq -1.346N^{2}+0.17N \nonumber
\end{align}
and
\begin{align}
%\label{eq: T2 bound }
\text{Re} \left[T_{2}\left(\rv\right)\right] & \leq 0.331N^{2}+0.556N.  \nonumber 
\end{align}
Therefore, we can finally conclude that 
%\begin{align}
%&\text{Re}\left[\frac{1}{\mu^{2}}\left(\bar{\fv}^{(2,0)}\left(\rv\right)\right)^{H} \bar{\fv}\left(\rv\right) \right] \leq \frac{1}{\mu^{2}}(-1.02N^{2}+0.726N) \leq -0.307. \nonumber
%\end{align}
\begin{align}
&\text{Re}\left[\frac{1}{\mu^{2}}\left(\bar{\fv}^{(2,0)}\left(\rv\right)\right)^{H} \bar{\fv}\left(\rv\right) \right] \nonumber\\
&\leq \frac{1}{\mu^{2}}(-1.02N^{2}+0.726N) \leq -0.307. \nonumber
\end{align}

\end{document}